\newif\iftwocol
\twocoltrue 

\iftwocol
\documentclass[journal,twoside]{IEEEtran}
\else
\documentclass[peerreviewca, draftclsnofoot, onecolumn]{IEEEtran}
\fi

\usepackage{tipa}

\usepackage{cite}
\ifCLASSINFOpdf
   \usepackage[pdftex]{graphicx}
\else
   \usepackage[dvips]{graphicx}
\fi
\usepackage{epstopdf}
\graphicspath{{./Images/}}

\usepackage[cmex10]{amsmath}
\usepackage{amssymb}
\usepackage{amsthm}
\usepackage{amsfonts}
\usepackage{textcomp}
\usepackage{bm}
\usepackage{cancel}
\usepackage{upgreek}
\usepackage{algorithm}
\usepackage{algorithmicx}
\usepackage{algpseudocode}
\usepackage{mathrsfs}
\usepackage{mathtools}

\usepackage{mleftright}
\mleftright

\usepackage{booktabs}

\usepackage{marvosym}

\usepackage{mathtools}
\DeclarePairedDelimiter\abs{\lvert}{\rvert}%
\DeclarePairedDelimiter\norm{\lVert}{\rVert}%

\makeatletter
\let\oldabs\abs
\def\abs{\@ifstar{\oldabs}{\oldabs*}}
\let\oldnorm\norm
\def\norm{\@ifstar{\oldnorm}{\oldnorm*}}
\makeatother

\newcommand{\set}[1]{\left\lbrace #1\right\rbrace}
\newcommand{\operator}[1]{\left\lbrace #1\right\rbrace}
\newcommand{\canonical}[1]{\left( #1\right)}

\newcommand{\vertOne}{\vphantom{0^0}}

\usepackage{qrcode}

\usepackage{multicol}
\usepackage{float}
\usepackage{url}
\usepackage{cite}
\usepackage[tight,footnotesize]{subfigure}

\usepackage{xcolor}
\definecolor{MyGrey5}{RGB}{150,150,150}
\definecolor{MyGrey}{RGB}{234,234,234}
\definecolor{MyBlue}{RGB}{0,0,255}
\definecolor{DarkSlateGray}{rgb}{0.1836,  0.3086  ,  0.3086}
\definecolor{Chocolate}{rgb}{0.8203    ,  0.4102  ,  0.1172}
\definecolor{FireBrick}{rgb}{ 0.6953   ,  0.1328  ,  0.1328}
\definecolor{MediumBlue}{rgb}{ 0       ,  0       ,  0.8008}
\definecolor{MyGreen}{rgb}{ 0          ,  0.5000  ,  0     }
\definecolor{MyRed}{rgb}{ 1        ,  0.0000  ,  0     }
\definecolor{MyPurple}{rgb}{0.5     ,    0  ,0.5}
\definecolor{WatermarkGrey}{rgb}{0.6,0.6,0.6}
\definecolor{WatermarkGrey2}{rgb}{0.8,0.8,0.8}
\definecolor{WatermarkGrey3}{rgb}{0.9,0.9,0.9}

\definecolor{MyBlack}{rgb}{ 0        ,  0  ,  0    }

\definecolor{myWhite}{rgb}{1,1,1}
\definecolor{mySnow}{rgb}{1,0.97917,0.97917}
\definecolor{myHoneydew}{rgb}{0.9375,1,0.9375}
\definecolor{myMintCream}{rgb}{0.95833,1,0.97917}
\definecolor{myAzure}{rgb}{0.9375,1,1}
\definecolor{myAliceBlue}{rgb}{0.9375,0.97083,1}
\definecolor{myGhostWhite}{rgb}{0.97083,0.97083,1}
\definecolor{myWhiteSmoke}{rgb}{0.95833,0.95833,0.95833}
\definecolor{mySeashell}{rgb}{1,0.95833,0.92969}
\definecolor{myBeige}{rgb}{0.95833,0.95833,0.85938}
\definecolor{myOldLace}{rgb}{0.99167,0.95833,0.89844}
\definecolor{myFloralWhite}{rgb}{1,0.97917,0.9375}
\definecolor{myIvory}{rgb}{1,1,0.9375}
\definecolor{myAntiqueWhite}{rgb}{0.97917,0.91797,0.83984}
\definecolor{myLinen}{rgb}{0.97917,0.9375,0.89844}
\definecolor{myLavenderBlush}{rgb}{1,0.9375,0.95833}
\definecolor{myMistyRose}{rgb}{1,0.89063,0.87891}
\definecolor{myGray}{rgb}{0.5,0.5,0.5}
\definecolor{myGainsboro}{rgb}{0.85938,0.85938,0.85938}
\definecolor{myLightGray}{rgb}{0.82422,0.82422,0.82422}
\definecolor{mySilver}{rgb}{0.75,0.75,0.75}
\definecolor{myDarkGray}{rgb}{0.66016,0.66016,0.66016}
\definecolor{myDimGray}{rgb}{0.41016,0.41016,0.41016}
\definecolor{myLightSlateGray}{rgb}{0.46484,0.53125,0.59766}
\definecolor{mySlateGray}{rgb}{0.4375,0.5,0.5625}
\definecolor{myDarkSlateGray}{rgb}{0.18359,0.30859,0.30859}
\definecolor{myBlack}{rgb}{0,0,0}
\definecolor{myRed}{rgb}{1,0,0}
\definecolor{myLightSalmon}{rgb}{1,0.625,0.47656}
\definecolor{mySalmon}{rgb}{0.97917,0.5,0.44531}
\definecolor{myDarkSalmon}{rgb}{0.91016,0.58594,0.47656}
\definecolor{myLightCoral}{rgb}{0.9375,0.5,0.5}
\definecolor{myIndianRed}{rgb}{0.80078,0.35938,0.35938}
\definecolor{myCrimson}{rgb}{0.85938,0.078125,0.23438}
\definecolor{myFireBrick}{rgb}{0.69531,0.13281,0.13281}
\definecolor{myDarkRed}{rgb}{0.54297,0,0}
\definecolor{myPink}{rgb}{1,0.75,0.79297}
\definecolor{myLightPink}{rgb}{1,0.71094,0.75391}
\definecolor{myHotPink}{rgb}{1,0.41016,0.70313}
\definecolor{myDeepPink}{rgb}{1,0.078125,0.57422}
\definecolor{myPaleVioletRed}{rgb}{0.85547,0.4375,0.57422}
\definecolor{myMediumVioletRed}{rgb}{0.77734,0.082031,0.51953}
\definecolor{myOrange}{rgb}{1,0.64453,0}
\definecolor{myDarkOrange}{rgb}{1,0.54688,0}
\definecolor{myCoral}{rgb}{1,0.49609,0.3125}
\definecolor{myTomato}{rgb}{1,0.38672,0.27734}
\definecolor{myOrangeRed}{rgb}{1,0.26953,0}
\definecolor{myYellow}{rgb}{1,1,0}
\definecolor{myLightYellow}{rgb}{1,1,0.875}
\definecolor{myLemonChiffon}{rgb}{1,0.97917,0.80078}
\definecolor{myLightGoldenrodYellow}{rgb}{0.97917,0.97917,0.82031}
\definecolor{myPapayaWhip}{rgb}{1,0.93359,0.83203}
\definecolor{myMoccasin}{rgb}{1,0.89063,0.70703}
\definecolor{myPeachPuff}{rgb}{1,0.85156,0.72266}
\definecolor{myPaleGoldenrod}{rgb}{0.92969,0.90625,0.66406}
\definecolor{myKhaki}{rgb}{0.9375,0.89844,0.54688}
\definecolor{myDarkKhaki}{rgb}{0.73828,0.71484,0.41797}
\definecolor{myGold}{rgb}{1,0.83984,0}
\definecolor{myBrown}{rgb}{0.64453,0.16406,0.16406}
\definecolor{myCornsilk}{rgb}{1,0.97083,0.85938}
\definecolor{myBlanchedAlmond}{rgb}{1,0.91797,0.80078}
\definecolor{myBisque}{rgb}{1,0.89063,0.76563}
\definecolor{myNavajoWhite}{rgb}{1,0.86719,0.67578}
\definecolor{myWheat}{rgb}{0.95833,0.86719,0.69922}
\definecolor{myBurlyWood}{rgb}{0.86719,0.71875,0.52734}
\definecolor{myTan}{rgb}{0.82031,0.70313,0.54688}
\definecolor{myRosyBrown}{rgb}{0.73438,0.55859,0.55859}
\definecolor{mySandyBrown}{rgb}{0.95417,0.64063,0.375}
\definecolor{myGoldenrod}{rgb}{0.85156,0.64453,0.125}
\definecolor{myDarkGoldenrod}{rgb}{0.71875,0.52344,0.042969}
\definecolor{myPeru}{rgb}{0.80078,0.51953,0.24609}
\definecolor{myChocolate}{rgb}{0.82031,0.41016,0.11719}
\definecolor{mySaddleBrown}{rgb}{0.54297,0.26953,0.074219}
\definecolor{mySienna}{rgb}{0.625,0.32031,0.17578}
\definecolor{myMaroon}{rgb}{0.5,0,0}
\definecolor{myGreen}{rgb}{0,0.5,0}
\definecolor{myPaleGreen}{rgb}{0.59375,0.98333,0.59375}
\definecolor{myLightGreen}{rgb}{0.5625,0.92969,0.5625}
\definecolor{myYellowGreen}{rgb}{0.60156,0.80078,0.19531}
\definecolor{myGreenYellow}{rgb}{0.67578,1,0.18359}
\definecolor{myChartreuse}{rgb}{0.49609,1,0}
\definecolor{myLawnGreen}{rgb}{0.48438,0.9875,0}
\definecolor{myLime}{rgb}{0,1,0}
\definecolor{myLimeGreen}{rgb}{0.19531,0.80078,0.19531}
\definecolor{myMediumSpringGreen}{rgb}{0,0.97917,0.60156}
\definecolor{mySpringGreen}{rgb}{0,1,0.49609}
\definecolor{myMediumAquamarine}{rgb}{0.39844,0.80078,0.66406}
\definecolor{myAquamarine}{rgb}{0.49609,1,0.82813}
\definecolor{myLightSeaGreen}{rgb}{0.125,0.69531,0.66406}
\definecolor{myMediumSeaGreen}{rgb}{0.23438,0.69922,0.44141}
\definecolor{mySeaGreen}{rgb}{0.17969,0.54297,0.33984}
\definecolor{myDarkSeaGreen}{rgb}{0.55859,0.73438,0.55859}
\definecolor{myForestGreen}{rgb}{0.13281,0.54297,0.13281}
\definecolor{myDarkGreen}{rgb}{0,0.39063,0}
\definecolor{myOliveDrab}{rgb}{0.41797,0.55469,0.13672}
\definecolor{myOlive}{rgb}{0.5,0.5,0}
\definecolor{myDarkOliveGreen}{rgb}{0.33203,0.41797,0.18359}
\definecolor{myTeal}{rgb}{0,0.5,0.5}
\definecolor{myBlue}{rgb}{0,0,1}
\definecolor{myLightBlue}{rgb}{0.67578,0.84375,0.89844}
\definecolor{myPowderBlue}{rgb}{0.6875,0.875,0.89844}
\definecolor{myPaleTurquoise}{rgb}{0.68359,0.92969,0.92969}
\definecolor{myTurquoise}{rgb}{0.25,0.875,0.8125}
\definecolor{myMediumTurquoise}{rgb}{0.28125,0.81641,0.79688}
\definecolor{myDarkTurquoise}{rgb}{0,0.80469,0.81641}
\definecolor{myLightCyan}{rgb}{0.875,1,1}
\definecolor{myCyan}{rgb}{0,1,1}
\definecolor{myAqua}{rgb}{0,1,1}
\definecolor{myDarkCyan}{rgb}{0,0.54297,0.54297}
\definecolor{myCadetBlue}{rgb}{0.37109,0.61719,0.625}
\definecolor{myLightSteelBlue}{rgb}{0.6875,0.76563,0.86719}
\definecolor{mySteelBlue}{rgb}{0.27344,0.50781,0.70313}
\definecolor{myLightSkyBlue}{rgb}{0.52734,0.80469,0.97917}
\definecolor{mySkyBlue}{rgb}{0.52734,0.80469,0.91797}
\definecolor{myDeepSkyBlue}{rgb}{0,0.74609,1}
\definecolor{myDodgerBlue}{rgb}{0.11719,0.5625,1}
\definecolor{myCornflowerBlue}{rgb}{0.39063,0.58203,0.92578}
\definecolor{myRoyalBlue}{rgb}{0.25391,0.41016,0.87891}
\definecolor{myMediumBlue}{rgb}{0,0,0.80078}
\definecolor{myDarkBlue}{rgb}{0,0,0.54297}
\definecolor{myNavy}{rgb}{0,0,0.5}
\definecolor{myMidnightBlue}{rgb}{0.097656,0.097656,0.4375}
\definecolor{myPurple}{rgb}{0.5,0,0.5}
\definecolor{myLavender}{rgb}{0.89844,0.89844,0.97917}
\definecolor{myThistle}{rgb}{0.84375,0.74609,0.84375}
\definecolor{myPlum}{rgb}{0.86328,0.625,0.86328}
\definecolor{myViolet}{rgb}{0.92969,0.50781,0.92969}
\definecolor{myOrchid}{rgb}{0.85156,0.4375,0.83594}
\definecolor{myFuchsia}{rgb}{1,0,1}
\definecolor{myMagenta}{rgb}{1,0,1}
\definecolor{myMediumOrchid}{rgb}{0.72656,0.33203,0.82422}
\definecolor{myMediumPurple}{rgb}{0.57422,0.4375,0.85547}
\definecolor{myAmethyst}{rgb}{0.59766,0.39844,0.79688}
\definecolor{myBlueViolet}{rgb}{0.53906,0.16797,0.88281}
\definecolor{myDarkViolet}{rgb}{0.57813,0,0.82422}
\definecolor{myDarkOrchid}{rgb}{0.59766,0.19531,0.79688}
\definecolor{myDarkMagenta}{rgb}{0.54297,0,0.54297}
\definecolor{mySlateBlue}{rgb}{0.41406,0.35156,0.80078}
\definecolor{myDarkSlateBlue}{rgb}{0.28125,0.23828,0.54297}
\definecolor{myMediumSlateBlue}{rgb}{0.48047,0.40625,0.92969}
\definecolor{myIndigo}{rgb}{0.29297,0,0.50781}
\definecolor{myGrey}{rgb}{0.5,0.5,0.5}
\definecolor{myLightGrey}{rgb}{0.82422,0.82422,0.82422}
\definecolor{myDarkGrey}{rgb}{0.66016,0.66016,0.66016}
\definecolor{myDimGrey}{rgb}{0.41016,0.41016,0.41016}
\definecolor{myLightSlateGrey}{rgb}{0.46484,0.53125,0.59766}
\definecolor{mySlateGrey}{rgb}{0.4375,0.5,0.5625}
\definecolor{myDarkSlateGrey}{rgb}{0.18359,0.30859,0.30859}

\definecolor{MyC1}{rgb}{0.2,0.2,0.2}
\definecolor{MyC2}{rgb}{0.4,0.4,0.4}
\definecolor{MyC3}{rgb}{0.6,0.6,0.6}
\definecolor{MyC4}{rgb}{0.5,0,0}

\definecolor{myCmap1}{rgb}{0.5151,0.0482,0.6697}
\definecolor{myCmap2}{rgb}{0.51986,0.15038,0.78122}
\definecolor{myCmap3}{rgb}{0.50624,0.24502,0.88013}
\definecolor{myCmap4}{rgb}{0.46845,0.32734,0.95715}
\definecolor{myCmap5}{rgb}{0.42262,0.40041,0.99163}
\definecolor{myCmap6}{rgb}{0.39171,0.47446,0.97876}
\definecolor{myCmap7}{rgb}{0.34868,0.54328,0.9273}
\definecolor{myCmap8}{rgb}{0.30536,0.60642,0.86689}
\definecolor{myCmap9}{rgb}{0.25736,0.66546,0.79834}
\definecolor{myCmap10}{rgb}{0.2215,0.71818,0.71517}
\definecolor{myCmap11}{rgb}{0.24346,0.75604,0.63608}
\definecolor{myCmap12}{rgb}{0.27135,0.79747,0.54425}
\definecolor{myCmap13}{rgb}{0.30037,0.83048,0.45329}
\definecolor{myCmap14}{rgb}{0.32929,0.86008,0.3625}
\definecolor{myCmap15}{rgb}{0.36336,0.89117,0.29194}
\definecolor{myCmap16}{rgb}{0.4385,0.91084,0.2995}
\definecolor{myCmap17}{rgb}{0.55047,0.92321,0.31979}
\definecolor{myCmap18}{rgb}{0.64985,0.9255,0.3352}
\definecolor{myCmap19}{rgb}{0.72705,0.9255,0.34385}
\definecolor{myCmap20}{rgb}{0.8,0.9255,0.3529}

\usepackage{hyperref}
\hypersetup{
	colorlinks,
	citecolor=MyBlack,
	filecolor=MyBlack,
	linkcolor=MyBlack,
	urlcolor=MyBlue,
}

\usepackage{nameref}
\makeatletter
\let\orgdescriptionlabel\descriptionlabel
\renewcommand*{\descriptionlabel}[1]{%
  \let\orglabel\label
  \let\label\@gobble
  \phantomsection
  \edef\@currentlabel{#1}%
  \let\label\orglabel
  \orgdescriptionlabel{#1}%
}
\makeatother

\let\oldsqrt\sqrt
\def\sqrt{\mathpalette\DHLhksqrt}
\def\DHLhksqrt#1#2{%
\setbox0=\hbox{$#1\oldsqrt{#2\,}$}\dimen0=\ht0
\advance\dimen0-0.2\ht0
\setbox2=\hbox{\vrule height\ht0 depth -\dimen0}%
{\box0\lower0.4pt\box2}}

\let\olduplus\uplus
\renewcommand{\uplus}{\raisebox{-0.25mm}{\resizebox{1.15\width}{!}{\hspace*{0.5mm}$\olduplus$}\hspace*{0.5mm}}}

\usepackage{tikz}
\usetikzlibrary{tikzmark}

\usetikzlibrary{positioning}
\usetikzlibrary{backgrounds}

\newcommand\solidSrule[1][1cm]{\rule[0.5ex]{#1}{.75pt}}
\newcommand\solidMrule[1][1cm]{\rule[0.4ex]{#1}{1.5pt}}
\newcommand\solidLrule[1][1cm]{\rule[0.5ex]{#1}{2.25pt}}
\newcommand\solidXXLrule[1][1cm]{\rule[-0.1ex]{#1}{5pt}}

\usetikzlibrary{arrows.meta}
\usetikzlibrary{shapes.misc}

\newcommand{\myColorLine}{\textcolor{myCmap1}{\solidMrule[0.175mm]}\textcolor{myCmap2}{\solidMrule[0.175mm]}\textcolor{myCmap3}{\solidMrule[0.175mm]}\textcolor{myCmap4}{\solidMrule[0.175mm]}\textcolor{myCmap5}{\solidMrule[0.175mm]}\textcolor{myCmap6}{\solidMrule[0.175mm]}\textcolor{myCmap7}{\solidMrule[0.175mm]}\textcolor{myCmap8}{\solidMrule[0.175mm]}\textcolor{myCmap9}{\solidMrule[0.175mm]}\textcolor{myCmap10}{\solidMrule[0.175mm]}\textcolor{myCmap11}{\solidMrule[0.175mm]}\textcolor{myCmap12}{\solidMrule[0.175mm]}\textcolor{myCmap13}{\solidMrule[0.175mm]}\textcolor{myCmap14}{\solidMrule[0.175mm]}\textcolor{myCmap15}{\solidMrule[0.175mm]}\textcolor{myCmap16}{\solidMrule[0.175mm]}\textcolor{myCmap17}{\solidMrule[0.175mm]}\textcolor{myCmap18}{\solidMrule[0.175mm]}\textcolor{myCmap19}{\solidMrule[0.175mm]}\textcolor{myCmap20}{\solidMrule[0.175mm]}}

\newcommand{\myColorBar}{\textcolor{myCmap1}{\solidXXLrule[0.175mm]}\textcolor{myCmap2}{\solidXXLrule[0.175mm]}\textcolor{myCmap3}{\solidXXLrule[0.175mm]}\textcolor{myCmap4}{\solidXXLrule[0.175mm]}\textcolor{myCmap5}{\solidXXLrule[0.175mm]}\textcolor{myCmap6}{\solidXXLrule[0.175mm]}\textcolor{myCmap7}{\solidXXLrule[0.175mm]}\textcolor{myCmap8}{\solidXXLrule[0.175mm]}\textcolor{myCmap9}{\solidXXLrule[0.175mm]}\textcolor{myCmap10}{\solidXXLrule[0.175mm]}\textcolor{myCmap11}{\solidXXLrule[0.175mm]}\textcolor{myCmap12}{\solidXXLrule[0.175mm]}\textcolor{myCmap13}{\solidXXLrule[0.175mm]}\textcolor{myCmap14}{\solidXXLrule[0.175mm]}\textcolor{myCmap15}{\solidXXLrule[0.175mm]}\textcolor{myCmap16}{\solidXXLrule[0.175mm]}\textcolor{myCmap17}{\solidXXLrule[0.175mm]}\textcolor{myCmap18}{\solidXXLrule[0.175mm]}\textcolor{myCmap19}{\solidXXLrule[0.175mm]}\textcolor{myCmap20}{\solidXXLrule[0.175mm]}}

\def\XXint#1#2#3{{\setbox0=\hbox{$#1{#2#3}{\int}$}
     \vcenter{\hbox{$#2#3$}}\kern-.5\wd0}}

\def\XXintS#1#2#3{{\setbox0=\hbox{$#1{#2#3}{\int}$}
     \vcenter{\hbox{$#2#3$}}\kern-.5\wd0}}

\DeclareTextSymbolDefault{\dh}{T1}

\newcommand{\closeomega}{{\text{\textcloseomega}}}

\newcommand{\w}{\text{\textit{w}}}

\newcommand{\myUpsilon}{\text{\scalebox{1.25}{\textupsilon}}}

\DeclareFontFamily{OT1}{pzc}{}
\DeclareFontShape{OT1}{pzc}{m}{it}{<-> s * [1.10] pzcmi7t}{}
\DeclareMathAlphabet{\mathpzc}{OT1}{pzc}{m}{it}

\newcommand{\intPMinfty}{\int_{-\infty}^{\infty}}

\makeatletter   
\DeclareRobustCommand{\equalSet}{\mathrel{\mathpalette\@verequiv=}}
\newcommand{\@verequiv}[2]{%
  \lower.5\p@\vbox{
    \lineskiplimit\maxdimen
    \lineskip+.1\p@
    \ialign{%
      $\m@th#1\hfil##\hfil$\crcr
      #2\crcr
      \equiv\crcr
    }%
  }%
}
\makeatother

\makeatletter   
\DeclareRobustCommand{\equivSet}{\mathrel{\mathpalette\@verequivv=}}
\newcommand{\@verequivv}[2]{%
  \lower.5\p@\vbox{
    \lineskiplimit\maxdimen
    \lineskip+.1\p@
    \ialign{%
      $\m@th#1\hfil##\hfil$\crcr
      #2\crcr
      =\crcr
    }%
  }%
}
\makeatother

\makeatletter
\newcommand*{\nequalSet}{%
  \mathrel{%
    \mathpalette\@nequalSet{\equalSet}%
  }%
}
\newcommand*{\@nequalSet}[2]{%
  \sbox0{\raisebox{\depth}{$#1\equalSet$}}%
  \sbox2{\raisebox{2\depth}{$#1/\m@th$}}%
  \ifdim\ht2>\ht0 %
    \sbox2{\resizebox{\nequalSetxscale\width}{\nequalSetyscale\ht0}{\unhbox2}}%
  \fi
  \sbox2{$\m@th#1\vcenter{\copy2}$}%
  \ooalign{%
    \hfil\phantom{\copy2}\hfil\cr
    \hfil$#1#2\m@th$\hfil\cr
    \hfil\copy2\hfil\cr
  }%
}
\newcommand*{\nequalSetxscale}{1}
\newcommand*{\nequalSetyscale}{2}
\makeatother

\makeatletter
\newcommand*{\nequivSet}{%
  \mathrel{%
    \mathpalette\@nequivSet{\equivSet}%
  }%
}
\newcommand*{\@nequivSet}[2]{%
  \sbox0{\raisebox{\depth}{$#1\equivSet$}}%
  \sbox2{\raisebox{2\depth}{$#1/\m@th$}}%
  \ifdim\ht2>\ht0 %
    \sbox2{\resizebox{\nequivSetxscale\width}{\nequivSetyscale\ht0}{\unhbox2}}%
  \fi
  \sbox2{$\m@th#1\vcenter{\copy2}$}%
  \ooalign{%
    \hfil\phantom{\copy2}\hfil\cr
    \hfil$#1#2\m@th$\hfil\cr
    \hfil\copy2\hfil\cr
  }%
}
\newcommand*{\nequivSetxscale}{1}
\newcommand*{\nequivSetyscale}{2}
\makeatother

\usepackage{pdflscape}
\usepackage{cuted}

\usepackage{nicefrac}

\newcommand{\dd}{\mathop{}\!\mathrm{d}}
\renewcommand{\IEEEproofindentspace}{0em}

\providecommand*{\eu}{\ensuremath{\mathrm{e}}}
\providecommand*{\ju}{\ensuremath{\mathrm{j}}}

\renewcommand{\Re}{\mathrm{Re}}

\usepackage{nomencl}
\makenomenclature

\usepackage[english]{babel}
\usepackage{blindtext}

\usepackage{float}

\usetikzlibrary{shapes.geometric}
\usetikzlibrary{calc}
\usetikzlibrary{decorations.pathreplacing}

\usetikzlibrary{fadings}

\usepackage{xfrac}

\begin{document}
\title{Unifying Common Signal Analyses with Instantaneous Time-Frequency Atoms}

\author{Steven Sandoval,~\IEEEmembership{Member,~IEEE,}
        Phillip~L.~De~Leon,~\IEEEmembership{Senior Member,~IEEE}

\thanks{Copyright \copyright 2025 IEEE. Personal use of this material is permitted. However, permission to use this material for any other purposes must be obtained from the IEEE by sending a request to pubs-permissions@ieee.org.}

\thanks{This paper has supplementary downloadable material available at {http://ieeexplore.ieee.org}, provided by the authors. This includes Figures 6 and 7 in the paper.}

\thanks{Steven Sandoval is with Klipsch School of Electrical and Computer Engineering, New Mexico State University (NMSU), Las Cruces NM 88003 USA. e-mail: spsandov@nmsu.edu }

\thanks{Phillip L.~De Leon is with the Department of Electrical and Computer Engineering, University of Colorado Denver, Denver CO 80204 USA. e-mail: Phillip.DeLeon@ucdenver.edu ~~~~~{\color{myFireBrick}\textbf{(Resubmitted:~Dec.~2025)}}
}


}%
\markboth{In Preparation for IEEE Transactions on Signal Processing}{Sandoval and De Leon: Unifying Common Signal Analyses with Time-Frequency Atoms}

\maketitle

\begin{abstract}   

In previous work, we presented a general framework for instantaneous time-frequency analysis but did not provide any specific details of how to compute a particular instantaneous spectrum (IS). In this work, we use instantaneous time-frequency atoms to obtain an IS associated with common signal analyses: time domain analysis, frequency domain analysis, fractional Fourier transform, synchrosqueezed short-time Fourier transform, and synchrosqueezed short-time fractional Fourier transform. By doing so, we demonstrate how the general framework can be used to unify these analyses and we develop closed-form expressions for the corresponding ISs. This is accomplished by viewing these analyses as decompositions into AM--FM components and recognizing that each uses a specialized (or limiting) form of a quadratic chirplet as a template during analysis. With a two-parameter quadratic chirplet atom, we can organize these ISs into a 2D continuum with points in the plane corresponding to a decomposition related to one of the signal analyses. Finally, using several example signals, we compute in closed-form the ISs for the various analyses.

\end{abstract}


\IEEEpeerreviewmaketitle

\section{Introduction}\label{sec:intro}

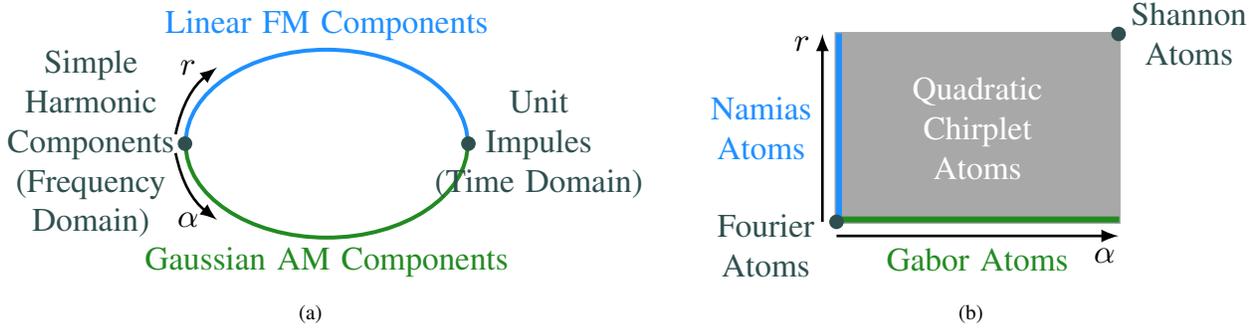
\begin{figure*}[htb!]
    \centering

\begin{minipage}[b]{0.5\textwidth}
\subfigure[]{
\scalebox{1.25}{\begin{tikzpicture}
    \node[ellipse,
        thin,
	draw = white,
	minimum width = 3cm, 
	minimum height = 2cm] (ellipse) at (0,0) {};

\draw[very thick, myDodgerBlue] (1.5,0) arc(0:180:1.5cm and 1cm);
\draw[very thick, myForestGreen] (-1.5,0) arc(180:360:1.5cm and 1cm);

\node[mark size=2pt,color=myDarkSlateGray] at (ellipse.west) {\pgfuseplotmark{*}};
\node[myDarkSlateGray,align=center,text width=2.25cm] at ($ (ellipse.west) + (-1,0) $) {Simple\\ Harmonic\\ Components\\(Frequency Domain)};

\node[myDodgerBlue, align=center,text width=4cm] at ($ (ellipse.north) + (0,0.25) $) {Linear FM Components};

\node[mark size=2pt,color=myDarkSlateGray] at (ellipse.east) {\pgfuseplotmark{*}};
\node[myDarkSlateGray,align=center,text width=2.25cm] at ($ (ellipse.east) + (0.75,0) $) {Unit\\ Impules\\(Time Domain)};


\node at ($ (ellipse.north west) + (-0.1,0.1) $) {$r~~~~~$};
\node at ($ (ellipse.south west) + (-0.1,-0.1) $) {$\alpha~~~~~$};
\draw [thick, -latex] ($ (ellipse.west) + (-0.1,0.1) $) to [bend left=20] ($ (ellipse.north west) + (-0.1,0.1) $);
\draw [thick, -latex] ($ (ellipse.west) + (-0.1,-0.1) $) to [bend right=20] ($ (ellipse.south west) + (-0.1,-0.1) $);

\node[myForestGreen, align=center,text width=4cm] at ($ (ellipse.south) + (0,-0.25) $) {Gaussian AM Components};

\end{tikzpicture}}
\label{fig:comps}}

\end{minipage}\begin{minipage}[b]{0.5\textwidth}
\subfigure[]{

\scalebox{1.25}{\begin{tikzpicture}
    \draw [myDarkGray, fill=myDarkGray, very thick] (0,0) rectangle (3,2);
    \draw [myForestGreen, fill=myForestGreen] (0,0) rectangle (3,0.05);
    \draw [myDodgerBlue, fill=myDodgerBlue] (0,0) rectangle (0.05,2);
    \node[myDodgerBlue, align=center,text width=2cm] (Namias) at (-0.8,1) {Namias\\ Atoms};
    \node[myForestGreen, align=center,text width=4cm] (Gabor) at (1.5,-0.4) {Gabor Atoms};

    \draw [thick, -latex] (-0.15,0) -- (-0.15,2) node[pos=0.95,left]{$r$};
    \draw [thick, -latex] (0,-0.15) -- (3,-0.15)
    node[pos=0.95,below]{$\alpha$};
    
    \node[mark size=2pt,color=myDarkSlateGray] at (0,0) {\pgfuseplotmark{*}};
    \node[myDarkSlateGray,align=center,text width=2cm] (FourierAtom) at (-0.75,-0.25) {Fourier\\ Atoms};

    \node[white, align=center,text width=2cm] (ChirpletAtom) at (1.5,1) {Quadratic\\ Chirplet\\ Atoms};

    \node[mark size=2pt,color=myDarkSlateGray] at (3.05,2.05) {\pgfuseplotmark{*}};
    \node[myDarkSlateGray,align=center,text width=2cm] (ShannonAtom) at (3.75,2) {Shannon\\Atoms};

\end{tikzpicture}}
\label{fig:atoms}}
\end{minipage}

    \caption{%
    (a) Illustration of the relationship between the components used in several classical analyses.     The end points correspond to time and frequency domain analyses (\textcolor{myDarkSlateGray}{\scalebox{1.5}{$\bullet$}}).     The 1D intermediate continuum on $r$ of linear FM components (\textcolor{myDodgerBlue}{\solidLrule[3mm]}) linking the time and frequency domain analyses.     The 1D intermediate continuum on $\alpha$ of Gaussian AM components (\textcolor{myForestGreen}{\solidLrule[3mm]}) linking the time and frequency domain analyses. %
    (b) Illustration of the relationship between various quadratic chirplet atoms. The corner points of Fourier and Shannon atoms correspond to frequency and time domain analyses (\textcolor{myDarkSlateGray}{\scalebox{1.5}{$\bullet$}}). While the 1D continuum on $r$ of Namias atoms (\textcolor{myDodgerBlue}{\solidLrule[3mm]}) and the 1D continuum on $\alpha$ of Gabor atoms does not provide a link between the time and frequency domain analyses, this may be accomplished in a 2D space of quadratic chirplets (\textcolor{myDarkGray}{\solidXXLrule[2mm]}) on $\alpha$ and $r$. 
    }
    \label{fig:compRelations}
\end{figure*}

The classical signal analyses are time domain (TD) and frequency domain (FD) where in the signal decomposition, the former utilizes the unit impulse components
\begin{IEEEeqnarray}{c}
    z(t) = \intPMinfty \psi_\tau(t) \dd \tau,~~  \psi_\tau(t) = z(\tau) \delta(t-\tau)
    \label{eqn:TDdecomposition}
\end{IEEEeqnarray}
and the latter simple harmonic components 
\begin{IEEEeqnarray}{c}
    z(t) =  \intPMinfty \psi_\omega(t) \dd \omega,~~ \psi_\omega(t) =  \frac{Z(\omega )}{\sqrt{2\pi}}  \eu^{\,\ju \omega t}.
    \label{eqn:FDdecomposition}
\end{IEEEeqnarray}
This is illustrated as the endpoints in Figure \ref{fig:comps}. Gabor interpreted these two decompositions as limiting forms of the Gaussian AM component with pulse width parameter $\alpha$ \cite{gaborTOC} 
\begin{IEEEeqnarray}{c}
    \textcolor{black}{\psi(t) =  a_0\exp\left[-\alpha^2(t-t_0)^2\right] \exp\big(\ju[ \omega_0 (t-t_0) +\varphi_0]\big).}~~~
\end{IEEEeqnarray}
The first limiting form, $\alpha\rightarrow0$ corresponds to \eqref{eqn:FDdecomposition} and the second limiting form, $\alpha\rightarrow \infty$ corresponds to \eqref{eqn:TDdecomposition}. This interpretation provides a continuum on $\alpha$ linking the two signal decompositions and is illustrated by the line in the lower half of Figure \ref{fig:comps}. Namias also interpreted these two decompositions as limiting forms but with the linear FM component and chirp rate parameter $r$ \cite{namias1980fractional}
\begin{IEEEeqnarray}{c}
    \textcolor{black}{\psi(t) = a_0  \exp\big(\ju[ -r(t-t_0)^2/2 +\omega_0 (t-t_0) + \varphi_0]\big).} 
\end{IEEEeqnarray}
The first limiting form, $r\rightarrow0$ corresponds to \eqref{eqn:FDdecomposition} and the second limiting form, $r\rightarrow \infty$ corresponds to \eqref{eqn:TDdecomposition}. This interpretation provides a continuum on $r$ linking the two signal decompositions and is illustrated by the line in the upper half of Figure \ref{fig:comps}.

Although the Gaussian AM and linear FM components each provide a continuum between the two classical signal decompositions, neither provides a mechanism for an instantaneous time-frequency analysis. In \cite{ISA2018_Sandoval} \textcolor{black}{(which is summarized in Section \ref{sec:ISA}),} we proposed a signal decomposition into AM--FM components of which the Gaussian AM and linear FM components are special cases. The AM--FM component $\psi_k(t)$ may be demodulated into a time-frequency object $\mathcal{A}_k(t,\omega)$ which we call an \textit{atom},\footnote{Our use of the term \textit{component} refers to a function of time and our use of the term \textit{atom} refers to a function of both time and frequency. Other works including our previous work, often use the terms \textit{component} and \textit{atom} interchangeably.} which allows construction of an instantaneous time-frequency spectrum. Simple harmonic components can be demodulated into well-defined Fourier atoms, Gaussian AM components can be demodulated into well-defined Gabor atoms, and linear FM components can be demodulated into well-defined Namias atoms. Unfortunately, for the unit impulse it is not clear how to demodulate as an AM--FM component into a well-defined atom. As Flandrin writes ``...representation in time ... makes no direct reference to a frequency content'' even though in reality it is present \cite{flandrin2018Explorations}. Similarly, ``...a frequency spectrum just ignores time'' even though in reality it too is present \cite{flandrin2018Explorations}. 

To obtain an atom for the unit impulse, one might consider taking the limit $\alpha\rightarrow\infty$ of a Gabor atom or the limit  $r\rightarrow\infty$ of a Namias atom. However, both of these approaches result in an atom with time-frequency properties that do not properly describe a unit impulse. This may be resolved, by recognizing that Gaussian AM components and linear FM components are special cases of the quadratic chirplet component and as a result, the Gabor and Namias atoms are special cases of the quadratic chirplet atom. Therefore, we propose an atom for the unit impulse as a limiting case of the quadratic chirplet atom when the limit is simultaneously taken in $\alpha$ and $r$. This approach, illustrated in Figure \ref{fig:atoms}, results in a Shannon atom that we will show has time-frequency properties that properly describe a unit impulse.

In this work, we use time-frequency atoms to obtain an instantaneous spectrum (IS) from common analysis methods by viewing these methods as decompositions into AM--FM components. We begin by choosing an analysis, e.g.~TD analysis, FD analysis, synchrosqueezed short-time Fourier transform (STFT), fractional Fourier transform (FrFT), synchrosqueezed short-time fractional Fourier transform (STFrFT), etc\footnote{Throughout this work, any references to ISs derived from STFT or STFrFT imply synchrosqueezing of the atoms to impose the structure necessary for a valid IS. See \cite{sandoval2022recasting} for further details.}.  For many applications, we do not need instantaneous time-frequency information and the analysis is useful as a decomposition. On the other hand, for applications where instantaneous time-frequency information is desired we must move beyond the component $\psi_k(t)$ to an atom $\mathcal{A}_k(t,\omega)$. 

Using instantaneous spectral analysis (ISA) theory introduced in \cite{ISA2018_Sandoval}, an atom may be obtained as follows. Each component $\psi_k(t)$ can be demodulated resulting in $\psi_k(t;\mathscr{C}_k)$. Then the collection of triplets forms a parameter set $\mathscr{S}=\{\mathscr{C}_k\}$. Each triplet $\mathscr{C}_k$ leads to a time-frequency atom $\mathcal{A}_k(t,\omega;\mathscr{C}_k)$ and from $\{\mathcal{A}_k(t,\omega;\mathscr{C}_k)\}$ we obtain the IS $\mathcal{S}(t,\omega;\mathscr{S})$. Finally, we show that the resulting IS $\mathcal{S}(t,\omega;\mathscr{S})$ corresponds to the signal under analysis by demonstrating that the AM--FM model $z(t;\mathscr{S})$ specializes to the synthesis equation for the choice of analysis.

The contributions of this paper are as follows.
\begin{enumerate}
    \item In \cite{ISA2018_Sandoval}, we developed the ISA framework based on a general AM--FM component but did not provide any specific details of how to compute a particular IS. In this work, we show how to associate a synthesis equation with an IS and by doing so we provide a two parameter ($\alpha$ and $r$) analysis method that leads to an estimated IS.
    \item \textcolor{black}{We express the Shannon atom as a limiting form of the quadratic chirplet atom  and prove it is the product of a unit impulse in time and a unit dispersion in frequency (defined in Section \ref{sec:ShannonAtom}). By defining the Shannon atom, we provide a means to demodulate the unit impulse leading to an IS for time domain analysis.}
    \item  We show that the common analyses listed above are specific choices for the two parameters and we give closed-form expressions for the corresponding IS. By doing so we demonstrate the IS as a unifying framework for the analyses under consideration.
\end{enumerate}

\section{ISA Background}\label{sec:ISA}

\subsection{Signal Synthesis using the ISA framework}

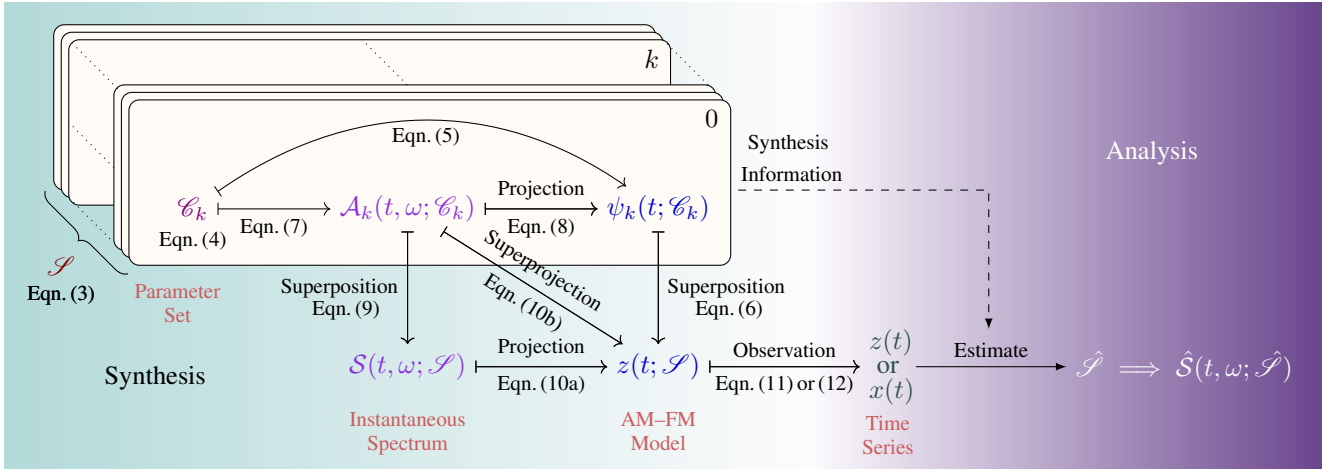
\begin{figure*}
    \centering

\begin{tikzpicture}

    \draw[rounded corners, fill=myAntiqueWhite!20] (-0.6,  0.5) rectangle (7.4, 2.7) {};   
    \draw[rounded corners, fill=myAntiqueWhite!20] (-0.5,  0.4) rectangle (7.5, 2.6) {};   
    \draw[rounded corners, fill=myAntiqueWhite!20] (-0.4,  0.3) rectangle (7.6, 2.5) {};   
    \draw[rounded corners, fill=myAntiqueWhite!20] (0.2, -0.3) rectangle (8.2, 1.9) {};
    \draw[rounded corners, fill=myAntiqueWhite!20] (0.3, -0.4) rectangle (8.3, 1.8) {};
    \draw[rounded corners, fill=myAntiqueWhite!20] (0.4, -0.5) rectangle (8.4, 1.7) {};

    \node(zero) at ($ (8.4,1.7) + (-0.25,-0.25) $){$0$};
    \node(kInd) at ($ (7.6, 2.5) + (-0.25,-0.25) $){$k$};
    
    \node[myDarkRed](ss) at (-0.5,-0.5){$\mathscr{S}$};
    \node(set) at (-0.5,-0.9){{\footnotesize Eqn.\,\eqref{eqn:set}}};
    \node(set) at (-0.5,-0.9){{\footnotesize Eqn.\,\eqref{eqn:set}}};

\draw[dotted] (0.2, 0.8) -- (-0.4, 1.4) {};
\draw[dotted] (0.25, 1.85) -- (-0.35, 2.45) {};
\draw[dotted] (4.1, 1.9) -- (3.5, 2.5) {};

\draw[dotted] (8.1, 1.9) -- (7.5, 2.5) {};

\draw[dotted] (0.15, -0.15) -- (-0.35, 0.35) {};

\draw[decorate, decoration={brace, amplitude=1ex, raise=1ex}]  (0.5, -0.5) -- (-0.6, 0.6) node[pos=.5, left=2.5ex] {};

    \node[myPurple](a) at (1.25,0.25){$\mathscr{C}_k$};
    \node(aaaa) at (1.25,-0.175){{\footnotesize Eqn.\,\eqref{eq:cannonical}}};

    \node[myBlueViolet,right=1.5cm of a](b){$\mathcal{A}_k(t,\omega;\mathscr{C}_k)$};
    \draw[|->] (a) -- (b)node[pos=0.5, below] {{\footnotesize Eqn.\,\eqref{eqn:TFatom}}};
        
    \node[myMediumBlue, right=1.5cm of b](c){$\psi_k(t;\mathscr{C}_k)$};
    \draw[|->] (b) -- (c) node[pos=0.5, below] {{\footnotesize Eqn.\,\eqref{eq:HSintOmegaComp}}};
    \draw[|->] (b) -- (c) node[pos=0.5, above] {\footnotesize Projection};

    \draw[|->] (a) to[out={90-55}, in={90+55}] (c);
    \node[above=0.4cm of b] {{\footnotesize ~~~~~Eqn.\,\eqref{eq:AMFMcompS}}};

    \node[myBlueViolet,below=1.5cm of b](e){$\mathcal{S}(t,\omega;\mathscr{S})$};

    \node[myMediumBlue, below=1.5cm of c](f){$z(t;\mathscr{S})$};
    \draw[|->] (e) -- (f) node[pos=0.5, below] {{\footnotesize Eqn.\,\eqref{eq:HSintOmega}}};
    \draw[|->] (e) -- (f) node[pos=0.5, above] {\footnotesize Projection};

    \draw[|->] (b) -- (f) node[pos=0.5, above, sloped] {\footnotesize Superprojection};
    \draw[|->] (b) -- (f) node[pos=0.5, below, sloped] {{\footnotesize Eqn.\,\eqref{eq:superProj}}};

    \node[myDarkSlateGray,right=2cm of f,align=center,text width=0.5cm](g){$z(t)$\\[-1mm] or\\[-1mm]$x(t)$};
    \draw[|->] (f) -- (g) node[pos=0.5, above] {{\footnotesize Observation}};
    \draw[|->] (f) -- (g) node[pos=0.5, below 
    ] {{\footnotesize Eqn.\,\eqref{eq:complexObs}\,or\,\eqref{eq:realObs}}};

    \node[black, right=2cm of g](goal){$\hat{\mathscr{S}}\implies\hat{\mathcal{S}}(t,\omega;\hat{\mathscr{S}})$};
    \draw[-latex,] (g) -- (goal) node[pos=0.5, above] {{\footnotesize Estimate}};

    \draw[|->] (c) -- (f) node[pos=0.5, right] {\footnotesize Superposition};
    \draw[|->] (c) -- (f) node[pos=0.7, right] {~~~{\footnotesize Eqn.\,\eqref{eq:AMFMmodel}}};

    \draw[|->] (b) -- (e) node[pos=0.5, left] {\footnotesize Superposition};
    \draw[|->] (b) -- (e) node[pos=0.7, left] {{\footnotesize Eqn.\,\eqref{eq:2DInstantaneousSpectrum}}~~~};

    \node[myIndianRed,below left =0.6cm and -1cm of a, align=center,text width=1.5cm](aLab){{\footnotesize Parameter\\[-1mm]Set}};
   
   \node[myIndianRed,below=0.175cm of e, align=center,text width=2cm](eLab){{\footnotesize Instantaneous\\[-1mm]Spectrum}};
   
   \node[myIndianRed,below=0.175cm of f, align=center,text width=1.5cm](fLab){{\footnotesize AM--FM\\[-1mm]Model}};
    
    \node[myIndianRed,below=-0.1cm of g, align=center,text width=1.5cm](gLab){{\footnotesize Time\\[-1mm]Series}};

 \begin{scope}[on background layer]
    \draw[draw opacity=0,shading = axis, 
    left color=myTeal!10, 
    right color=myThistle!5,
    shading angle=90, 
    ] (-1.25,3) rectangle  (9.5,-3.2);
    \draw[draw opacity=0,shading = axis, 
    right color=myIndigo!20, 
    left color=myThistle!5,
    shading angle=90, 
    ] (9.5,3) rectangle  (16.25,-3.2);    
 \end{scope}
    \node at (0.75,-2) (g) {Synthesis};
    \node[black] at (14,1) (g) {Analysis};
    
    \draw[-latex, dashed] (8.5,0.5) -| ($ (goal) + (-2.6,0.5) $) node[pos=0.1, above, align=center,text width=1.25cm] {{\footnotesize Synthesis\\ Information}};

\end{tikzpicture}

   \caption{A pictorial description of the ISA framework. The ISA framework provides an ideal model for \textcolor{black}{construction} of an IS from a parameter set. A signal may be projected from the IS and observation corresponds to loss of the parameter set and possibly the loss of the imaginary signal part. The goal of analysis is to estimate the IS from synthesis. The problem is under-determined, thus synthesis information external to the signal is necessary to guide the choice of analysis.}
    \label{fig:ISAsummary}
\end{figure*}

In \cite{ISA2018_Sandoval}, we introduced ISA as a general framework for time-frequency analysis depicted in Figure~\ref{fig:ISAsummary}\textcolor{black}{,} where: 1) each canonical triplet $\mathscr{C}_k$ has a single-valued mapping to a signal component $\mathscr{C}_k\mapsto\psi_k(t;\mathscr{C}_k)$, 2) each canonical triplet has a single-valued mapping to a time-frequency atom $\mathscr{C}_k\mapsto\mathcal{A}_k(t,\omega;\mathscr{C}_k)$, and 3) each time-frequency atom has a single-valued mapping to a  signal component \textcolor{black}{$\mathcal{A}_k(t,\omega;\mathscr{C}_k)\mapsto \psi_k(t;\mathscr{C}_k)$}. A signal is then represented by means of a set of canonical triplets 
\begin{IEEEeqnarray}{c}
    \mathscr{S} \triangleq \set{\mathscr{C}_0, \mathscr{C}_1, \cdots,\mathscr{C}_{K-1}}
    \label{eqn:set}
\end{IEEEeqnarray}
specifying an IS $\mathcal{S}(t, \omega;\mathscr{S})$ that is \textcolor{black}{constructed} from a superposition of time-frequency atoms. Then, the signal $z(t;\mathscr{S})$ may be projected from the IS.

Summarizing the key results in \cite{ISA2018_Sandoval}, we define the $k$th canonical triplet by 
\begin{IEEEeqnarray}{c}
    \mathscr{C}_k\triangleq\canonical{a_k(t),{\omega}_k(t),\phi_k\vertOne}\label{eq:cannonical}
\end{IEEEeqnarray}
where $a_k(t)$ is the instantaneous amplitude (IA), $\omega_k(t)$ is the instantaneous frequency (IF), and $\phi_k$ is the phase reference. The $k$th complex AM--FM component is then given in AM--FM, polar, and Cartesian forms by 
\begin{IEEEeqnarray}{rCl}
	\IEEEyesnumber\label{eq:AMFMcompS}\IEEEyessubnumber*
	    \phantom{~}\psi_k\left( t ; \mathscr{C}_k \vertOne\right) & \triangleq & a_k(t) \eu^{ \,\ju \left[\int_{-\infty}^{t} \omega_k(\uptau)\dd\uptau +\phi_k\right] }\label{eq:AMFMcompA}\\
	  	& = & a_k(t) \eu^{\,\ju \theta_k(t)}\label{eq:AMFMcompB}\\
	  	& = & s_k(t)+\ju \sigma_k(t)\label{eq:AMFMcompC}
\end{IEEEeqnarray}
where $\theta_k(t)$ is the phase function, $s_k(t)$ is the real part, and $\sigma_k(t)$ is the imaginary part. The complex signal $z(t; \mathscr{S} )$ is represented as a superposition of $K$ (possibly infinite) complex AM--FM components
\begin{IEEEeqnarray}{rCl}
	\IEEEyesnumber\label{eq:AMFMmodel}\IEEEyessubnumber*
	    z\left( t ; \mathscr{S} \vertOne\right)  & \triangleq & \sum\limits_{k=0}^{K-1}\psi_k\left( t ; \mathscr{C}_k \vertOne\right)\label{eq:AMFM_A}\\
	  	& = & x(t)+\ju y(t).\label{eq:AMFM_D}
\end{IEEEeqnarray}
We define the $k$th time-frequency atom\footnote{The normalization constants differ from our previous work. This reflects our switch to a unitary definition of the Fourier transform.} as 
\begin{IEEEeqnarray}{rCl}
   \mathcal{A}_k(t, \omega; \mathscr{C}_k) \triangleq  \sqrt{2 \pi}\,  \psi_k\left( t ; \mathscr{C}_k \vertOne\right)\,\delta\left(\vertOne\omega-{\omega}_k(t)\vertOne\right)\label{eqn:TFatom}
\end{IEEEeqnarray}
which maps to the $k$th complex AM--FM component with
\begin{IEEEeqnarray}{c}
    \frac{1}{\sqrt{2 \pi}}\intPMinfty\mathcal{A}_k(t,\omega;\mathscr{C}_k) \dd\omega=\psi_k(t;\mathscr{C}_k).
    \label{eq:HSintOmegaComp}
\end{IEEEeqnarray}
Referring to Figure~\ref{fig:ISAsummary}, the IS is \textcolor{black}{constructed} as a superposition of $K$ (possibly infinite) time-frequency atoms
\begin{IEEEeqnarray}{rCl}
    \mathcal{S}(t,\omega;\mathscr{S}) 
      & \triangleq &  \sum\limits_{k=0}^{K-1}\mathcal{A}_k(t, \omega; \mathscr{C}_k)\label{eq:2DInstantaneousSpectrum} 
\end{IEEEeqnarray}
and the complex signal $z(t;\mathscr{S})$ is synthesized as a superposition of time-frequency atoms and a projection onto the time axis with
\begin{IEEEeqnarray}{rCl}\IEEEyesnumber*\IEEEyessubnumber*
     z(t;\mathscr{S})   &=& \frac{1}{\sqrt{2 \pi}}\intPMinfty\mathcal{S}(t,\omega;\mathscr{S}) \dd\omega \label{eq:HSintOmega}\\
                        &=& \frac{1}{\sqrt{2 \pi}}\intPMinfty\sum\limits_{k=0}^{K-1}\mathcal{A}_k(t, \omega; \mathscr{C}_k) \dd\omega.
    \label{eq:superProj}
\end{IEEEeqnarray}
We call \eqref{eq:superProj} the \emph{superprojection} equation.

The act of observation corresponds to the loss of exact knowledge of the set $\mathscr{S}$
\begin{IEEEeqnarray}{c}
    z(t;\mathscr{S})\mapsto z(t).
    \label{eq:complexObs}
\end{IEEEeqnarray}
Moreover, in many applications only the real part $x(t)$, is observed (or measured) and the imaginary part $y(t)$, is \textit{latent}, i.e.~the act of observation corresponds to 
\begin{IEEEeqnarray}{c}
    z(t;\mathscr{S})\mapsto x(t) ~~~\text{according to}~~~ x(t) = \Re\operator{z(t;\mathscr{S})}.~~~
    \label{eq:realObs}
\end{IEEEeqnarray}



\subsection{Signal Analysis Using the ISA Framework}

In \cite{ISA2018_Sandoval}, we specified  the ``assembly rules'' that allow, by specification of canonical triplets, the synthesis of signals using AM--FM components and \textcolor{black}{construction} of ISs using time-frequency atoms. The AM--FM components are the basic elements by which a signal $z(t)$ is synthesized. Moreover, AM--FM components are projections of time-frequency atoms which are the basic elements by which an IS $S(t,\omega)$ is constructed. As shown in the right half of Figure~\ref{fig:ISAsummary}, the objective of the theory is to estimate, for a given signal $z(t)$ in the usual function spaces, the IS $S(t,\omega;\mathscr{S})$.

Although IS theory provides what may be considered as an \textit{ideal model} for ISs and AM--FM signals, the many-to-one mappings are sources of information loss which allow an infinite number of ISs and component sets to map to the same signal. As a result, the reverse process is under-determined and \textit{no unique analysis model exists}. 

\begin{figure*}[t]
    \centering

\begin{tikzpicture}

    \node[myDarkSlateGray,align=center,text width=0.5cm] (sig) at (0,0) {$z(t)$\\[-1mm] or\\[-1mm]$x(t)$};

    \node[right=3cm of sig](comps){$\{\hat{\psi}_k(t)\}$};

    \draw[|->] (sig) -- (comps)node[pos=0.5, above] {{\footnotesize Decomposition}};
    
\draw[|->] (sig) -- (comps)node[pos=0.5, below] {{\footnotesize in form of Eqn.\,(\ref{eq:AMFM_A})}};

    \node[ right=3cm of comps](triplet){$\{\hat{\mathscr{C}}_k(t)\}=\hat{\mathscr{S}}\implies \hat{\mathcal{S}}(t,\omega;\hat{\mathscr{S}})$};

    \draw[|->] (comps) -- (triplet)node[pos=0.5, above] {{\footnotesize Demodulatation}};

    \draw[|->] (comps) -- (triplet)node[pos=0.5, below] {{\footnotesize with Eqns.\,(\ref{eq:IAest}) - (\ref{eq:DemodTriplet}) }};


    right color=myAmethyst!70!black, 
    left color=myThistle!5,
    shading angle=90, 
    ] (-0.5,0.75) rectangle  (12.75,-0.75);    
   
\begin{scope}[on background layer]
   
    \draw[draw opacity=0,shading = axis, 
    left color=myFireBrick!15, 
    right color=myThistle!5,
    shading angle=90, 
    ] (-0.5,0.75) rectangle  (3.5,-1.25);
    
    \draw[draw opacity=0,shading = axis, 
    right color=myDodgerBlue!30!white, 
    left color=myThistle!5,
    shading angle=90, 
    ] (3.5,0.75) rectangle  (12.75,-1.25);  

 \end{scope}
    \node at (1.85,-1) (g) {Ambiguous};
    \node at (8.5,-1) (g) {Exact Localization For Each Time-Frequency Atom};    
\end{tikzpicture}
    \caption{A pictorial description of an approach to Analysis that begins with a decomposition of a signal into components which can be expressed in AM-FM form (polar form with differentiable phase) as described in \eqref{eq:AMFMcompS}.}
    \label{fig:ISAanalysis}
\end{figure*}
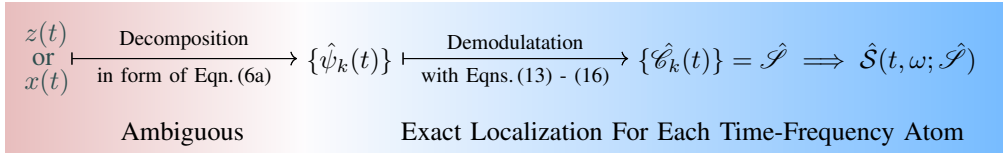

One approach to analysis is to consider two stages: 1) signal decomposition and 2) component demodulation as illustrated in Figure \ref{fig:ISAanalysis}. With this approach, all ambiguity lies in how the signal $z(t)$ is decomposed into a set of components---there exist an infinite number of ways to express a whole as a sum of parts. However, every decomposition has the advantage that it can be associated with an IS in which each component is exactly localized. 

Suppose that a signal $z(t)$ is observed which was synthesized with unknown parameter set $\mathscr{S}$ leading to an unknown IS $\mathcal{S}(t,\omega\textcolor{black}{;\mathscr{S}})$. Decomposing the signal $z(t)$ into a superposition of a set of components $\{\hat{\psi}_k(t)\}$ each with differentiable phases [as expressed in \eqref{eq:AMFM_A}] allows each component to be interpreted as AM--FM and demodulated with
\begin{IEEEeqnarray}{rCl} 
    \hat{a}_k(t) &=& |\hat{\psi}_k(t)\big|\label{eq:IAest},\\
    \hat{\omega}_k(t) &=& \frac{\dd}{\dd t}\arg\big\{\hat{\psi}_k(t)\big\},\\
    \hat{\phi}_k &=& \arg\big\{\hat{\psi}_k(0)\big\}.
\end{IEEEeqnarray}
With the parameters, we have the triplet
\begin{IEEEeqnarray}{c}
    \hat{\mathscr{C}}_k\triangleq\canonical{\hat{a}_k(t),\hat{\omega}_k(t),\hat{\phi}_k\vertOne}
    \label{eq:DemodTriplet}
\end{IEEEeqnarray}
and the estimated IS \textcolor{black}{$\hat{\mathcal{S}}(t,\omega;\hat{\mathscr{S}})$} follows from \eqref{eq:2DInstantaneousSpectrum}.
\color{black}

Although this work is a contribution to time-frequency analysis theory, for examples and tools using the ISA framework for real-world signals, including speech and a bat chirp, we refer the reader to \cite{ASRU2015,sandoval2022recasting,sandoval2022isa}.


\color{black}


\subsection{Decomposition using a Quadratic Chirplet as a Template}
The general form of a canonical triplet was given in \eqref{eq:cannonical}. Choosing the IA as a Gaussian pulse 
\begin{IEEEeqnarray}{c}
    a(t)=\exp\left(-\alpha^2t^2\right)
\end{IEEEeqnarray}
and a linear IF 
\begin{IEEEeqnarray}{c}
    \omega(t) = -rt
\end{IEEEeqnarray}
gives the triplet
\begin{IEEEeqnarray}{c}
    \mathscr{C}=\canonical{\exp(-\alpha^2t^2),-rt,0\vertOne}.
\end{IEEEeqnarray}
Using \eqref{eq:AMFMcompS} and \eqref{eqn:TFatom} gives the quadratic chirplet component
\begin{IEEEeqnarray}{c}
    \psi(t)  = \exp\left(-\alpha^2 t^2\right) \exp\left(-\ju r t^2/2 \vertOne\right)\label{eqn:quadChirpComp}
\end{IEEEeqnarray}
and quadratic chirplet atom
\begin{IEEEeqnarray}{c} 
    \mathcal{A}(t,\omega) = \sqrt{2\pi} \exp\left(-\alpha^2 t^2\right)   \exp\left(-\ju r t^2/2 \vertOne\right) \delta\left(\omega +r t  \right).\IEEEeqnarraynumspace
\end{IEEEeqnarray}

Choosing $\alpha$ and $r$ allows one to specialize the quadratic chirplet component which is illustrated in Figure \ref{fig:comps}.  
Moreover, for each choice we can identify a common analysis which uses the component as a template. Specifically, the template is used to generate a family of components through frequency shifting. Depending on the analysis, the signal is either projected onto the family of components or the signal is filtered using the family of components as impulse responses for each channel. For example, choosing $\alpha=0$, $r=0$, and frequency shifting by $\closeomega$, yields the family of simple harmonic components at all frequencies
\begin{IEEEeqnarray}{c} 
    \psi_\closeomega(t) =  \eu^{\,\ju \closeomega t }.
\end{IEEEeqnarray}
As another example, choosing $0<\alpha<\infty$, $r=0$, and frequency shifting by channelizer frequency $\nu$, yields a family of Gaussian impulse responses for a filterbank.
\begin{IEEEeqnarray}{c} 
    \psi_\nu(t;\alpha) =  \exp\left(-\alpha^2 t^2\right)  \eu^{\,\ju \nu t }.
\end{IEEEeqnarray}

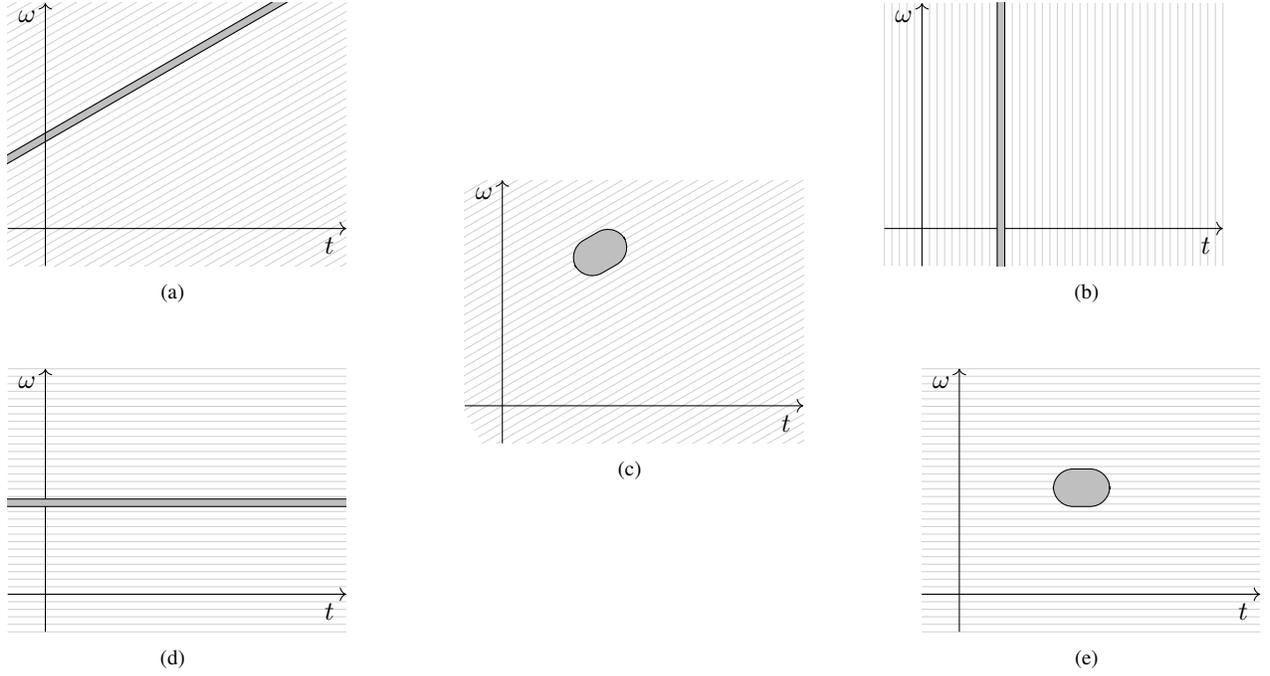
\begin{figure*}
    \centering

    \begin{minipage}[b]{0.33\linewidth}
    \centering	
    \subfigure[]{
        \begin{tikzpicture}
        \begin{scope}
        \clip(-0.5,-0.5) rectangle (4,3);
            \begin{scope}[rotate=30]
            \foreach \n in {-25,...,30}{
                \draw[-,myLightGray,thin] (-1,{\n/10}) -- (5,{\n/10});
            }
                \begin{scope}[shift={(0,0)}]
                    \draw[black,thin,fill=mySilver,rounded corners=0.05] (-0.5,1) rectangle ({5},{1+0.1});
                \end{scope}
            \end{scope}
        \end{scope}
        \draw[->] (-0.5,0) -- (4,0) node[pos=0.95, below] {$t$};
        \draw[->] (0,-0.5) -- (0,3) node[pos=0.95, left] {$\omega$};
    \end{tikzpicture}
    \label{fig:labA}}
    \end{minipage}%
    \hfill%
    \begin{minipage}[b]{0.33\linewidth}
    \centering	
    \subfigure[]{
        \begin{tikzpicture}
        \foreach \n in {-5,...,40}{
            \draw[-,myLightGray,thin] ({\n/10},-0.5) -- ({\n/10},3);
        }
        \draw[->] (-0.5,0) -- (4,0) node[pos=0.95, below] {$t$};
        \draw[->] (0,-0.5) -- (0,3) node[pos=0.95, left] {$\omega$};

        \begin{scope}[shift={(1,0)}]
        \clip(-0.5,-0.5) rectangle (4,3);
            \draw[black,thin,fill=mySilver,rounded corners=0.05] (0,-0.75) rectangle (0.1,3.5);
        \end{scope}
        
    \end{tikzpicture}
    \label{fig:labB}}
    \end{minipage}

    \vspace*{-0.1\linewidth}

    \begin{minipage}[b]{0.33\linewidth}
    \centering	
    \subfigure[]{
        \begin{tikzpicture}
        \pgfmathsetmacro{\myWidth}{0.75}
        \pgfmathsetmacro{\myHeight}{0.5}
        \begin{scope}
        \clip(-0.5,-0.5) rectangle (4,3);
        \begin{scope}[shift={(0.6,0)}]
        \begin{scope}[rotate=30]
            \foreach \n in {-25,...,30}{
                \draw[-,myLightGray,thin] (-1,{\n/10}) -- (5,{\n/10});
            }
            \begin{scope}[shift={(0.25,{.3333/2})}]
            \draw[black,thin,fill=mySilver,rounded corners=7.5] (1,1) rectangle ({1+\myWidth},{1+\myHeight});
        \end{scope}
            \end{scope}
        \end{scope}
        \end{scope}
        \draw[->] (-0.5,0) -- (4,0) node[pos=0.95, below] {$t$};
        \draw[->] (0,-0.5) -- (0,3) node[pos=0.95, left] {$\omega$};
    \end{tikzpicture}
    \label{fig:labBB}}
    \end{minipage}

    \vspace*{-0.1\linewidth}
    
    \begin{minipage}[b]{0.33\linewidth}
    \centering	
    \subfigure[]{
        \begin{tikzpicture}
        \foreach \n in {-5,...,30}{
            \draw[-,myLightGray,thin] (-0.5,{\n/10}) -- (4,{\n/10});
        }
        \draw[->] (-0.5,0) -- (4,0) node[pos=0.95, below] {$t$};
        \draw[->] (0,-0.5) -- (0,3) node[pos=0.95, left] {$\omega$};

        \begin{scope}[shift={(0,{.3333/2})}]
                \clip(-0.5,-0.5) rectangle (4,3);
            \draw[black,thin,fill=mySilver,rounded corners=0.05] (-0.75,1) rectangle (4.5,{1+0.1});
        \end{scope}
        
    \end{tikzpicture}
    \label{fig:labD}}
    \end{minipage}%
    \hfill%
    \pgfmathsetmacro{\myWidth}{0.75}
    \pgfmathsetmacro{\myHeight}{0.5}
    \begin{minipage}[b]{0.33\linewidth}
    \centering	
    \subfigure[]{
        \begin{tikzpicture}
        \begin{scope}
        \clip(-0.5,-0.5) rectangle (4,3);
        \foreach \n in {-5,...,30}{
            \draw[-,myLightGray,thin] (-0.5,{\n/10}) -- (4,{\n/10});
        }
        \begin{scope}[shift={(0.25,{.3333/2})}]
        \draw[black,thin,fill=mySilver,rounded corners=7.5] (1,1) rectangle ({1+\myWidth},{1+\myHeight});
        \end{scope}
        \end{scope}
        \draw[->] (-0.5,0) -- (4,0) node[pos=0.95, below] {$t$};
        \draw[->] (0,-0.5) -- (0,3) node[pos=0.95, left] {$\omega$};
    \end{tikzpicture}
    \label{fig:labE}}
    \end{minipage}%

    \caption{Channelizing the time-frequency plane for common analyses: (a) FrFT, (b) TD (c) STFrFT, (d) FD, and (e) STFT. There is a shaded region (\textcolor{mySilver}{\solidXXLrule[2mm]}) representing the effective duration and effective (fractional) bandwidth of the template (time-shifted and/or frequency-shifted). While the duration-bandwidth product lower bounds the area of the shaded region, it does not limit the calculation of instantaneous parameters of the signal component in each channel. The filterbank methods may be understood to have overlapping passbands each covering a range of channels.}
    \label{fig:TFplane}
\end{figure*}

Additionally, choosing $\alpha$ and $r$ allows one to specialize the quadratic chirplet atom. This is illustrated in Figure \ref{fig:atoms}, where each choice of atom defines a channelizing of the time-frequency plane illustrated in Figure \ref{fig:TFplane}. For example, choosing $\alpha=0$, $r=0$ leads to the Fourier atom with channelizing as illustrated in Figure \ref{fig:TFplane}(d). As another example, choosing $0<\alpha<\infty$, $r=0$ leads to the Gabor atom with channelizing as illustrated in Figure \ref{fig:TFplane}(e).






In the following sections, we consider different choices of $\alpha$ and $r$ and show that several common analyses correspond to specific choices for the two parameters and we give closed-form expressions for the corresponding IS.  By doing so we demonstrate the IS as a unifying framework for the analyses under consideration.

\section{Frequency Domain Analysis}\label{sec:FD}
Frequency domain analysis is carried out through the Fourier transform (FT)
\begin{IEEEeqnarray}{rCl}\IEEEyesnumber
    Z(\omega) 
    &=& \frac{1}{\sqrt{2\pi}} \intPMinfty z(t) \eu^{-\ju\omega t} \dd t\label{eq:FTanal}
\end{IEEEeqnarray}
and the corresponding signal synthesis is given by
\begin{IEEEeqnarray}{rCl}\IEEEyesnumber     
    z(t) 
    &=& \frac{1}{\sqrt{2\pi}} \intPMinfty Z(\omega )  \eu^{\,\ju \omega t} \dd \omega\label{eq:FTsynth}
\end{IEEEeqnarray}
where we adopt the unitary convention for convenience. Note that \eqref{eq:FTanal} can be viewed as projecting onto the family of components generated by choosing $\alpha=0$ and $r=0$. Additionally, this decomposition specifies a family of components of the form
\begin{IEEEeqnarray}{c} 
    \psi_\closeomega(t;\mathscr{C}_\closeomega^\text{\sc f}) = \frac{Z(\closeomega)}{\sqrt{2\pi}} \eu^{\,\ju \closeomega t }
\end{IEEEeqnarray}
parameterized by a Fourier triplet
\begin{IEEEeqnarray}{c}
    \mathscr{C}_\closeomega^\text{\sc f} \triangleq\left( 
    \frac{|Z(\closeomega)|}{\sqrt{2\pi}},~~
    \closeomega,~~\arg\{Z(\closeomega)\}
    \right).\label{eqn:fourierContinuum}
\end{IEEEeqnarray}
Therefore, with \eqref{eqn:TFatom}, a family of Fourier atoms is given by 
\begin{IEEEeqnarray}{c} 
    \mathcal{A}_\closeomega^\text{\sc f}(t,\omega;\mathscr{C}_\closeomega^\text{\sc f}) = Z(\closeomega)\eu^{\,\ju\closeomega t} \delta(\omega-\closeomega)
    \label{eqn:fourierAtom}
\end{IEEEeqnarray}
and the IS of an atom is illustrated in Figure~\ref{fig:DiracA}. Superimposing these atoms gives the IS 
\begin{IEEEeqnarray}{rCl}    \IEEEyesnumber\label{eq:FtISclosedform} \IEEEyessubnumber*
    \mathcal{S}^\text{\sc  fd}(t,\omega) 
    &=&\intPMinfty  \mathcal{A}^\text{\sc f}_\closeomega(t,\omega;\mathscr{C}_\closeomega^\text{\sc f})  \dd\closeomega \\
    &=&\intPMinfty \sqrt{2\pi} \frac{|Z(\closeomega)|}{\sqrt{2\pi}} \exp\big(\ju [\closeomega t+\arg\{Z(\closeomega)\}]\big)\nonumber\\
    & & \quad\times\delta(\omega-\closeomega)  \dd\closeomega \IEEEyessubnumber*\\
    &=&\intPMinfty Z(\closeomega )  \eu^{\,\ju \closeomega t} \delta(\omega-\closeomega)  \dd\closeomega \\
    &=& Z(\omega )  \eu^{\,\ju \omega t} \label{eqn:ISshc}. 
\end{IEEEeqnarray}
Finally, we show that the AM--FM model in \eqref{eq:AMFM_A} specializes to the synthesis equation for Fourier analysis by applying the projection equation in (\ref{eq:HSintOmega}) to \eqref{eqn:ISshc}
\begin{IEEEeqnarray}{rCl}    \IEEEyesnumber\label{eq:Ftclosedform}
    z(t) = \frac{1}{\sqrt{2\pi}} \intPMinfty Z(\omega )  \eu^{\,\ju \omega t} \dd \omega.
\end{IEEEeqnarray}


\section{Fractional Fourier Analysis}\label{sec:FrFT}
\begin{landscape}
\begin{figure*}[p]
\centering
  	\hspace*{-2.5in}\begin{minipage}[b]{0.6\linewidth}
  		\centering	
  		\subfigure[]{
  		\includegraphics[trim=115 50 70 50,clip,width = 0.99\linewidth]{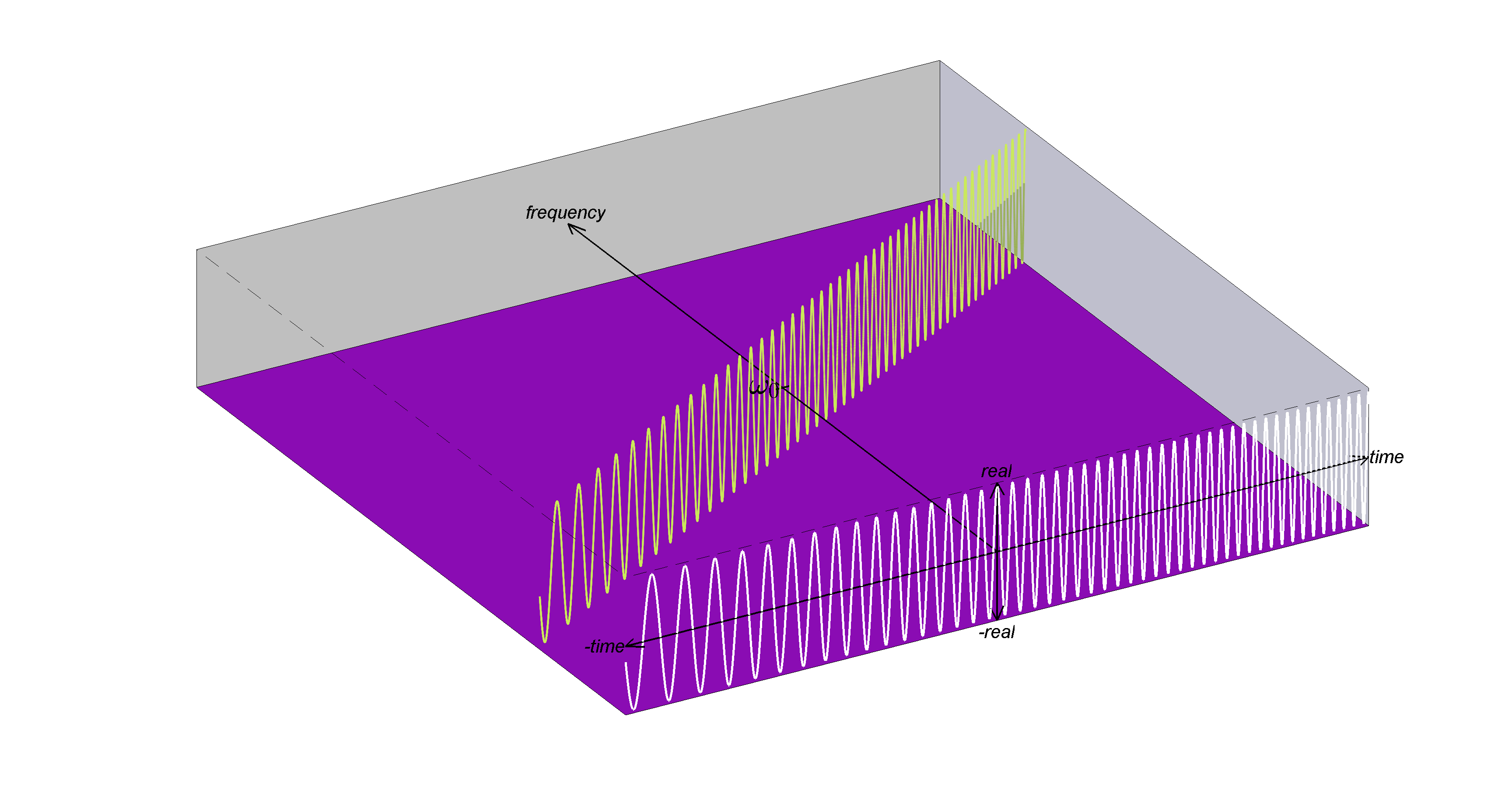}
  		\label{fig:DiracE}
  	}
  	\end{minipage}\hfill\begin{minipage}[b]{0.6\linewidth}
  		\centering	
  		\subfigure[]{
  		\includegraphics[trim=115 50 70 50,clip,width = 0.99\linewidth]{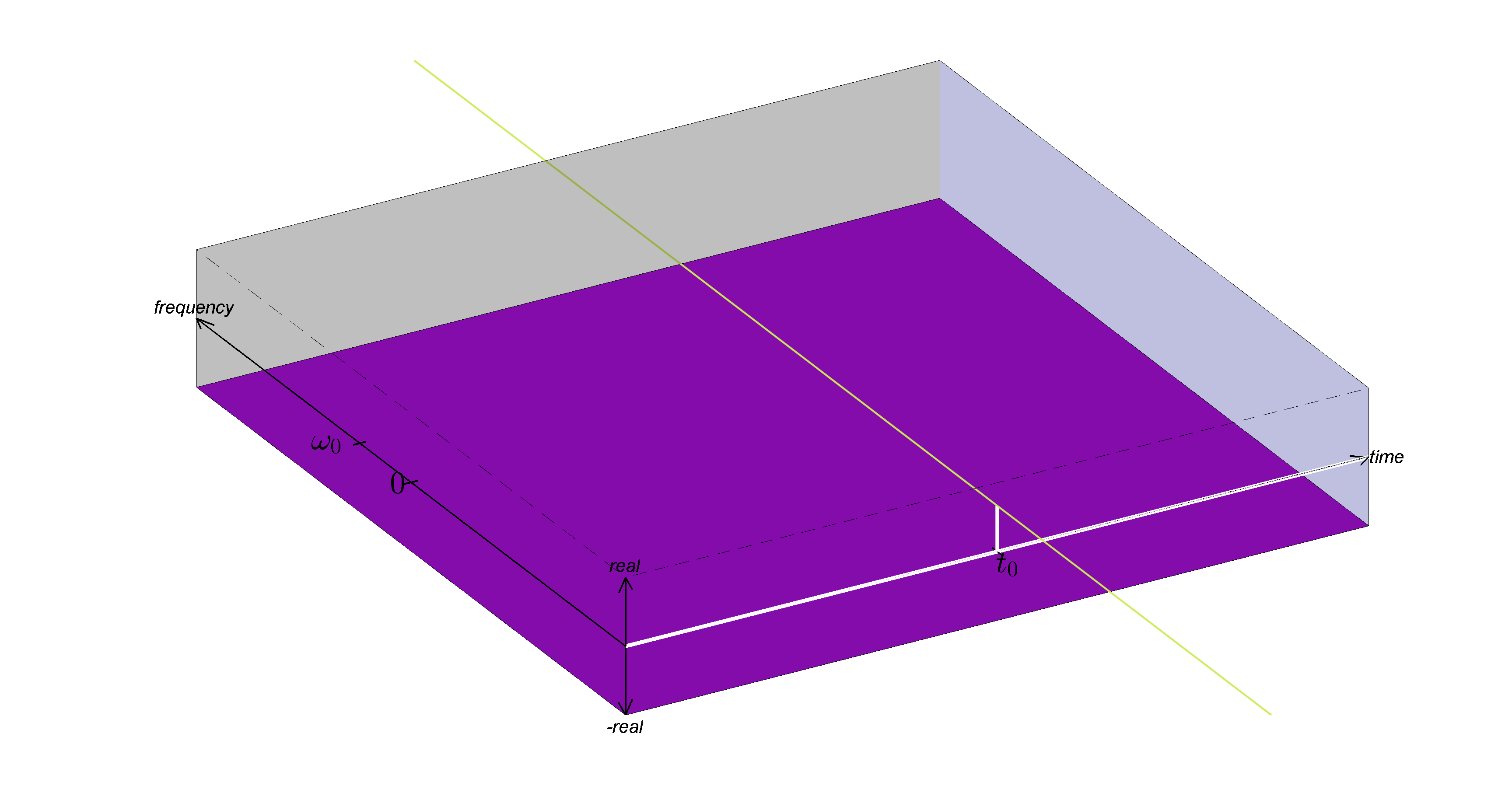}
  		\label{fig:DiracD}
  	}
  	\end{minipage}
   
 \vspace*{-1cm}

   \hspace*{-2.5in}\hfill\begin{minipage}[b]{0.6\linewidth}
  		\centering	
  		\subfigure[]{
  		\includegraphics[trim=115 50 70 50,clip,width = 0.99\linewidth]{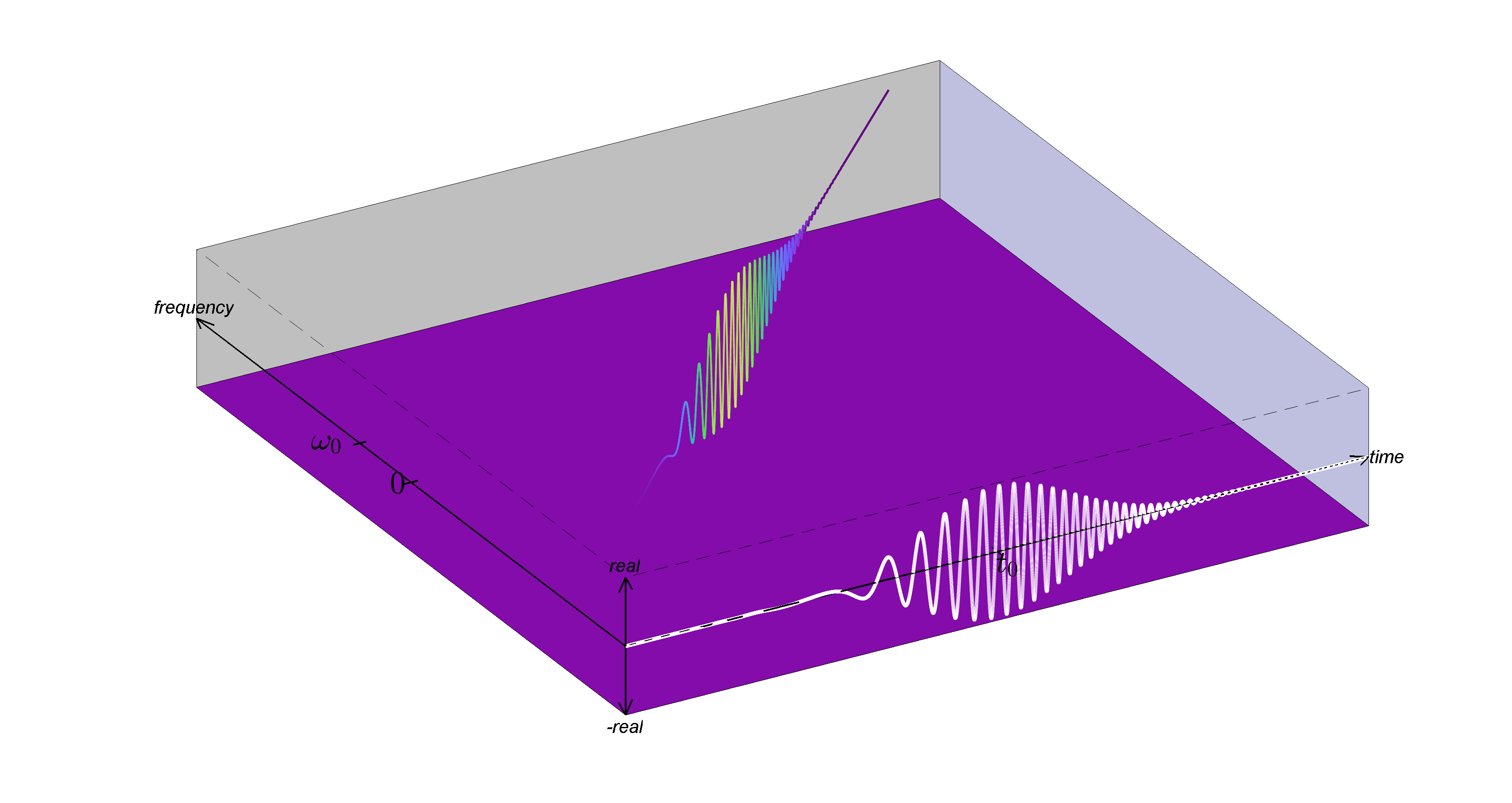}
  		\label{fig:DiracC}
  	}
  	\end{minipage}\hfill\null

\vspace*{-1cm}
   
   \hspace*{-2.5in}\begin{minipage}[b]{0.6\linewidth}
  		\centering
  		\subfigure[]{
  		{\includegraphics[trim=115 50 70 50,clip,width = 0.99\linewidth]{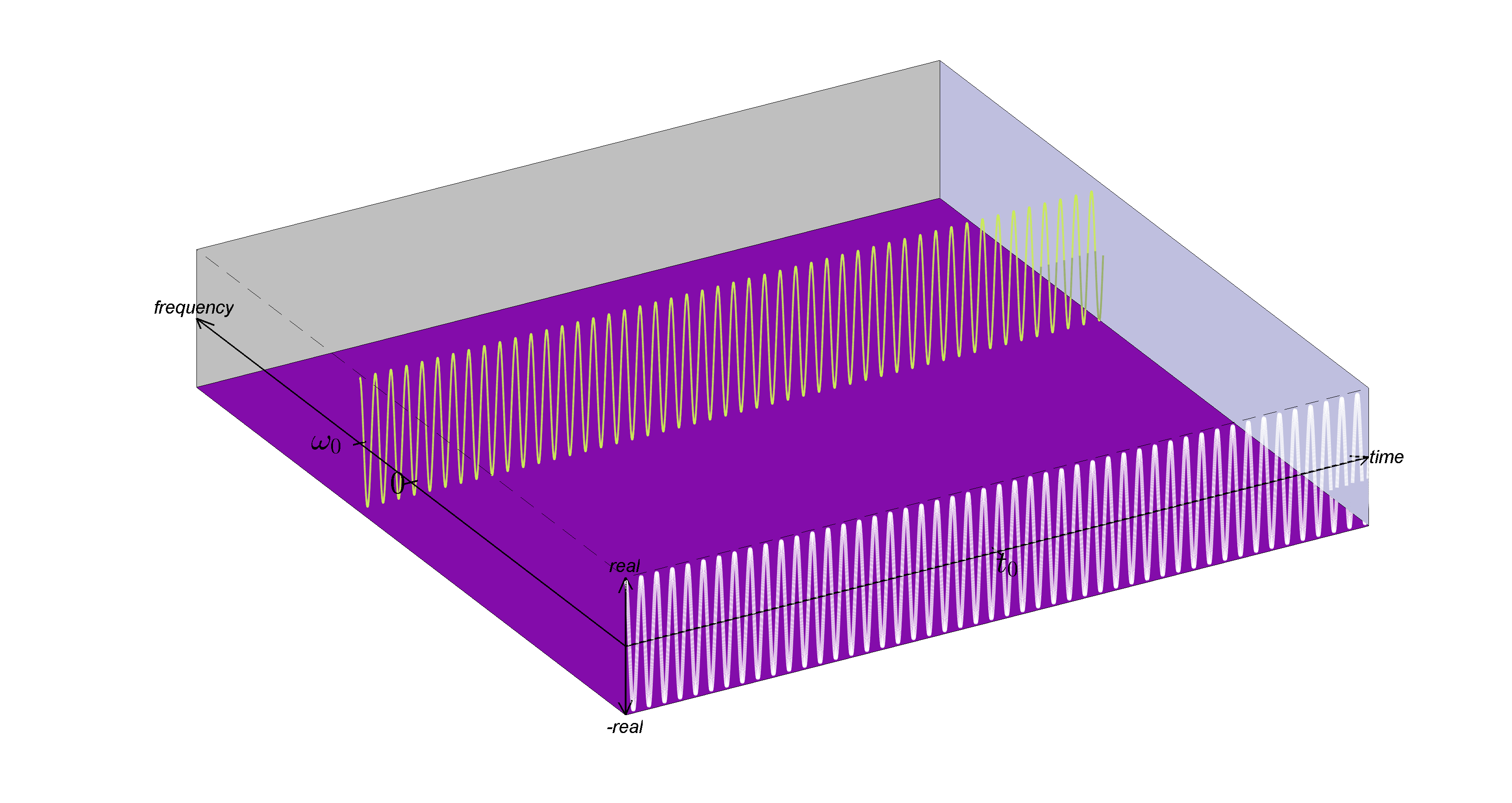}}
  		\label{fig:DiracA}
  	}
  	\end{minipage}\hfill\begin{minipage}[b]{0.6\linewidth}
  		\centering
  		\subfigure[]{
  		{\includegraphics[trim=115 50 70 50,clip,width = 0.99\linewidth]{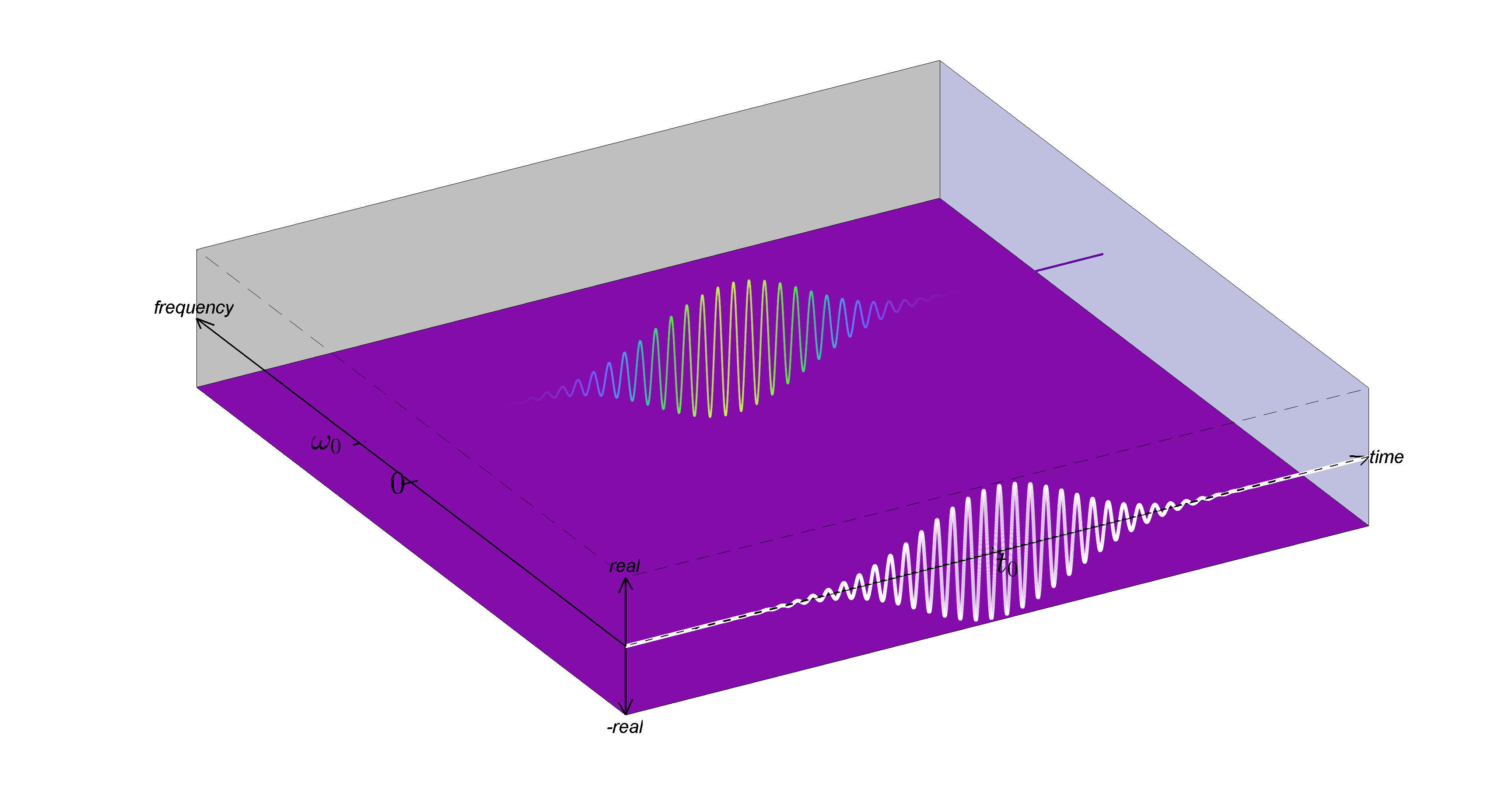}}
  		\label{fig:DiracB}
  	}
  	\end{minipage}

  	\hspace*{-2.5in}\begin{minipage}[b]{9in}\caption{The IS (\myColorLine) associated with %
   (a) Namias atom, %
   (b) Shannon atom, %
   (c) quadratic chirplet atom,
   (d) Fourier atom, and %
   (e) Gabor atom. %
   }\end{minipage}
  	  \label{fig:Dirac}	
\end{figure*}
\end{landscape}

Fractional Fourier analysis \cite{Ozaktas2001fractionalfourier,ozaktas1999introduction,namias1980fractional,mcbride1987namias,almeida1993introduction,condon1937immersion} is carried out through the fractional FT of order $0<p<2$ and angle $\gamma = p\pi/2$
\begin{IEEEeqnarray}{c}\IEEEyesnumber
    Z_\gamma (u)   \triangleq \frac{1}{\sqrt{2\pi}} \intPMinfty z(t)C_\gamma \eu^{-\ju\theta_\gamma(t;u)} \dd t\label{eq:FrFTanal}
\end{IEEEeqnarray}
where 
\begin{IEEEeqnarray}{c}\IEEEyesnumber
    \textcolor{black}{C_\gamma  = \frac{\exp({-\ju[\pi/4-\gamma/2]})}{\sqrt{\sin(\gamma)}}}
\end{IEEEeqnarray}
with phase $\theta_\gamma(t;u) = -\frac{r}{2}t^2 +c_\mathrm{x} u t - \frac{r}{2}u^2$, chirp rate $r = \cot(\gamma)$, and crossing constant $c_\mathrm{x}=\csc(\gamma)$. Note that \eqref{eq:FrFTanal} can be viewed as projecting onto the family of components generated by choosing $\alpha=0$ and the chirp rate $r$ as described above. The corresponding signal synthesis is given by
\begin{IEEEeqnarray}{rCl}\IEEEyesnumber     
    z(t) 
    &=&  \frac{1}{\sqrt{2\pi}} \intPMinfty   Z_\gamma (u) C_\gamma^\dagger  \eu^{\,\ju \theta_\gamma(t;u)} \dd u.\label{eq:FrFtsynth}
\end{IEEEeqnarray}
The fractional FT decomposes a signal into a family of Namias components each of the form ($\,{}^\dagger$ denotes complex conjugate) 
\begin{IEEEeqnarray}{c} 
    \psi_u\big(t;\mathscr{C}_u^\text{\sc n}(\gamma)\big) = \frac{Z_\gamma (u) C_\gamma^\dagger}{\sqrt{2\pi}}  \eu^{\,\ju \theta_\gamma(t;u)} \label{eq:NamiasComponent}
\end{IEEEeqnarray}
parameterized by a Namias triplet
\begin{IEEEeqnarray}{c}
    \mathscr{C}_u^\text{\sc n}(\gamma) \!\triangleq\! \left( 
     \frac{|Z_\gamma (u)C_\gamma^\dagger|}{\sqrt{2\pi}} ,~\!
    -r t \!+\! c_\mathrm{x} u,~\!-\frac{ru^2}{2}
    \!+\! \arg\{Z_\gamma (u)C_\gamma^\dagger\}\right).~~~~~~\label{eqn:frftTriplet}
\end{IEEEeqnarray}
Therefore, with \eqref{eqn:TFatom}, a family of Namias atoms is given by 
\begin{IEEEeqnarray}{rCl} 
    \mathcal{A}_{\gamma ,u}^\text{\sc n}(t,\omega;\mathscr{C}_u^\text{\sc n}(\gamma)) &=& Z_\gamma (u) C_\gamma^\dagger  \eu^{\,\ju \theta_\gamma(t;u)}\nonumber\\
    &~&\quad\times \delta\left(\omega - \vertOne \left(-r t + c_\mathrm{x} u\right)   \vertOne\right)
    \label{eqn:FrFTAtom}
\end{IEEEeqnarray}
and the IS of an atom is illustrated in Figure~\ref{fig:DiracE}. Superimposing these atoms gives the IS
\begin{IEEEeqnarray}{rCl}\IEEEyesnumber\IEEEyessubnumber*     
    \mathcal{S}_\gamma ^\text{\sc ff}(t,\omega) &=&\intPMinfty  \mathcal{A}_{\gamma ,u}^\text{\sc n}(t,\omega;\mathscr{C}_u^\text{\sc n}(\gamma))  \dd u\\
    &=&\intPMinfty \!\!\!  Z_\gamma (u) C_\gamma^\dagger  \eu^{\,\ju \theta_\gamma(t;u)} \delta\left(\omega - \vertOne \left(-r t + c_\mathrm{x} u\right)   \vertOne\right) \dd u ~~~~~~~ \label{eq:namiasReconB}\\
    &=&  Z_\gamma \big(   [ \omega  +r t ]/ c_\mathrm{x} \big) C_\gamma^\dagger  \exp\big(\ju \theta_\gamma(t;[ \omega  +r t ]/ c_\mathrm{x})\big).\label{eq:namiasRecon}
\end{IEEEeqnarray}

Finally, we show that the AM--FM model in \eqref{eq:AMFM_A} specializes to the synthesis equation for fractional Fourier analysis by applying the projection equation in (\ref{eq:HSintOmega}) to \eqref{eq:namiasReconB}
\begin{IEEEeqnarray}{rCl}\IEEEyesnumber     
    \frac{1}{\sqrt{2\pi}}\intPMinfty\mathcal{S}_\gamma ^\text{\sc ff}(t,\omega) \dd \omega 
    &=&  \frac{1}{\sqrt{2\pi}}\intPMinfty  \intPMinfty   Z_\gamma (u) C_\gamma^\dagger  \eu^{\,\ju \theta_\gamma(t;u)}\nonumber\\&~&\quad \times\delta\left(\omega - \vertOne \left(-r t + c_\mathrm{x} u\right)   \vertOne\right) \dd u  \dd \omega \nonumber\\
    &=&  \frac{1}{\sqrt{2\pi}} \intPMinfty   Z_\gamma (u) C_\gamma^\dagger  \eu^{\,\ju \theta_\gamma(t;u)} \dd u  \nonumber\\
    &=& z(t).
\end{IEEEeqnarray}

\section{STFT Analysis}\label{sec:STFT}


Short-time Fourier transform analysis \cite{Allen1977ShortTime, lim1987advanced} using the filterbank interpretation is carried out with
\begin{IEEEeqnarray}{rCl}\IEEEyesnumber\IEEEyessubnumber*
     Z_\w(t;\nu)  &=& z(t) * \textcolor{black}{[\w(t)\eu^{\,\ju\nu t}]} \label{eqn:STFTanal}\\ 
     &=&  a_\w(t;\nu) \eu^{\,\ju \theta_\w(t;\nu)}
     \label{eqn:polarSTFT}
\end{IEEEeqnarray}
with channelizer frequency $\nu$ and real even-symmetric window function $\w(t)$ and the corresponding signal synthesis is given by
\begin{IEEEeqnarray}{rCl}\IEEEyesnumber\IEEEyessubnumber*
     z(t) &\triangleq& \dfrac{1}{2\pi\w(0)}\intPMinfty Z_\w(t;\nu)  \dd\nu\label{eq:FBanalysisSynth}\\
     &=&\dfrac{1}{2\pi\w(0)}\intPMinfty a_\w(t;\nu) \eu^{\,\ju \theta_\w(t;\nu)} \dd\nu\label{eq:FBanalysisSynthB}.
\end{IEEEeqnarray}
Note that \eqref{eqn:STFTanal} can be viewed as the signal filtered using the family of components as impulse responses for each channel. The special case of a Gaussian window corresponds to the family of components generated with $0<\alpha<\infty$ and $r=0$. However, in the interest of generality we proceed with a general window function. The STFT decomposes a signal into a family of AM--FM components each of the form
\begin{IEEEeqnarray}{c} 
    \psi_\nu(t;\mathscr{C}_\nu^\text{\sc  stft}) = \frac{1}{2\pi\w(0)}a_\w(t;\nu) \eu^{\,\ju \theta_\w(t;\nu) \vertOne}
\end{IEEEeqnarray}
parameterized by a triplet
\begin{IEEEeqnarray}{c}
    \mathscr{C}_\nu^\text{\sc  stft} \triangleq\left( 
    \frac{1}{2\pi\w(0)}a_\w(t;\nu),~
    \frac{\dd}{\dd t}\theta_\w(t;\nu),~\theta_\w(0;\nu)
    \right).~~~~\label{eqn:stftTriplet}
\end{IEEEeqnarray}
In \cite{sandoval2022recasting}, we used the ISA framework to recast the filterbank interpretation of the STFT in terms of an IS  and showed that synchrosqueezing is necessary for a valid IS. With \eqref{eqn:TFatom}, a family of synchrosqueezed STFT atoms is given by 
\begin{IEEEeqnarray}{rCl} 
    \mathcal{A}_\nu^\text{\sc  stft}(t,\omega;\mathscr{C}_\nu^\text{\sc  stft}) &=&\frac{1}{\sqrt{2\pi}\w(0)} Z_\w(t;\nu) \nonumber\\&~&\quad\times\delta\left(\omega - \vertOne \frac{\dd}{\dd t}\theta_\w(t;\nu)   \vertOne\right).
    \label{eqn:STFTAtom}
\end{IEEEeqnarray}
Superimposing the synchrosqueezed \cite{kodera1976new, kodera1978analysis,meignen2019synchrosqueezing, kodera1978analysis, auger1994and, auger1995improving, auger2013time, sandoval2022recasting} STFT atoms gives the IS
\begin{IEEEeqnarray}{rCl}\IEEEyesnumber\IEEEyessubnumber*     
    \mathcal{S}^\text{\sc  stft}(t,\omega) &=&\intPMinfty  \mathcal{A}^\text{\sc  stft}(t,\omega;\mathscr{C}_\nu^\text{\sc  stft})  \dd\nu\\
                                        &=&\intPMinfty  \frac{1}{\sqrt{2\pi}\w(0)} Z_\w(t;\nu)  \nonumber\\
                                        &~&\quad\times\delta\left(\omega - \vertOne \frac{\dd}{\dd t}\theta_\w(t;\nu)   \vertOne\right)
  \dd\nu.~~~~~~\label{eq:STFTis}
\end{IEEEeqnarray}
Finally, we show that the AM--FM model in \eqref{eq:AMFM_A} specializes to the filterbank summation method for short-time synthesis by applying the projection equation in (\ref{eq:HSintOmega}) to \eqref{eq:STFTis}
\begin{IEEEeqnarray}{rCl}\IEEEyesnumber     
    \frac{1}{\sqrt{2\pi}}\intPMinfty\mathcal{S}^\text{\sc  stft}(t,\omega;\mathscr{S}) \dd\omega
    &=&   \dfrac{1}{2\pi\w(0)}\intPMinfty Z_\w(t;\nu)  \dd\nu\nonumber\\
    &=& z(t).
\end{IEEEeqnarray}

\section{STFrFT Analysis}\label{sec:STFrFT}
Short-time fractional Fourier transform  \cite{shi2020novel} analysis using the filterbank interpretation is carried out with
\begin{IEEEeqnarray}{rCl}\IEEEyesnumber\IEEEyessubnumber*
      Z_{\gamma,\w}(t,u) &=& z(t)*_\gamma \textcolor{black}{[\w(t) \eu^{\,\ju c_\mathrm{x} u t}]} \label{eqn:STFrFTanal}\\
     &=&  a_{\gamma,\w}(t;u) \eu^{\ju \theta_{\gamma,\w}(t;u)}
     \label{eqn:polarSTFrFT}
\end{IEEEeqnarray}
with fractional convolution  $*_\gamma$ defined as
\begin{IEEEeqnarray}{rCl}\IEEEyesnumber\IEEEyessubnumber*
    f(t) *_\gamma g(t) &=& \intPMinfty f(\tau)g(t-\tau)\eu^{-\ju r/2(t^2-\tau^2) } \dd\tau\\
     &=& \eu^{-\ju \frac{r}{2} t^2} \big[\big(f(t)\eu^{\,\ju \frac{r}{2} t^2}\big)*g(t)\big].
\end{IEEEeqnarray}
Similar to the STFT, \eqref{eqn:STFrFTanal} can be viewed as the signal filtered using the family of components as impulse responses for each channel. The special case of a Gaussian window corresponds to the family of components generated with $0<\alpha<\infty$ and with chirp rate $r$ as previously described. However, in the interest of generality we proceed with a general window function. The corresponding signal synthesis is given by
\begin{IEEEeqnarray}{rCl}\IEEEyesnumber\IEEEyessubnumber*
    z(t) &=&  \frac{c_\mathrm{x}}{2\pi\w(0)}\intPMinfty   Z_{\gamma,\w}(t;u)  \eu^{\,\ju c_\mathrm{x} u t}  \dd u \\
         &=&  \frac{c_\mathrm{x}}{2\pi\w(0)}\intPMinfty   a_{\gamma,\w}(t;u) \eu^{\ju \theta_{\gamma,\w}(t;u)}  \eu^{\,\ju c_\mathrm{x} u t}  \dd u .~~~\label{eq:STFrFTsynthB}
\end{IEEEeqnarray}
The STFrFT decomposes a signal into a family of AM--FM components each of the form
\begin{IEEEeqnarray}{c} 
    \psi_u(t;\mathscr{C}_u^\text{\sc  stfrft}) = \frac{c_\mathrm{x}a_{\gamma,\w}(t;u)}{2\pi\w(0)} \eu^{\,\ju \theta_{\gamma,\w}(t;u)}  \eu^{\,\ju c_\mathrm{x} u t} 
\end{IEEEeqnarray}
parameterized by a triplet
\begin{IEEEeqnarray}{c}
    \mathscr{C}_u^\text{\sc  stfrft} \!\triangleq\! \left( 
    \frac{c_\mathrm{x}a_{\gamma,\w}(t;u)}{2\pi\w(0)},~\!
        \frac{\dd}{\dd t}\theta_{\gamma,\w}(t;u)+c_\mathrm{x}u,~\!\theta_{\gamma,\w}(0;u)
    \right).~~~~~~\label{eqn:stfrftTriplet}
\end{IEEEeqnarray}
With \eqref{eqn:TFatom}, a family of synchrosqueezed STFrFT atoms is given by 
\begin{IEEEeqnarray}{rCl} 
    \mathcal{A}_u^\text{\sc  stfrft}(t,\omega;\mathscr{C}_u^\text{\sc  stfrft}) &=& \frac{c_\mathrm{x}}{\sqrt{2\pi}\w(0)} Z_{\gamma,\w}(t;u)  \eu^{\,\ju c_\mathrm{x} u t} \nonumber\\
    &~&\quad\times \delta\left(\omega - \vertOne \frac{\dd}{\dd t}\theta_{\gamma,\w}(t;u)   \vertOne\right).
    \label{eqn:STFrFTAtom}
\end{IEEEeqnarray}
Following the development of the IS corresponding to the synchrosqueezed STFT \cite{sandoval2022recasting}, we proceed as follows. The IS is \textcolor{black}{constructed} from synchrosqueezed STFrFT atoms
\begin{IEEEeqnarray}{rCl}\IEEEyesnumber\IEEEyessubnumber*     
    \mathcal{S}^\text{\sc  stfrft}(t,\omega) &=&\intPMinfty  \mathcal{A}^\text{\sc  stfrft}(t,\omega;\mathscr{C}_u^\text{\sc  stfrft})  \dd u \\
   &=&\intPMinfty  \frac{c_\mathrm{x}}{\sqrt{2\pi}\w(0)} Z_{\gamma,\w}(t;u)  \eu^{\,\ju c_\mathrm{x} u t} \nonumber\\&~&\quad\times\,\delta\left(\omega - \vertOne \frac{\dd}{\dd t}\theta_{\gamma,\w}(t;u)   \vertOne\right)  \dd u .\label{eq:isFRSTFT}
\end{IEEEeqnarray}
Finally, we show that the AM--FM model in \eqref{eq:AMFM_A} specializes to the filterbank summation method for fractional short-time synthesis by applying the projection equation in (\ref{eq:HSintOmega}) to \eqref{eq:STFTis}
\begin{IEEEeqnarray}{l}\IEEEyesnumber\IEEEyessubnumber*     
\frac{1}{\sqrt{2\pi}}\intPMinfty\mathcal{S}^\text{\sc  stfrft}(t,\omega;\mathscr{S}) \dd\omega\\
   \quad =  \frac{1}{2\pi}\intPMinfty\intPMinfty  \frac{c_\mathrm{x}}{\w(0)}   \eu^{\,\ju c_\mathrm{x} u t} \,\delta\left(\omega - \vertOne \frac{\dd}{\dd t}\theta_{\gamma,\w}(t;u)   \vertOne\right)  \dd u \dd\omega\nonumber\\
   \quad =  \frac{c_\mathrm{x}}{2\pi\w(0)}\intPMinfty   Z_{\gamma,\w}(t;u)  \eu^{\,\ju c_\mathrm{x} u t}  \dd u \\
   \quad = z(t).
\end{IEEEeqnarray}


\section{Discussion} \label{sec:Discussion}
In Sections III -- VI, all the analysis methods described can be viewed as a choice of a quadratic chirplet template in the parameters $\alpha$ and $r$. We point out that the analyses under consideration can be categorized into two groups: a projection type where the same component is used in analysis and synthesis, and a filterbank type where the synthesis components do not have the same form as the analysis components (and thus require synchrosqueezing). For example, the FT uses the same family of components $\psi_\closeomega(t;\mathscr{C}_\closeomega^\text{\sc f})$ in synthesis \eqref{eq:FTsynth} that is projected onto in analysis \eqref{eq:FTanal}. Likewise, the FrFT uses the same family of components $\psi_u\big(t;\mathscr{C}_u^\text{\sc n}(\gamma)\big)$ in synthesis \eqref{eq:FrFtsynth} that is projected onto in analysis \eqref{eq:FrFTanal}. On the other hand, the STFT uses bandpass AM--FM components $\psi_\nu(t;\mathscr{C}_\nu^\text{\sc  stft})$ in synthesis \eqref{eq:FBanalysisSynthB} which do not match the family obtained from the template $\w(t)\eu^{\,\ju\nu t}$ used in analysis \eqref{eqn:STFTanal}. Similarly, the STFrFT uses fractional bandpass AM--FM components $\psi_u(t;\mathscr{C}_u^\text{\sc  stfrft})$ in synthesis \eqref{eq:STFrFTsynthB} which do not match the family obtained from the template $\w(t)\eu^{\,\ju c_\mathrm{x} u t}$ used in analysis \eqref{eqn:STFrFTanal}.  Furthermore with the STFT and STFrFT, due to the mismatch in the analysis and synthesis components, the IF of the synthesis components are unknown prior to performing analysis. This unknown forces a phase derivative to be calculated to determine the atom [see \eqref{eqn:STFTAtom} and \eqref{eqn:STFrFTAtom}], which we have shown in \cite{sandoval2022recasting} to be equivalent to synchrosqueezing.

While the IS corresponding to the analyses thus far are easily developed, this is not true for the TD analysis which considers unit impulses as the basic components. This is because the unit impulse is not directly an AM-FM component and thus cannot be expressed with a component triplet. \textit{Therefore we ask, does a time-frequency atom in a limiting form of \eqref{eqn:TFatom} exist, corresponding to the unit impulse?} In the following section, we propose an atom corresponding to the unit impulse which we term the Shannon atom. This atom is then used to specify the IS corresponding to TD analysis.


\section{The Shannon Atom}\label{sec:ShannonAtom}

In Gabor's seminal paper \cite{gaborTOC}, a time-frequency analysis was described in which signals were decomposed into a set of Gaussian AM components each of the form\footnote{Gabor actually referred to these basic components as logons.}
\begin{IEEEeqnarray}{c}
    \psi_k(t;\mathscr{C}_k^\text{\sc g}) =  \exp\left[-\alpha^2(t-t_k)^2\right] \eu^{\,\ju( \omega_k t +\phi_k)} \label{eqn:gaborComp}
\end{IEEEeqnarray}
parameterized by the Gabor triplet
\begin{IEEEeqnarray}{c}
    \mathscr{C}_k^\text{\sc g} \triangleq\left( 
     \exp\left[-\alpha^2(t-t_k)^2\right],~~
    \omega_k,~~\phi_k
    \right)\label{eqn:gaborAM}
\end{IEEEeqnarray}
where $t_k$, $\omega_k$, and $\phi_k$ are constants. The corresponding time-frequency atom, termed a Gabor atom, is formed by using \eqref{eqn:gaborAM} in \eqref{eqn:TFatom} and given by
\begin{IEEEeqnarray}{rCl}
     \mathcal{A}_k^\text{\sc g}(t,\omega;\mathscr{C}_k^\text{\sc g}) &=& \sqrt{2\pi}  \exp\left(-\alpha^2(t-t_k)^2\right) \eu^{\,\ju( \omega_k t +\phi_k)} \nonumber \\
     &~& \quad\times\delta(\omega-\omega_k)
     \label{eqn:gaborAtom}
\end{IEEEeqnarray}
and the IS is illustrated in Figure~\ref{fig:DiracB}. Gabor interpreted the two limiting forms, $\alpha\rightarrow \infty$ and $\alpha\rightarrow0$ of the Gaussian AM component, to correspond to TD analysis and FD analysis, respectively \cite{gaborTOC} \textcolor{black}{as was illustrated in Figure \ref{fig:compRelations}(a).} Although modeling a unit impulse as a limiting form of a (normalized) Gaussian AM component when $\alpha\rightarrow\infty$ passes to a unit impulse, it does not lead to an atom that is meaningful because there is no meaningful choice for $\omega_k$ in \eqref{eqn:gaborAM} and \eqref{eqn:gaborAtom}. \textcolor{black}{As was illustrated in Figure \ref{fig:compRelations}(b), starting at the Fourier atom and moving right ($\alpha\rightarrow\infty$) does not take us to the Shannon atom.}

On the other hand, Namias interpreted the two limiting forms of $r\rightarrow \infty$ and $r\rightarrow0$ of the Namias component in \eqref{eq:NamiasComponent} to correspond to TD analysis and FD analysis, respectively \cite{namias1980fractional, mcbride1987namias}\textcolor{black}{~as was illustrated in Figure \ref{fig:compRelations}(a).} Although for Fractional Fourier analysis with finite $r$ the component triplet in \eqref{eqn:frftTriplet} is well-defined, in the limiting case it does not lead to an atom that is meaningful because there is no meaningful choice for the IA in \eqref{eqn:frftTriplet} and \eqref{eqn:FrFTAtom}. \textcolor{black}{As was illustrated in Figure \ref{fig:compRelations}(b), starting at the Fourier atom and moving up ($r\rightarrow\infty$) does not take us to the Shannon atom.}


\color{black}

As we will see, the combination of Gabor's IA approach and Namias' IF approach provides a solution to understanding TD analysis as an IS. To this end, we define the quadratic chiplet triplet as
\begin{IEEEeqnarray}{c}
    \mathscr{C}^\text{\sc q}(\beta) \triangleq\left( 
    \dfrac{\beta}{ \sqrt{2\pi}} \exp\left(\dfrac{-\beta^2t^2}{2}\right),~~
    -\beta t+\omega_\mathrm{x},~~0
    \right)
    \label{eqn:limitingComponent}
\end{IEEEeqnarray}
where $\beta$ is a parameter and $\omega_\mathrm{x}$ is the IF crossing \cite{bracewell1980fourier, mann1995chirplet}. 
From this parameterization and using (\ref{eq:AMFMcompS}), we have the quadratic chirplet component
\begin{IEEEeqnarray}{c}
    \psi\big(t;\mathscr{C}^\text{\sc q}(\beta)\big)  = \dfrac{\beta}{ \sqrt{2\pi}} \exp\left(\dfrac{-\beta^2t^2}{2}\right) \exp\left[\ju \theta(t;\beta) \vertOne\right]\label{eqn:GAMLFM}
\end{IEEEeqnarray}
where $\theta(t;\beta) = \int_{-\infty}^{t}\left( -\beta \uptau+\omega_\mathrm{x}\right)\dd\uptau $. From the parameterization in \eqref{eqn:limitingComponent} and using (\ref{eqn:TFatom}), the quadratic chirplet atom is
\begin{IEEEeqnarray}{rCl}
     \mathcal{A}^\text{\sc q}\big(t,\omega;\mathscr{C}^\text{\sc q}(\beta)\big) &=&  \beta \exp\left(\dfrac{-\beta^2t^2}{2}\right) \exp\left[\ju \theta(t;\beta) \vertOne\right] \nonumber \\ 
     &~& \quad\times    \delta(\omega-[-\beta t+\omega_\mathrm{x}])
     \label{eqn:QuadraticChirpletAtom}
\end{IEEEeqnarray}
and the IS is illustrated in Figure~\ref{fig:DiracC}.

Next, we propose the Shannon atom as a \textit{limiting} form of the quadratic chirplet atom. Our procedure, in the following subsections, is outlined as follows:
\begin{enumerate}
    \item[\textit{A.}] define the unit impulse and the unit dispersion (which is dual to the unit impulse) as hyperreal functions,
    \item[\textit{B.}] show that the limiting form of the quadratic chirplet component is a unit impulse,
    \item[\textit{C.}] show that the limiting form of the quadratic chirplet atom is a product of a unit impulse and unit dispersion.
\end{enumerate}

\subsection{Defining the Unit Impulse and Unit Dispersion Utilizing Non-Standard Analysis}

We utilize non-standard analysis \cite{robinson1974non,hoskins2009delta} to define the unit impulse and unit dispersion, the latter of which has properties that are dual to the unit impulse, as hyperreal functions. Let $\!{}^{\scalebox{0.7}{$\divideontimes$}}\!\!\:\mathbb{R}$ be the hyperreal number system containing the standard real numbers $\mathbb{R}$, non-zero infinitesimals ($\epsilon \approx 0$), and unlimited numbers ($\myUpsilon=1/\epsilon\approx \infty$).

We define the hyperreal Gaussian function $\mathrm{G}_\beta(u)$ using a hyperreal parameter $\beta$
\begin{IEEEeqnarray}{c}
    \mathrm{G}_\beta(u) = \frac{\beta}{\sqrt{2\pi}} \exp\left(\frac{-\beta^2 u^2}{2}\right).
    \label{eq:hyperreal_gaussian}
\end{IEEEeqnarray}
The action of the hyperreal function $\mathrm{G}_\beta(u)$ as a distribution $D$ is the standard part of the hyperreal integral
\begin{IEEEeqnarray}{c}
    \langle D, f \rangle = \text{st}\left( \int_{-\infty}^\infty \mathrm{G}_\beta(u) f(u) \dd u \right)
\end{IEEEeqnarray}
where $f(u)$ is a standard test function.

\subsubsection{The Unit Impulse}
The unit impulse $\delta(\cdot)$ can be obtained by setting the parameter $\beta$ to an unlimited number $\myUpsilon$
\begin{IEEEeqnarray}{c}
    \delta(u) \equiv \mathrm{G}_\myUpsilon(u).
\end{IEEEeqnarray}
This corresponds to the limit where the Gaussian's width approaches zero ($\beta \to \infty$). The standard part of the integral yields the Dirac delta distribution $D_\delta$
\begin{IEEEeqnarray}{c}
    \langle D_\delta, f \rangle = \text{st}\left( \int_{-\infty}^\infty \delta(u) f(u) \dd u \right) = f(0).
\end{IEEEeqnarray}
Furthermore, the hyperreal function $\delta(u)$ has unit area
\begin{IEEEeqnarray}{c}
     \int_{-\infty}^\infty \delta(u) \dd u  = 1
     \label{eqn:deltaProp2}
\end{IEEEeqnarray}
and has zero value
\begin{IEEEeqnarray}{c}
     \delta(u) = 0
     \label{eqn:deltaProp1}
\end{IEEEeqnarray}
for standard non-zero $u\in \mathbb{R}$.

\subsubsection{The Unit Dispersion}
The unit dispersion $\varepsilon(\cdot)$ can be  obtained by setting the parameter $\beta$ to an infinitesimal $\epsilon$
\begin{IEEEeqnarray}{c}
    \varepsilon(u) \equiv \mathrm{G}_\epsilon(u).
\end{IEEEeqnarray}
This corresponds to the limit where the Gaussian's width approaches infinity ($\beta \to 0$). The standard part of the integral is zero, yielding the zero distribution $D_0$
\begin{IEEEeqnarray}{c}
    \langle D_0, f \rangle = \text{st}\left( \int_{-\infty}^\infty \varepsilon(u) f(u) \dd u \right) = 0.
\end{IEEEeqnarray}
Furthermore, the hyperreal function $\varepsilon(u)$ has unit area
\begin{IEEEeqnarray}{c}
     \int_{-\infty}^\infty \varepsilon(u) \dd u  = 1
     \label{eq:DispersionProp}
\end{IEEEeqnarray}
and has non-zero infinitesimal value
\begin{IEEEeqnarray}{c}
     \varepsilon(u) = \epsilon
\end{IEEEeqnarray}
for all standard $u \in \mathbb{R}$.

\subsection{Unit Impulse as a Limiting Form of the Quadratic Chirplet Component}
\label{ssec:representingunitimpulse}


\vspace{1em}\noindent\textbf{Lemma}
Let $\myUpsilon\in\!{}^{\scalebox{0.7}{$\divideontimes$}}\!\:\mathbb{R}$ be an unlimited number. The limiting form of the quadratic chirplet component in \eqref{eqn:GAMLFM} is the unit impulse. That is
\begin{IEEEeqnarray}{c}
    \lim_{\beta\rightarrow \myUpsilon}\psi\big(t;\mathscr{C}^\text{\sc q}(\beta)\big)  = \delta(t).
    \label{eq:lemma}
\end{IEEEeqnarray}
\begin{IEEEproof} 
First, the IA has Gaussian form so in the limit as ${\beta\rightarrow\myUpsilon}$, (\ref{eqn:deltaProp1})  is satisfied \cite{bracewell1980fourier}. Next, we show (\ref{eqn:deltaProp2}) as follows 
\begin{IEEEeqnarray*}{+rCl+x*}
    \IEEEeqnarraymulticol{3}{l}{ \displaystyle\lim_{\beta\rightarrow\myUpsilon}\int_{-\infty}^{\infty}\psi\big(t;\mathscr{C}^\text{\sc q}(\beta)\big)\dd t} \nonumber\\\quad
    &=& \displaystyle\lim_{\beta\rightarrow\myUpsilon}\int_{-\infty}^{\infty} \dfrac{\beta}{ \sqrt{2\pi}} \exp\left(\dfrac{-\beta^2t^2}{2}\right) \exp\left[\ju\theta(t;\beta)\vertOne\right]\dd t\nonumber\\
    &=& \displaystyle\lim_{\beta\rightarrow\myUpsilon} \frac{\beta}{\sqrt{2\pi}} \intPMinfty  \exp\left(\frac{-\beta^2-\mathrm{j}\beta}{2} t^2\right) \exp\left(\mathrm{j} \omega_\mathrm{x} t\vertOne\right)  \dd t\\
    &=& \displaystyle\lim_{\beta\rightarrow\myUpsilon} {\beta} \mathcal{F}\left\{\exp\left(\frac{-\beta^2-\mathrm{j}\beta}{2} t^2\right)\right\}_{\,\omega=\omega_\mathrm{x}}\\
    &=& 1&\nonumber
\end{IEEEeqnarray*}
where $\mathcal{F}\{\cdot\}_{\,\omega=\omega_\mathrm{x}}$ denotes the FT evaluated at $\omega=\omega_\mathrm{x}$.  Thus, both (\ref{eqn:deltaProp2}) and (\ref{eqn:deltaProp1}) are satisfied.%
\end{IEEEproof}%

\subsection{Shannon Atom as a Limiting Form of the Quadratic Chirplet Atom}
\label{ssec:ShannonAtom}

\vspace{1em}\noindent\textbf{Theorem}
The Shannon atom is the product of a unit impulse in time and a unit dispersion in frequency. That is
\begin{IEEEeqnarray}{c}
    \mathcal{A}^\text{\sc s}(t,\omega)  = \sqrt{2\pi}\delta(t)\varepsilon\left(\omega\right).
\end{IEEEeqnarray}
\begin{IEEEproof} 
Let $\myUpsilon\in\!{}^{\scalebox{0.7}{$\divideontimes$}}\!\:\mathbb{R}$ be an unlimited number. We express the Shannon atom as the limiting form of the quadratic chirplet \footnote{We drop the triplet parameter for the Shannon atom because the triplet in \eqref{eqn:limitingComponent} becomes meaningless in the limiting form.}
\begin{IEEEeqnarray}{rCl}
    \mathcal{A}^{\text{\sc s}}(t,\omega) &\equiv& \displaystyle\lim_{\beta\rightarrow\myUpsilon}\mathcal{A}^\text{\sc q}\big(t,\omega;\mathscr{C}^\text{\sc q}(\beta)\big)\IEEEyessubnumber\\
    &=& \displaystyle\lim_{\beta\rightarrow\myUpsilon}
    \sqrt{2 \pi}\,\psi\big(t;\mathscr{C}^\text{\sc q}(\beta)\big)
        \delta(\omega-[-\beta t+\omega_\mathrm{x}]).\IEEEyessubnumber~~~~~~
        \label{eqn:ShannonAtomLimiting} 
\end{IEEEeqnarray}
In the limit, the quadratic chirplet component $\psi\big(t;\mathscr{C}^\text{\sc q}(\beta)\big)$ tends to the unit impulse as shown in \eqref{eq:lemma}. Simultaneously, the slope of the IF goes to $-\myUpsilon$ and the IF ${\omega(t;\myUpsilon)=-\myUpsilon t + \omega_\mathrm{x}}$ pivots about $\omega_\mathrm{x}$ at  $t=0$. Let $\eta\in\!{}^{\scalebox{0.7}{$\divideontimes$}}\!\:\mathbb{R}$ be an infinitesimal such that $|\eta\myUpsilon|$ is unlimited, then during the infinitesimal time interval from $t=-\eta$ to $t=\eta$, the atom traverses from $\omega\approx\infty$ to $\omega\approx-\infty$, thus
\begin{IEEEeqnarray}{c}
    \mathcal{A}^\text{\sc s}(t,\omega)  = \sqrt{2\pi}\delta(t)\varepsilon\left(\omega\right)\label{eq:ShannonAtomMother}
\end{IEEEeqnarray}
and the resulting IS for the Shannon atom is illustrated in Figure~\ref{fig:DiracD}. This illustrates an appealing result: \textit{the IF of the Shannon atom passes through all frequencies during an infinitesimal time interval about} $t=0$. 
\end{IEEEproof}

Finally, $\delta(t)\varepsilon\left(\omega\right)$ is interpreted as a 2D hyperreal function  which is concentrated at time $t=0$ and has been \textit{spread out} over all $\omega$ so applying \eqref{eq:HSintOmegaComp}  and \eqref{eq:DispersionProp} gives the component
\begin{IEEEeqnarray}{c}
 \frac{1}{\sqrt{2\pi}}{\int_{-\infty}^{\infty}\mathcal{A}^\text{\sc s}(t,\omega) \dd\omega=\delta(t)}.\label{eq:myprop}
\end{IEEEeqnarray}
Therefore, \eqref{eq:ShannonAtomMother} provides an answer to our original question about the existence of a time-frequency atom corresponding to the unit impulse.

\color{black}

\section{Time Domain Analysis}\label{sec:TD}
Time domain analysis is carried out through association of $z(t)$ with weights of a linear superposition of shifted unit impulses and the corresponding signal synthesis is given by
\begin{IEEEeqnarray}{c}
    z(t)   = \intPMinfty z(\tau)\delta(t-\tau) \dd\tau.   \label{eqn:sifting}
\end{IEEEeqnarray}
This decomposition specifies a family of components which are a set of scaled, time-shifted unit impulses
\begin{IEEEeqnarray}{c}
    \psi_\tau(t) = z(\tau)\delta(t-\tau)
\end{IEEEeqnarray}
parameterized by the measurement $z(\tau)$. Because the unit impulse is not directly an AM-FM component and thus cannot be expressed with a component triplet, we specify a family of Shannon atoms given by 
\begin{IEEEeqnarray}{c}
     \mathcal{A}_\tau^\text{\sc s}(t,\omega)  = z(\tau) \sqrt{2\pi}\delta(t-\tau)\epsilon\left(\omega\right).\label{eq:ShannonAtom}
\end{IEEEeqnarray}
Superimposing these atoms gives the IS 
\begin{IEEEeqnarray}{rCl}  \IEEEyesnumber\IEEEyessubnumber*   
    \mathcal{S}^\text{\sc  td}(t,\omega) &\triangleq&\intPMinfty  \mathcal{A}_\tau^\text{\sc s}(t,\omega) \dd\tau \label{eqn:HSlimitTD}\\
                &=&    \sqrt{2\pi}\intPMinfty  z(\tau)\delta(t-\tau)\epsilon\left(\omega\right)  \dd\tau.
    \label{eq:ISdirac}
\end{IEEEeqnarray}
Finally, we show that the AM--FM model in \eqref{eq:AMFM_A} specializes to the synthesis equation for time domain analysis by applying the projection equation  in (\ref{eq:HSintOmega}) to \eqref{eq:ISdirac}
\begin{IEEEeqnarray}{rCl}  \IEEEyesnumber
    z(t) = \intPMinfty  z(\tau)\delta(t-\tau)  \dd\tau.
\end{IEEEeqnarray}


\section[Monocomponent Analysis and The Analytic Signal]{Monocomponent Analysis and \\The Analytic Signal}\label{sec:MC}
Finally, we present two analyses that do not fit into the plane given in Figure~\ref{fig:compRelations}(b), namely, that of an IS consisting of a single time-frequency atom moving though the time-frequency space.

\subsection{Monocomponent Analysis}
Assuming a monocomponent signal $K=1$, the IS obtained through direct AM--FM demodulation of the signal $z(t)$ is given by 
\begin{IEEEeqnarray}{rCl} 
    \mathcal{S}^\text{\sc mc}(t,\omega) &=& \sqrt{2\pi}z(t)\delta\left(\omega-\frac{\dd}{\dd t}\arg\big\{z(t)\big\}\right).
    \label{eq:ISmc}
\end{IEEEeqnarray}

\subsection{Monocomponent Analysis of the Analytic Signal}
When the signal $z(t)$ is latent, i.e.~only the real part $x(t)$ of the complex signal $z(t)$ is observed as in \eqref{eq:realObs}, the imaginary part is ambiguous and must be estimated to extend the signal back to the complex domain. The analytic signal (AS) \cite{ville, bedrosian1962analytic, cohen1995time, xia1999analytic, boashash2003time} refers to a complex extension of a real signal $x(t)$ and is often expressed in terms of the Hilbert transform $\mathcal{H}\operator{\cdot}$ as
\begin{IEEEeqnarray}{rCl}\IEEEyesnumber
    z_\mathsf{a}\!\!\:(t) = x(t) +\ju\mathcal{H}\operator{x(t)}.
\end{IEEEeqnarray}
Using ISA and analyzing $x(t)$ with \eqref{eq:FTanal} and \eqref{eqn:ISshc}, this is equivalent to 
\begin{IEEEeqnarray}{rCl}\IEEEyesnumber
     z_\mathsf{a}(t)   &=& \frac{1}{\sqrt{2 \pi}} \int_{-\infty}^{\infty} 2\mathrm{u}(\omega) \mathcal{S}^\text{\sc  fd}(t,\omega) \dd\omega
     \label{eq:AS}
\end{IEEEeqnarray}
where $\mathrm{u}(\omega)$ is the Heaviside unit step function with $\mathrm{u}(0)=1/2$ and using \eqref{eq:ISmc} the corresponding IS is
\begin{IEEEeqnarray}{rCl} 
    \mathcal{S}^\text{\sc a}(t,\omega) &=& \sqrt{2\pi}z_\mathrm{a}(t)\delta\left(\omega-\frac{\dd}{\dd t}\arg\big\{z_\mathrm{a}(t)\big\}\right).
    \label{eq:IS_AS}
\end{IEEEeqnarray}
While \eqref{eq:AS} appears to imply that the AS is a multicomponent signal model, \eqref{eq:IS_AS} makes explicit that the AS is a monocomponent analysis that forms a single signal component by fusing together the projection of Fourier atoms with nonnegative IF, before demodulating as a single atom. In \cite{ISA2018_Sandoval}, we highlighted that the AS method may artificially impose frequency symmetries and by giving up the AS, we allowed for alternative complex extensions and hence alternative IA/IF parameterizations for a real signal that may be more useful or provide better estimation accuracy.  Thus, we urge the practitioner to use the AS only after careful examination of the context in which it is to be used, because in general \eqref{eq:IS_AS} may differ significantly from \eqref{eq:ISmc}.


\section{Examples} \label{sec:Examples}
To demonstrate the IS associated with various choices for analysis, we consider five example signals: 1) simple harmonic, 2) unit impulse, 3) linear FM, 4) Gaussian AM, and 5) Sinusoidal AM.  For convenience, Table \ref{tab:AnalysesISs}  lists the various analyses developed in Sections \ref{sec:FD} - \ref{sec:STFrFT}, \ref{sec:TD}, and \ref{sec:MC} and the corresponding equation reference for the IS. For the STFT, we use a Gaussian window with pulse width parameter $\alpha$
\begin{IEEEeqnarray}{c}
    \w(t) = \eu^{-\alpha^2t^2}.
\end{IEEEeqnarray}
Finally, for brevity, we do not consider the STFrFT in these examples and leave that to the reader.

\begin{table}[htb]
    \centering
    \caption{Analyses and the Corresponding ISs}
    \begin{tabular}{ccc}
    \toprule
        \textbf{Analysis} & \textbf{IS Equation} & \textbf{Decomposition}\\
    \midrule
        Frequency Domain & \eqref{eqn:ISshc} & Projection\\ 
        Fractional Fourier  & \eqref{eq:namiasRecon} & Projection\\
        Short-time Fourier Transform  & \eqref{eq:STFTis} & Filterbank\\
        Short-time Fractional Fourier Transform & \eqref{eq:isFRSTFT} & Filterbank\\
        Time Domain & \eqref{eq:ISdirac} & Projection\\
        Monocomponent & \eqref{eq:ISmc} & None\\
        Analytic Signal & \eqref{eq:IS_AS} & None\\
        \bottomrule
    \end{tabular}
    \label{tab:AnalysesISs}
\end{table}

\subsection{Simple Harmonic Signal}
Referring the reader to the synthesis in Figure \ref{fig:ISAsummary}, we use the parameter set
$\mathscr{S} = \{\mathscr{C}_0\}$ consisting of the triplet
\begin{IEEEeqnarray}{c}
    \mathscr{C}_0\triangleq\canonical{1,\omega_0,0\vertOne}.
\end{IEEEeqnarray}
This parameter set defines the IS
\begin{IEEEeqnarray}{rCl} 
    \mathcal{S}(t,\omega) &=& \sqrt{2\pi} \eu^{\,\ju\omega_0 t}  \delta(\omega-\omega_0 )  
    \label{eq:CEgroundtruth}
\end{IEEEeqnarray}
and the signal
\begin{IEEEeqnarray}{rCl} 
    z(t) &=&  \eu^{\,\ju\omega_0 t} .
\end{IEEEeqnarray}

Table \ref{tab:ISs Simple Harmonic Signal} provides the true IS (synthesis) and the ISs for each of the analyses. For the simple harmonic signal, the IS associated with FD analysis [see Figure~\ref{fig:shc}] and monocomponent analysis perfectly estimates the true IS. Similarly, the IS associated with (synchrosqueezed) STFT analysis  estimates the true IS to within a constant. Both fractional Fourier and TD analyses [see Figure~\ref{fig:shcTD}] spread energy equally over the time-frequency plane and rely on constructive/deconstructive interference to project the correct signal.

\begin{table*}[tbh]
    \centering
    \caption{ISs for the Simple Harmonic Signal}
    \begin{tabular}{cl}
    \toprule
        \textbf{Description} & \textbf{Instantaneous Spectrum}\\
    \midrule
    Synthesis & $    \mathcal{S}(t,\omega) = \sqrt{2\pi} \eu^{\,\ju\omega_0 t}  \delta(\omega-\omega_0 )  
$\\[1.5mm]
    Frequency Domain Analysis%
         &  $\mathcal{S}^\text{\sc  fd}(t,\omega) = \sqrt{2\pi} \eu^{\,\ju\omega_0 t} \delta(\omega-\omega_0 )$\\[1.5mm]
     Fractional Fourier Analysis%
        & $\mathcal{S}_\gamma ^\text{\sc ff}(t,\omega) 
    = Z_\gamma \big(   [ \omega  +r t ]/ c_\mathrm{x} \big) C_\gamma^\dagger  \exp\big(\ju \theta_\gamma(t;[ \omega  +r t ]/ c_\mathrm{x})\big)$, where\\
        & $Z_\gamma (u) =  \sqrt{1+\ju\tan(\gamma)} \exp\left( -\ju \pi(u^2\tan(\gamma) - 2 u \frac{\omega}{2\pi} \sec(\gamma) + (\frac{\omega_0}{2\pi})^2 \tan(\gamma)\right)$\\[1.5mm]
 
      Short-time Fourier Analysis%
       &  $\mathcal{S}^\text{\sc  stft}(t,\omega)= c_\mathrm{a} \eu^{\,\ju\omega_0 t}  \delta(\omega-\omega_0 )$, where $c_\mathrm{a}$ is a constant\\[1.5mm]
       Time Domain Analysis & $\mathcal{S}^\text{\sc td}(t,\omega)= \sqrt{2\pi}\eu^{\,\ju\omega_0 t}\epsilon(\omega)$\\[1.5mm]
       Monocomponent Analysis  & $    \mathcal{S}^\text{\sc mc}(t,\omega) = \sqrt{2\pi} \eu^{\,\ju\omega_0 t}  \delta(\omega-\omega_0 )  
$\\
       \bottomrule
    \end{tabular}
    \label{tab:ISs Simple Harmonic Signal}
\end{table*}

\subsection{Unit Impulse Signal}
As noted earlier, because the unit impulse is not directly an AM-FM component it cannot be expressed with a component triplet. Therefore, we begin by directly considering the IS consisting of a single Shannon atom
\begin{IEEEeqnarray}{rCl} \IEEEyessubnumber*
    \mathcal{S}(t,\omega) &=& \mathcal{A}^\text{\sc s}(t,\omega)\\
                            &=&\sqrt{2\pi}\delta(t)\epsilon\left(\omega\right)
\end{IEEEeqnarray}
and the signal
\begin{IEEEeqnarray}{rCl} 
    z(t) &=& \delta(t).
\end{IEEEeqnarray}

\begin{landscape}

\begin{figure*}[p]
\centering
  	\hspace*{-2.5in}\begin{minipage}[b]{4.75in}
  		\centering
  		\subfigure[]{
  		{\includegraphics[trim=115 50 70 50,clip,width = 0.99\linewidth]{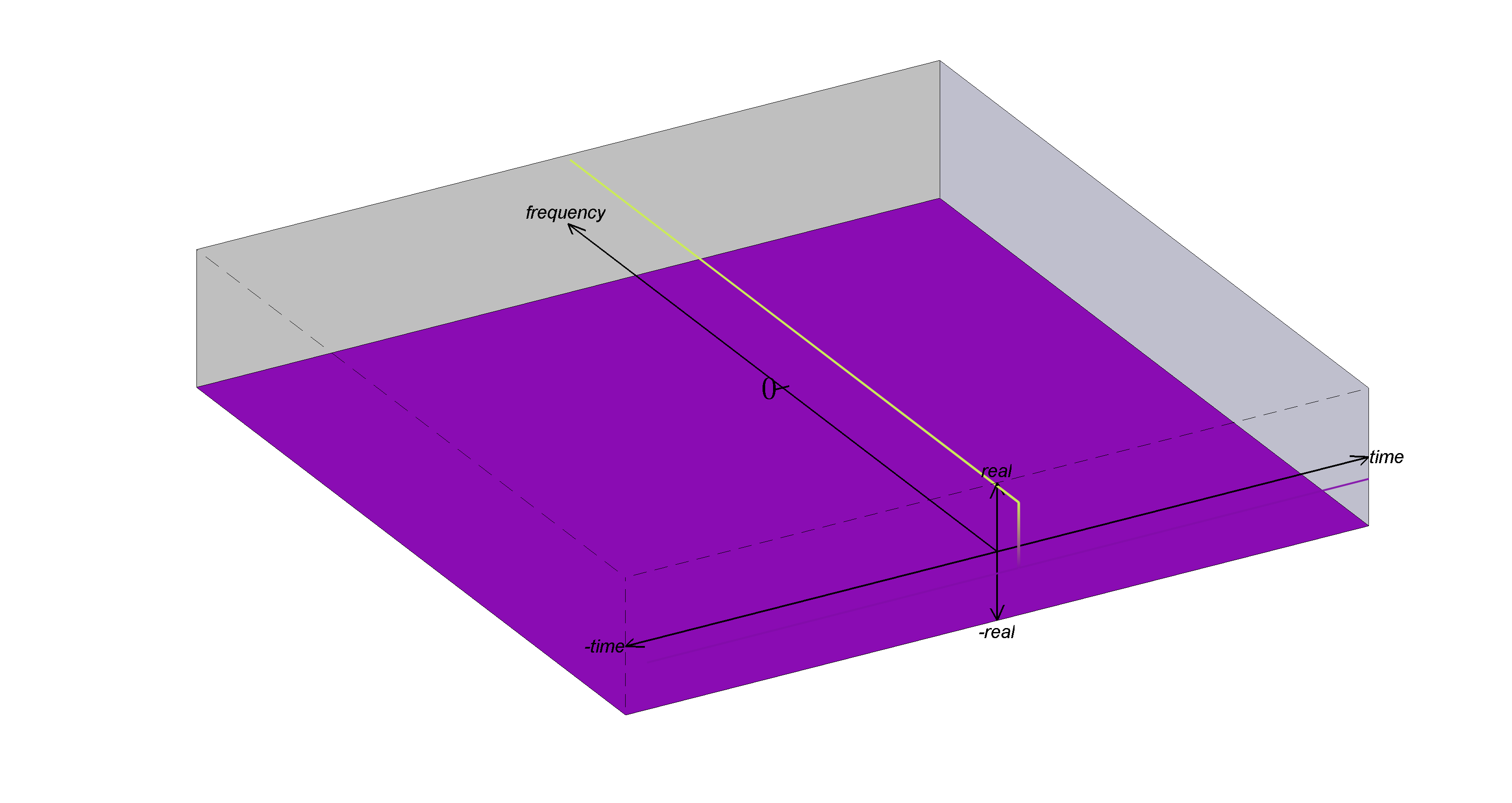}}
  		\label{fig:impulseTD}
  	}
  	\end{minipage}\begin{minipage}[b]{4.75in}
  		\centering
  		\subfigure[]{
  		{\includegraphics[trim=115 50 70 50,clip,width = 0.99\linewidth]{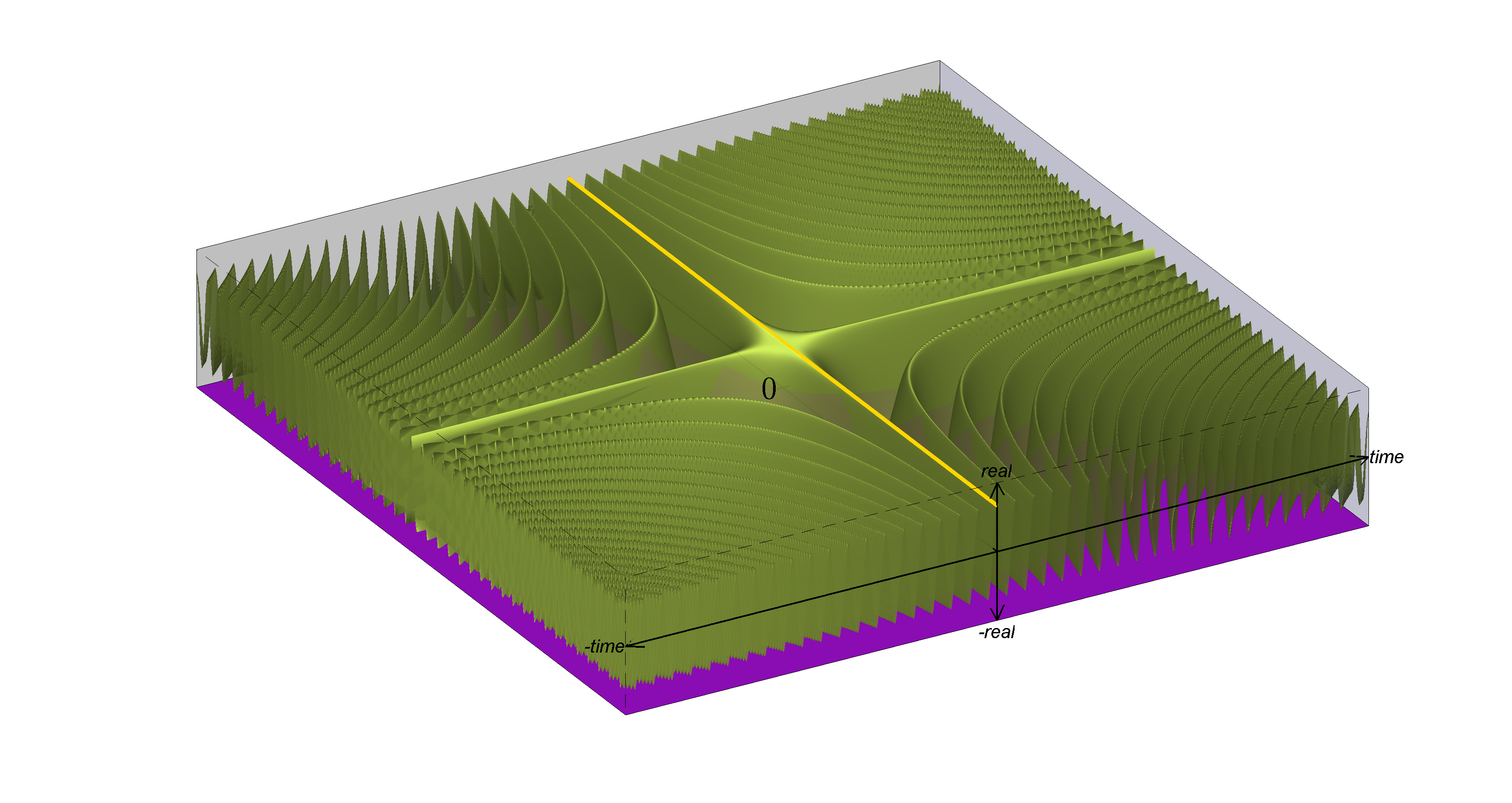}}
  		\label{fig:impulse}
  	}
  	\end{minipage}
  	\hspace*{-2.5in}\begin{minipage}[b]{4.75in}
  		\centering	
  		\subfigure[]{
  		\includegraphics[trim=115 50 70 50,clip,width = 0.99\linewidth]{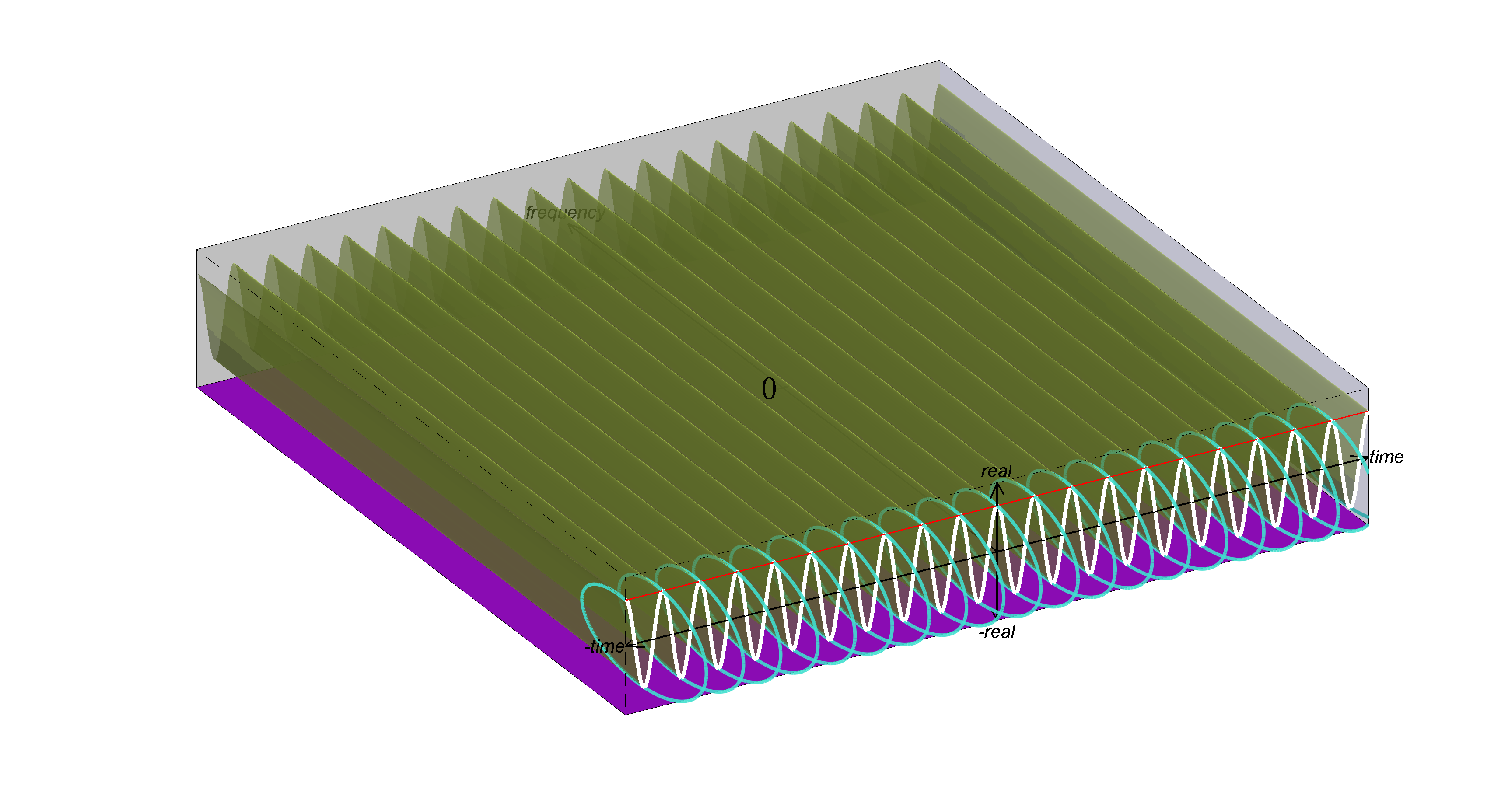}
  		\label{fig:shcTD}
  	}
  	\end{minipage}\begin{minipage}[b]{4.75in}
  		\centering	
  		\subfigure[]{
  		\includegraphics[trim=115 50 70 50,clip,width = 0.99\linewidth]{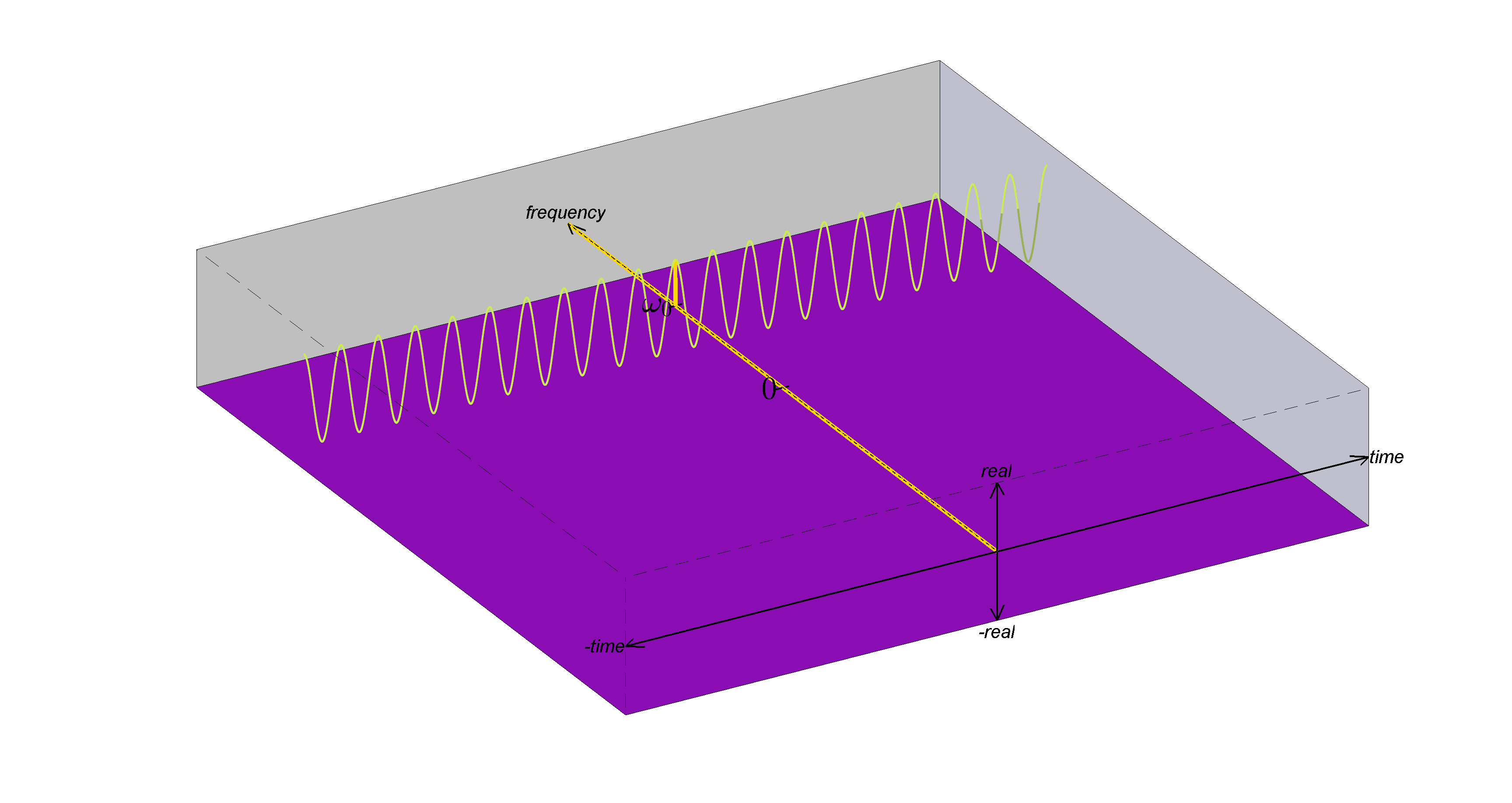}
  		\label{fig:shc}
  	}
  	\end{minipage}
  	\hspace*{-2.5in}\begin{minipage}[b]{9in}\caption{The IS (\myColorBar) corresponding to TD analysis (left column) is shown for (a) unit impulse and (c) complex exponential. In these plots the real waveform (\textcolor{myGainsboro}{\solidMrule[3.5mm]}), complex waveform (\textcolor{myTurquoise}{\solidMrule[3.5mm]}), and signal magnitude (\textcolor{myRed}{\solidSrule[3.5mm]}) may be shown along the time axis. The IS (\myColorBar) corresponding to FD analysis (right column) is shown for (b) unit impulse and (d) complex exponential. In these plots the Fourier transform (\textcolor{myGold}{\solidMrule[3.5mm]}) is highlighted at $t=0$. }\end{minipage}
  	  \label{fig:ISexamples3}	
\end{figure*}

\end{landscape}

Table \ref{tab:ISs Unit Impulse Signal} provides the true IS (synthesis) and the ISs for each of the analyses. For the unit impulse signal, the IS associated with TD analysis  [see Figure~\ref{fig:impulseTD}] perfectly estimates the true IS. The IS associated with FD  [see Figure~\ref{fig:impulse}] and fractional Fourier analyses spread energy equally over the time-frequency plane and rely on constructive/deconstructive interference to project the correct signal. The IS associated with (synchrosqueezed) STFT analysis spreads energy equally across all frequencies, but localizes in time according to the parameter $\alpha$. The monocomponent analysis perfectly localizes the energy in time, but incorrectly assigns the energy to $\omega=0$. 


\begin{table*}[tbh]
    \centering
    \caption{ISs for the Unit Impulse Signal}
    \begin{tabular}{cl}
    \toprule
        \textbf{Description} & \textbf{Instantaneous Spectrum}\\
    \midrule
    Synthesis & $\mathcal{S}(t,\omega) = \sqrt{2\pi}\delta(t)\epsilon\left(\omega\right)$\\[1.5mm]
    Frequency Domain Analysis%
         &  $\mathcal{S}^\text{\sc  fd}(t,\omega) = \sqrt{2\pi}  \eu^{\,\ju \omega t} $\\[1.5mm]
     Fractional Fourier Analysis%
         & $\mathcal{S}_\gamma ^\text{\sc ff}(t,\omega) 
    = \sqrt{1-\ju \cot(\gamma) \exp\big( \ju \pi [( \omega  +r t )/ c_\mathrm{x}]^2 \cot(\gamma) \big)}  C_\gamma^\dagger  \eu^{\,\ju \theta_\gamma(t;[ \omega  +r t ]/ c_\mathrm{x})}$\\[1.5mm]
      Short-time Fourier Analysis%
       &  $\mathcal{S}^\text{\sc  stft}(t,\omega)=  \frac{1}{\sqrt{2\pi}} \eu^{-\alpha^2t^2}\eu^{\,\ju\omega t} $\\
       Time Domain Analysis & $\mathcal{S}^\text{\sc td}(t,\omega)= \sqrt{2\pi}\delta(t)\epsilon(\omega)$\\[1.5mm]
       Monocomponent Analysis  & $\mathcal{S}^\text{\sc mc}(t,\omega) = \sqrt{2\pi}\delta(t)\delta(\omega)$\\
       \bottomrule
    \end{tabular}
    \label{tab:ISs Unit Impulse Signal}
\end{table*}

\subsection{Linear FM Signal}
Referring the reader to the synthesis stage of Figure \ref{fig:ISAsummary}, we use the parameter set
$\mathscr{S} = \{\mathscr{C}_0\}$ consisting of the triplet
\begin{IEEEeqnarray}{c}
    \mathscr{C}_0\triangleq\canonical{1,\omega_0- r_0 t,0\vertOne}
\end{IEEEeqnarray}
where $r_0$ is the chirp rate. This parameter set defines the IS
\begin{IEEEeqnarray}{rCl} 
    \mathcal{S}(t,\omega) &=& \sqrt{2\pi} \eu^{\,\ju(\omega_0 t- r_0 t^2/2)}  \delta\big(\omega-(\omega_0- r_0 t) \big)      
\end{IEEEeqnarray}
and the signal
\begin{IEEEeqnarray}{rCl} 
    z(t) &=&  \eu^{\,\ju(\omega_0 t- r_0 t^2/2)} .
\end{IEEEeqnarray}

Table \ref{tab:ISs Linear FM Signal} provides the true IS (synthesis) and the ISs for each of the analyses. For the linear FM signal, the IS associated with monocomponent analysis perfectly estimates the true IS. When the chirp rate $r$ of the Fractional FT is equal to the chirp rate of the signal $r_0$, the IS associated with fractional Fourier analysis also perfectly estimates the true IS to within a constant. However, when $r\neq r_0$ energy is spread equally across the time-frequency plane. The IS associated with both FD and TD analyses spread energy evenly over the entire time-frequency plane. As shown in \cite{sandoval2022recasting}, the IS associated with (synchrosqueezed) STFT analysis concentrates the energy very closely around the true IF, but does not perfectly estimate the IS.

\begin{table*}[tbh]
    \centering
    \caption{ISs for the Linear FM Signal}
    \begin{tabular}{cl}
    \toprule
        \textbf{Description} & \textbf{Instantaneous Spectrum}\\
    \midrule
    Synthesis & $\mathcal{S}(t,\omega) = \sqrt{2\pi} \eu^{\,\ju(\omega_0 t- r_0 t^2/2)}  \delta\big(\omega-(\omega_0- r_0 t) \big)$\\[1.5mm]
    Frequency Domain Analysis%
         &  $\mathcal{S}^\text{\sc  fd}(t,\omega) = \frac{1}{\sqrt{r_0}} \exp\left( \ju\left[\frac{(\omega-\omega_0)^2}{2 r_0}-\frac{\pi}{4} \right]\right) \eu^{\,\ju \omega t} $\\[1.5mm]
     Fractional Fourier Analysis%
         & $\mathcal{S}_\gamma ^\text{\sc ff}(t,\omega) 
    =  Z_\gamma \big(   [ \omega  +r t ]/ c_\mathrm{x} \big) C_\gamma^\dagger  \exp\big(\ju \theta_\gamma(t;[ \omega  +r t ]/ c_\mathrm{x})\big)$, where\\
            & $Z_\gamma (u) =  \begin{cases}
                        \sqrt{\frac{1+\ju \tan(\gamma)}{1-r_0 \tan(\gamma)/(2\pi)}} \exp\left( j\pi \frac{u^2[-r_0/(2\pi)-\tan(\gamma)]+ 2 u f_0 \sec(\gamma) -f_0^2 \tan(\gamma)}{1-r_0\tan(\gamma)/(2\pi)} \right)            ,& r\neq r_0\\
                         C_\gamma\eu^{\,\ju r_0u^2/2}/\sqrt{2\pi}\delta\big(u+\omega_0 / c_\mathrm{x} \big)      ,& r= r_0
                    \end{cases}$\\
             & with $f_0=\omega_0/(2\pi)$ and phase $\theta_\gamma(t;u) = -\frac{r}{2}t^2 +c_\mathrm{x} u t - \frac{r}{2}u^2$\\       
      Short-time Fourier Analysis%
       &  $\mathcal{S}^\text{\sc  stft}(t,\omega)$ is given in \cite{sandoval2022recasting}\\[1.5mm]
       Time Domain Analysis & $\mathcal{S}^\text{\sc td}(t,\omega)= \sqrt{2\pi} \eu^{\,\ju(\omega_0 t-  r_0 t^2/2)}  \epsilon(\omega)$\\[1.5mm]
    Monocomponent Analysis  & $\mathcal{S}^\text{\sc mc}(t,\omega) = \sqrt{2\pi} \eu^{\,\ju(\omega_0 t-  r_0 t^2/2)}  \delta\big(\omega-(\omega_0- r_0 t) \big)$\\
       \bottomrule
    \end{tabular}
    \label{tab:ISs Linear FM Signal}
\end{table*}

\subsection{Gaussian AM Signal}
Referring the reader to the synthesis stage of Figure \ref{fig:ISAsummary}, we use the parameter set
$\mathscr{S} = \{\mathscr{C}_0\}$ consisting of the triplet
\begin{IEEEeqnarray}{c}
    \mathscr{C}_0\triangleq\canonical{\eu^{-\alpha_0^2t^2},\omega_0,0\vertOne}
\end{IEEEeqnarray}
where $\alpha_0$ is a duration parameter. This parameter set defines the IS
\begin{IEEEeqnarray}{rCl} 
    \mathcal{S}(t,\omega) &=& \sqrt{2\pi} \eu^{-\alpha_0^2 t^2} \eu^{\,\ju\omega_0 t}  \delta\big(\omega-\omega_0 \big)       
\end{IEEEeqnarray}
and the signal
\begin{IEEEeqnarray}{rCl} 
    z(t) &=&  \eu^{-\alpha_0^2 t^2} \eu^{\,\ju\omega_0 t} .
\end{IEEEeqnarray}

\begin{figure*}[p]
\centering
\begin{minipage}[b]{0.7\linewidth}
  		\centering
  		\subfigure[]{
  		{\includegraphics[trim=115 50 70 50,clip,width = \linewidth]{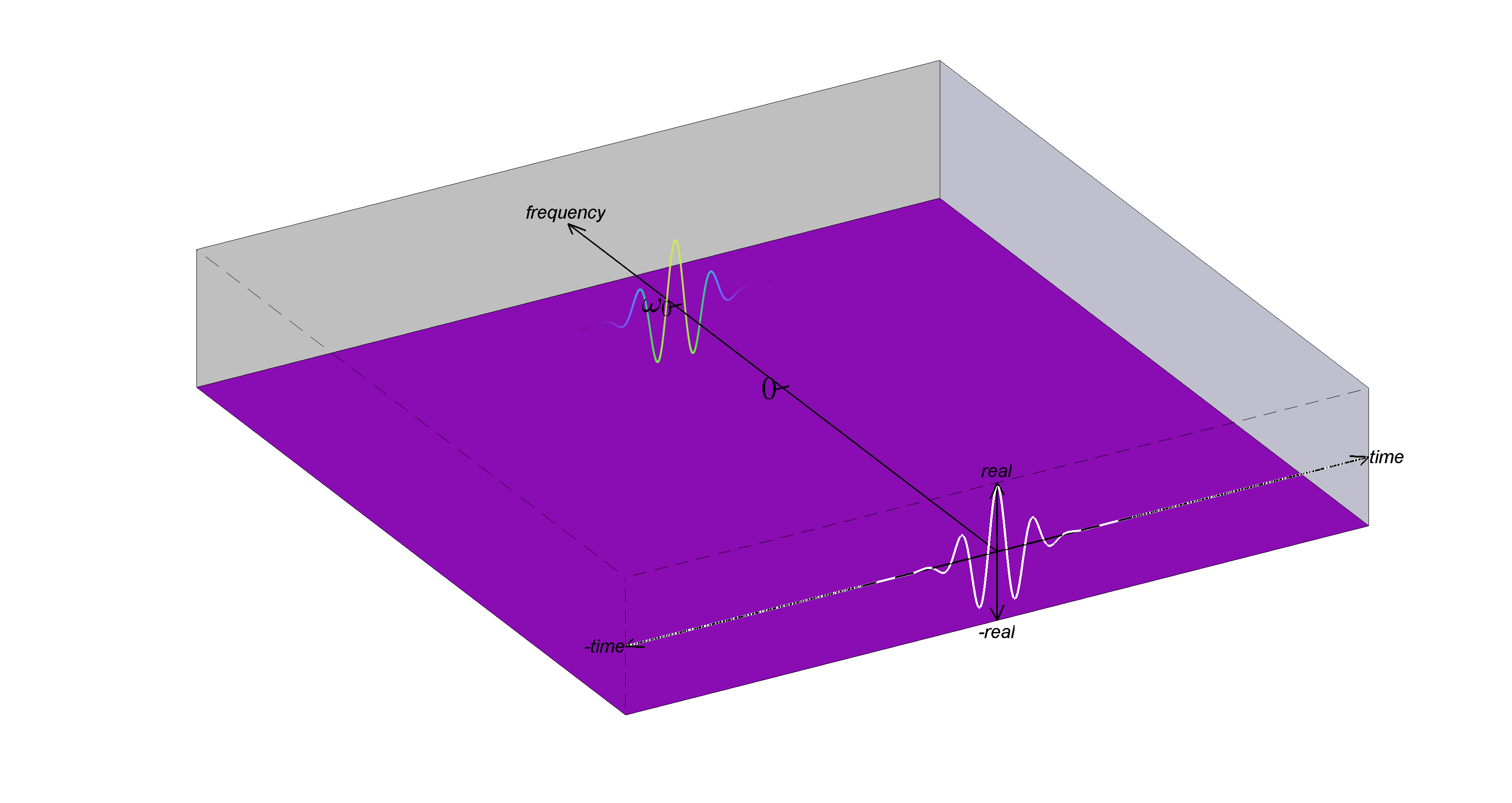}}
  		\label{fig:GaussAM_A}
  	}
  	\end{minipage}
  	\begin{minipage}[b]{0.7\linewidth}
  		\centering
  		\subfigure[]{
  		{\includegraphics[trim=115 50 70 50,clip,width = \linewidth]{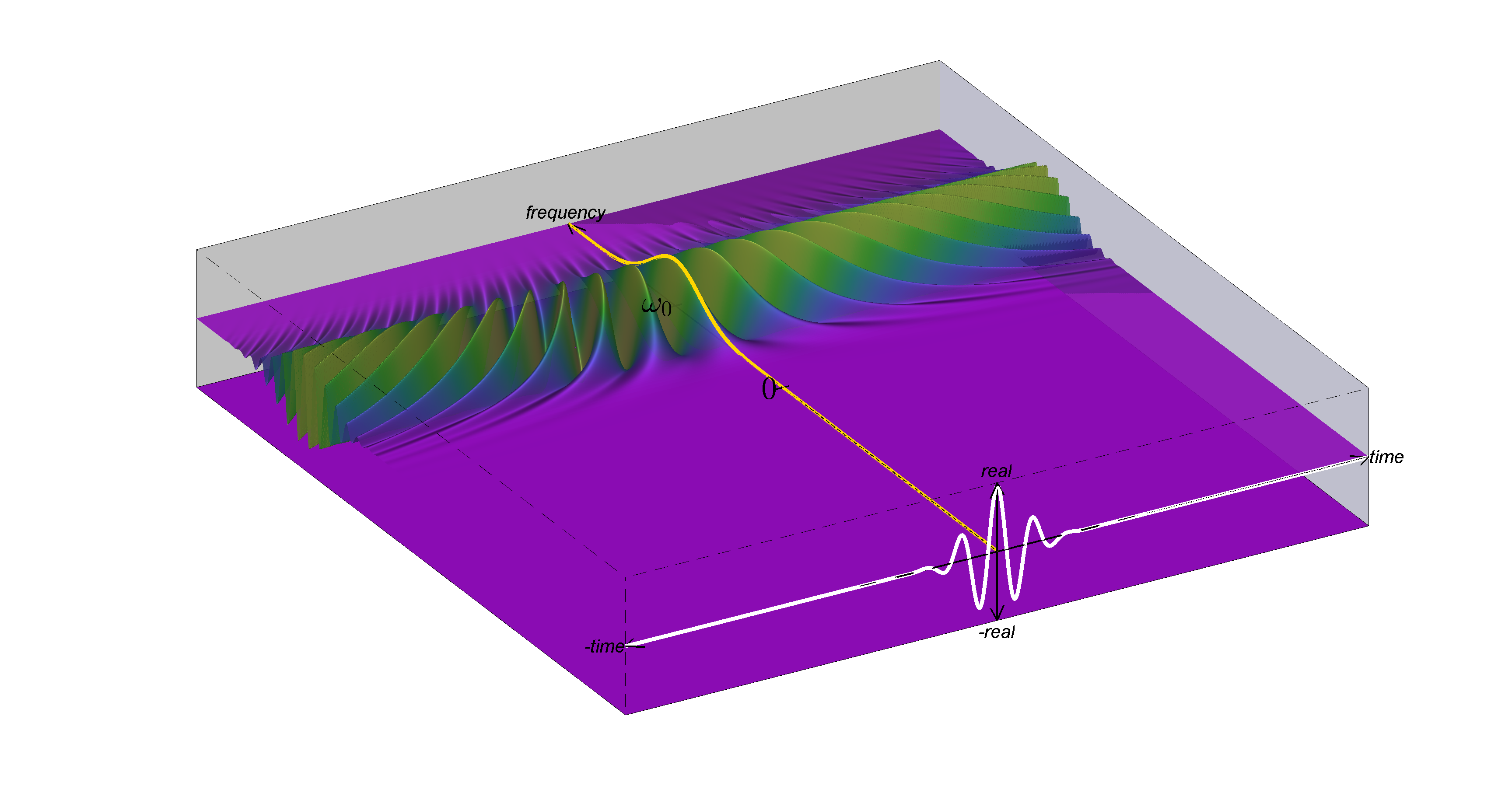}}
  		\label{fig:GaussAM_C}
  	}
  	\end{minipage}
  	\begin{minipage}[b]{0.7\linewidth}
  		\centering
  		\subfigure[]{
  		{\includegraphics[trim=115 50 70 50,clip,width = \linewidth]{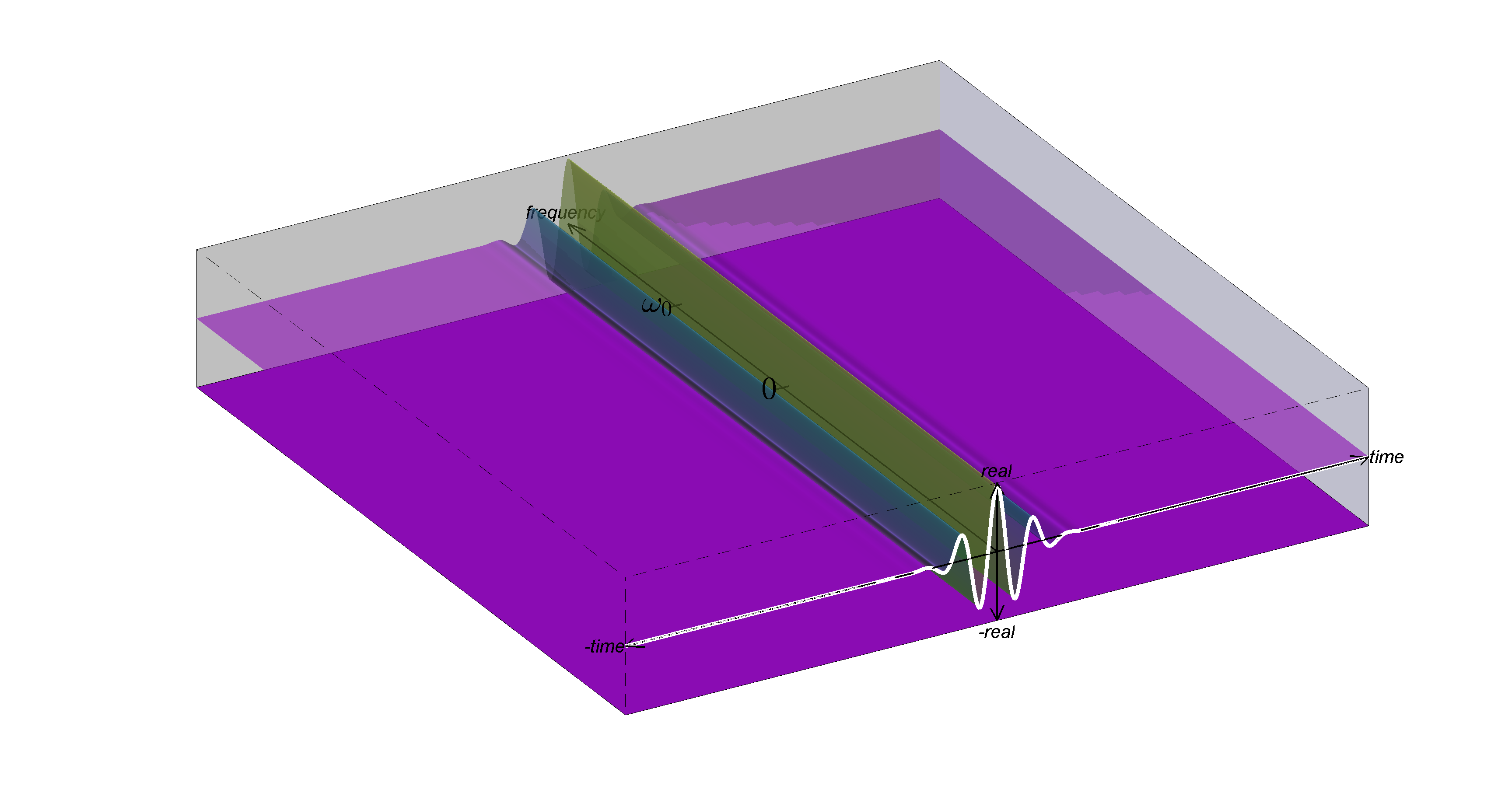}}
  		\label{fig:GaussAM_B}
  	}
  	\end{minipage}
\caption{The ISs (\myColorBar) for the Gaussian AM signal using (a) monocomponent analysis, (b) FD analysis, and (c) TD analysis. The IS associated with monocomponent analysis in (a) perfectly estimates the IS and is exactly localized in time and frequency, even though the product of the effective bandwidth of the IS in (b) and the effective duration of the IS in (c) has a lower bound.}
\label{fig:GaussAM}
\end{figure*}

Table \ref{tab:ISs Gauss AM Signal} provides the true IS (synthesis) and the ISs for each of the analyses. For the Gaussian AM signal, the IS associated with monocomponent analysis perfectly estimates the true IS [see Figure \ref{fig:GaussAM_A}]. The IS associated with FD analysis localizes energy about $\omega_0$ with effective bandwidth $\sigma_\omega^2 = 4\alpha_0^2$ for all time [see Figure \ref{fig:GaussAM_B}]. On the other hand, the IS associated with TD analysis localizes energy in time with effective duration $\sigma_t^2 =1/\alpha_0^2$ for all frequencies [see Figure \ref{fig:GaussAM_C}]. The IS associated with (synchrosqueezed) STFT analysis localizes in both time and frequency as a bivariate Gaussian centered at $t=0$ and $\omega=\omega_0$ with $\sigma_t^2 = 1/(2\kappa^2)$ and frequency variance $2\alpha_0^2\kappa^2/\alpha^2$. To understand the IS associated with fractional Fourier analysis we consider the last term of $Z_\gamma(u)$ in Table \ref{tab:ISs Gauss AM Signal}. In this case, the energy is spread as a Gaussian along the angle $\gamma$ and at the extremes of $\gamma$ i.e.~$0$ and $\pi/2$, the energy spread matches FD and TD analysis.  

This example highlights a common misinterpretation in time-frequency analysis \cite{cohen1995time, papandreou2002applications, stankovic2014time, boashash2015time, flandrin2018Explorations, grochenig2001foundations,moca2021time,hlawatsch2013time,boashash2015time,carmona1998practical,flandrin1998time,mallat1999wavelet,bremaud2002mathematical,bracewell1980fourier,hlawatsch2013time}. While the product of $\sigma_t^2$ and $\sigma_\omega^2$ is certainly lower bounded, the uncertainty principle is solely a statement about $\mathcal{S}^\text{\sc  td}(t,\omega)$ and $\mathcal{S}^\text{\sc  fd}(t,\omega)$---it does not imply that a ground truth IS does not exist for this signal (or that it is not computable). In fact, we have clearly calculated a closed form expression for the IS using the monocomponent analysis which without question, perfectly localizes in time and frequency.

\begin{table*}[tbh]
    \centering
    \caption{ISs for the Gauss AM Signal}
    \begin{tabular}{cl}
    \toprule
        \textbf{Description} & \textbf{Instantaneous Spectrum}\\
    \midrule
    Synthesis & $\mathcal{S}(t,\omega) = \sqrt{2\pi} \eu^{-\alpha_0^2 t^2} \eu^{\,\ju\omega_0 t}  \delta\big(\omega-\omega_0 \big)    $\\[1.5mm]
    Frequency Domain Analysis%
         &  $\mathcal{S}^\text{\sc  fd}(t,\omega) = \frac{1}{\sqrt{2}\alpha_0}  \exp\left(-\frac{(\omega-\omega_0)^2}{4\alpha_0^2}\right) \eu^{\,\ju \omega t} $\\[1.5mm]
     Fractional Fourier Analysis%
         & $\mathcal{S}_\gamma ^\text{\sc ff}(t,\omega) 
    =  Z_\gamma \big(   [ \omega  +r t ]/ c_\mathrm{x} \big) C_\gamma^\dagger  \exp\big(\ju \theta_\gamma(t;[ \omega  +r t ]/ c_\mathrm{x})\big)$, where\\
            & $Z_\gamma (u) =  
    \exp\big(-\ju\omega_0^2/(4\pi)\sin(\gamma)\cos(\gamma)\big) 
    \exp\big(\ju  u \omega_0\cos(\gamma)\big)
    \sqrt{\frac{1-\ju \cot(\gamma)}{\chi-\ju\cot(\gamma)}} 
    $\\    & $\quad\times
    \exp\left( \ju\pi (u-[\omega_0/(2\pi)]\sin(\gamma))^2\frac{\cot\big(\gamma(\chi^2-1)\big)}{\chi^2+\cot^2(\gamma)} \right) 
    \exp\left[ -\pi (u-[\omega_0/(2\pi)]\sin(\gamma))^2 \frac{\chi\csc^2(\gamma)}{\chi^2+\cot^2(\gamma)} \right]$,\\
    &\textcolor{black}{and $\chi=r_0/(2\pi)$}\\
      Short-time Fourier Analysis%
       &  $\mathcal{S}^\text{\sc  stft}(t,\omega) = \frac{1}{\sqrt{2\pi}}  c 
                \eu^{-\kappa^2t^2/2}
                \eu^{-\frac{\alpha^2(\omega-\omega_0)^2}{2\alpha_0^2\kappa^2}}     
                 \eu^{\,\ju \omega t}$, where \\
                 & $\kappa^2  = \frac{1}{\frac{1}{2\alpha_0^2}+\frac{1}{2\alpha^2}} = \frac{2\alpha_0^2\alpha^2}{\alpha_0^2+\alpha^2}$\\
                & and $c$ is a constant that depends only and $\alpha$ and $\alpha_0$.\\[1.5mm]
       Time Domain Analysis & $\mathcal{S}^\text{\sc td}(t,\omega)= \sqrt{2\pi} \eu^{-\alpha_0^2 t^2} \eu^{\,\ju\omega_0 t}\epsilon(\omega)$\\[1.5mm]
       Monocomponent Analysis  & $\mathcal{S}^\text{\sc mc}(t,\omega) = \sqrt{2 \pi} \eu^{-\alpha_0^2 t^2} \eu^{\,\ju \omega_0 t}\,\delta\left(\vertOne\omega-{\omega}_0\vertOne\right)$. \\
       \bottomrule
    \end{tabular}
    \label{tab:ISs Gauss AM Signal}
\end{table*}

\color{black}

\subsection{Sinusoidal AM Signal}

\begin{table*}[tbh]
    \centering
    \caption{ISs for the Sinusiodal AM Signal}
    \begin{tabular}{cl}
    \toprule
        \textbf{Description} & \textbf{Instantaneous Spectrum}\\
    \midrule
    Synthesis &  $\mathcal{S}_1(t,\omega) =\sqrt{2\pi}\big[ 
    \eu^{\,\ju\omega_\mathrm{a} t} \delta(\omega-\omega_\mathrm{a})
    + \eu^{\,\ju\omega_\mathrm{b} t}\delta(\omega-\omega_\mathrm{b})\big]   $\\[1.5mm]
    ~ &$\mathcal{S}_2(t,\omega) = \sqrt{2\pi} 2\cos\left(\omega_\Delta t\vertOne\right) \exp(\,\ju \omega_0 t ) \delta(\omega-\omega_0 ])   $\\[1.5mm]
    Frequency Domain Analysis%
         &  $\mathcal{S}^\text{\sc  fd}(t,\omega) =\sqrt{2\pi}\big[ 
    \eu^{\,\ju\omega_\mathrm{a} t} \delta(\omega-\omega_\mathrm{a})
    + \eu^{\,\ju\omega_\mathrm{b} t}\delta(\omega-\omega_\mathrm{b})\big]  $\\[1.5mm]
       Fractional Fourier Analysis%
         & $\mathcal{S}_\gamma ^\text{\sc ff}(t,\omega) 
    =  Z_\gamma \big(   [ \omega  +r t ]/ c_\mathrm{x} \big) C_\gamma^\dagger  \exp\big(\ju \theta_\gamma(t;[ \omega  +r t ]/ c_\mathrm{x})\big)$, where\\
            & $Z_\gamma (u) = \frac{2}{\sqrt{\cos(\gamma)}} \left| \cos\big(\phi_\Delta(u)\big) \right|   \exp\left( \ju \left[ \bar{\phi}(u) + \frac{\gamma}{2} + \varphi(u) \right] \right)$\\
    & $\bar{\phi}(u) = \frac{1}{r} \left( \left[ \frac{\omega_{\mathrm{a}} + \omega_{\mathrm{b}}}{2} \right] c_\mathrm{x} u - \frac{u^2}{2} - \left[ \frac{\omega_{\mathrm{a}}^2 + \omega_{\mathrm{b}}^2}{4} \right] \right)$ and $\varphi(u)=\begin{cases}
    0,   & u \geq 0\\
    \pi, & u<0
\end{cases}$ \\
      Short-time Fourier Analysis%
       &  $\mathcal{S}^\text{\sc  stft}(t,\omega)= \intPMinfty  \frac{1}{\sqrt{2\pi}} Z_\w(t;\nu) \delta\left(\omega - \vertOne \frac{\dd}{\dd t}\theta_\w(t;\nu)   \vertOne\right)\dd\nu$\\ 
       & where $Z_\w(t;\nu)
       =  C_{\mathrm{a}}(\nu) \eu^{\,\ju\omega_{\mathrm{a}} t} + C_{\mathrm{b}}(\nu) \eu^{\,\ju\omega_{\mathrm{b}} t} = a_\w(t;\nu) \eu^{\,\ju \theta_\w(t;\nu)} $,\\
     & $a_\w(t;\nu)= \sqrt{C_{\mathrm{a}}^2(\nu) + C_{\mathrm{b}}^2(\nu) + 2C_{\mathrm{a}}(\nu)C_{\mathrm{b}}(\nu) \cos(2\omega_\Delta t)} $, \\
     & $ \theta_\w(t;\nu) = \omega_0 t + \arctan\left(\left(\frac{C_{\mathrm{b}}(\nu) - C_{\mathrm{a}}(\nu)}{C_{\mathrm{a}}(\nu) + C_{\mathrm{b}}(\nu)}\right) \tan(\omega_\Delta t)\right)$\\
    & $\frac{\dd}{\dd t}\theta_\w(t;\nu) = \omega_0 + \omega_\Delta \frac{C_{\mathrm{b}}^2(\nu) - C_{\mathrm{a}}^2(\nu)}{C_{\mathrm{a}}^2(\nu) + C_{\mathrm{b}}^2(\nu) + 2C_{\mathrm{a}}(\nu)C_{\mathrm{b}}(\nu) \cos(2\omega_\Delta t)}$\\
     & $C_{\mathrm{a}}(\nu) =   \exp\left(-\frac{(\omega_{\mathrm{a}} - \nu)^2}{2\alpha^2}\right)$, 
     and $C_{\mathrm{b}}(\nu) =  \exp\left(-\frac{(\omega_{\mathrm{b}} - \nu)^2}{2\alpha^2}\right)$\\
       Time Domain Analysis & $\mathcal{S}^\text{\sc td}(t,\omega)=\sqrt{2\pi} \left[\eu^{\,\ju\omega_\mathrm{a} t} + \eu^{\,\ju\omega_\mathrm{b} t}\right]\varepsilon(\omega)= \sqrt{2\pi}2\cos\left(\omega_\Delta t\vertOne\right) \eu^{\,\ju\left[ \omega_0 t \right]}\varepsilon(\omega)$\\[1.5mm]
       Monocomponent Analysis  & $\mathcal{S}^\text{\sc mc}(t,\omega) = \sqrt{2\pi} 2\cos\left(\omega_\Delta t\vertOne\right) \exp(\,\ju\omega_0 t) \delta(\omega-\omega_0 )$ \\
       \bottomrule
    \end{tabular}
    \label{tab:ISs Sinusoidal AM Signal}
\end{table*}

Finally, we expand on the sinusoidal AM example in \cite{ISA2018_Sandoval} which can be developed with both a component set consisting of a single triplet or a component set consisting of two triplets, either of which result in the same signal $z(t)$. To begin each case, we refer the reader to the synthesis stage of Figure \ref{fig:ISAsummary}.

\textit{CASE 1:}
We use the parameter set $\mathscr{S}_1= \{\mathscr{C}_0,\mathscr{C}_1\}$ consisting of two components  
\begin{IEEEeqnarray}{c}
     \mathscr{C}_0 = \left( 1,~ \omega_\mathrm{a},~ 0 \vertOne\right)~~~~\text{and}~~~~\mathscr{C}_1 = \left( 1,~ \omega_\mathrm{b},~ 0 \vertOne\right).
\end{IEEEeqnarray}
This parameter set defines the IS
\begin{IEEEeqnarray}{rCl} 
    \mathcal{S}_1(t,\omega) &=&\sqrt{2\pi}\big[ 
    \eu^{\,\ju\omega_\mathrm{a} t} \delta(\omega-\omega_\mathrm{a})
    + \eu^{\,\ju\omega_\mathrm{b} t}\delta(\omega-\omega_\mathrm{b})\big]
\end{IEEEeqnarray}
and the signal
\begin{IEEEeqnarray}{rCl}\IEEEyesnumber
	z_1 (t) &=& \eu^{\,\ju\omega_\mathrm{a} t} + \eu^{\,\ju\omega_\mathrm{b} t}.\label{eq:EXz1b}
\end{IEEEeqnarray}

\textit{CASE 2:}
We use the parameter set
$\mathscr{S}_2 = \{\mathscr{C}_0\}$ consisting of the triplet
\begin{IEEEeqnarray}{c}
     \mathscr{C}_0 = \left( 2\cos\left(\omega_\Delta t\vertOne\right),~~      \omega_0,~~ 0 \vertOne\right)
\end{IEEEeqnarray}
where $\omega_0 = \frac{1}{2}(\omega_{\mathrm{a}} + \omega_{\mathrm{b}})$ and $\omega_\Delta = \frac{1}{2}(\omega_{\mathrm{b}} - \omega_{\mathrm{a}})$. This parameter set defines the IS
\begin{IEEEeqnarray}{rCl} 
    \mathcal{S}_2(t,\omega) &=& \sqrt{2\pi} 2\cos\left(\omega_\Delta t\vertOne\right) \exp(\,\ju \omega_0 t ) \delta(\omega-\omega_0)
\end{IEEEeqnarray}
and the signal
\begin{IEEEeqnarray}{rCl}\IEEEyesnumber
	z_2 (t) &=&  2\cos\left(\omega_\Delta t\vertOne\right) \eu^{\,\ju\omega_0 t },\label{eq:EXz1a}
\end{IEEEeqnarray}
which is the same sinusoidal AM signal, i.e.~$z_1(t) = z_2(t)$, even though $\mathcal{S}_1(t,\omega)\neq\mathcal{S}_2(t,\omega)$.


Table \ref{tab:ISs Sinusoidal AM Signal} provides the true ISs (synthesis) and the ISs for each of the analyses.  For the sinusoidal AM signal, the IS associated with FD analysis perfectly estimates the true IS if the signal was synthesized as in case 1.  On the other hand, the IS associated with monocomponent analysis perfectly estimates the true IS if the signal was synthesized as in case 2. For the IS associated with TD analysis the energy oscillates in time at the beat frequency across all frequencies. To understand the IS associated with fractional Fourier analysis we consider magnitude of $Z_\gamma(u)$ in Table \ref{tab:ISs Sinusoidal AM Signal}. In this case, the magnitude $|Z_\gamma(u)|$ is a cosine-modulated function $2 |\cos(\phi_\Delta(u))|$ as the result of the interference of the two distinct complex exponential terms. 

To understand the IS associated with (synchrosqueezed) STFT analysis in Table \ref{tab:ISs Sinusoidal AM Signal}, consider choices of $\alpha$ and $\omega_\Delta$:
\begin{enumerate}

\item When $\alpha\rightarrow 0$, the convolution in \eqref{eqn:STFTanal} acts as a perfectly selective frequency filter (a narrow band-pass filter centered at $\nu$). Thus, the IS tends toward the IS for FD analysis
\begin{IEEEeqnarray}{rCl} 
     \mathcal{S}^\text{\sc  stft}(t,\omega) \approx   \sqrt{2\pi}\big[ 
    \eu^{\,\ju\omega_\mathrm{a} t} \delta(\omega-\omega_\mathrm{a})
    + \eu^{\,\ju\omega_\mathrm{b} t}\delta(\omega-\omega_\mathrm{b})\big]\nonumber\\
\end{IEEEeqnarray}

\item When $\alpha\rightarrow \infty$, the window $\w(t)$ becomes very narrow in time and the convolution in \eqref{eqn:STFTanal} is approximately 
\begin{IEEEeqnarray}{rCl} 
    Z_\w(t;\nu) \approx z(t) * \delta(t) = z(t).
\end{IEEEeqnarray}
Thus, taking the integral in \eqref{eq:STFTis} over the channelizer frequency from $\nu=-\myUpsilon$ to $\nu=\myUpsilon$, the IS tends toward
\begin{IEEEeqnarray}{rCl} 
     \mathcal{S}^\text{\sc  stft}(t,\omega) \approx   \sqrt{2\pi}4\myUpsilon \cos\left(\omega_\Delta t\vertOne\right) \exp(\,\ju\omega_0 t) \delta(\omega-\omega_0 ).\nonumber\\
\end{IEEEeqnarray}

\item When $0<\alpha<\infty$, we consider the relative size of  $\omega_\Delta$. First, note that 
\begin{IEEEeqnarray}{rCl} 
    Z_\w(t;\nu) = C_{\mathrm{a}}(\nu) \eu^{\,\ju\omega_\mathrm{a} t} + C_{\mathrm{b}}(\nu) \eu^{\,\ju\omega_\mathrm{b} t}.
\end{IEEEeqnarray}

~~a) When the bandwidth associated with the choice of $\alpha$ is small compared to $\omega_\Delta$, i.e.~the two frequency components of $z(t)$ are widely separated. As a result, $C_{\mathrm{a}}(\nu)$ will be large relative to $C_{\mathrm{b}}(\nu)$ when $\nu$ is near $\omega_\mathrm{a}$ (and vice versa), and where the window has allowed spectral leakage. Even though this leakage is well known in STFT analysis, we note that the synchrosqueezing reassigns the energy in each channel to the IF, greatly reducing the leakage. This can be seen as follows. When $\nu \approx \omega_\mathrm{a}$, $\frac{\dd}{\dd t}\theta_\w(t;\nu) \approx \omega_0 - \omega_\Delta =\omega_\mathrm{a}$ and when $\nu\approx\omega_\mathrm{b}$, $\frac{\dd}{\dd t}\theta_\w(t;\nu) \approx \omega_0 + \omega_\Delta =\omega_\mathrm{b}$. Thus, the IS tends to the IS for FD analysis
\begin{IEEEeqnarray}{rCl} 
     \mathcal{S}^\text{\sc  stft}(t,\omega) &\approx& A_{\mathrm{a}} \eu^{\,\ju\omega_\mathrm{a} t}\delta(\omega-\omega_\mathrm{a}) + A_{\mathrm{b}} \eu^{\,\ju\omega_\mathrm{b} t}\delta(\omega-\omega_\mathrm{b})\nonumber\\
\end{IEEEeqnarray}
where $A_{\mathrm{a}}$ and $A_{\mathrm{b}}$ are constants.

~~b) When the bandwidth associated with the choice of $\alpha$ is large compared to $\omega_\Delta$, 
$C_{\mathrm{a}}(\nu) \approx C_{\mathrm{b}}(\nu) \approx C$ and
\begin{IEEEeqnarray}{rCl} \IEEEyesnumber
    Z_\w(t;\nu) &\approx& C \eu^{\,\ju\omega_\mathrm{a} t} + C \eu^{\,\ju\omega_\mathrm{b} t}\IEEEyessubnumber\\
    &=& 2C \cos(\omega_\Delta t) \eu^{\,\ju\omega_0 t}.\IEEEyessubnumber
\end{IEEEeqnarray}
The synchrosqueezing can be understood as follows. The IF $\frac{\dd}{\dd t}\theta_\w(t;\nu)$ in Table \ref{tab:ISs Sinusoidal AM Signal} oscillates around $\omega_0$ due to the $\cos(2\omega_\Delta t)$ term. When $\omega_\Delta$ is small, $\frac{\dd}{\dd t}\theta_\w(t;\nu)$ approaches $\omega_0$. Thus, the IS tends to the IS for monocomponent analysis
\begin{IEEEeqnarray}{rCl} 
    \mathcal{S}^\text{\sc  stft}(t,\omega) \approx A \cos(\omega_\Delta t) \eu^{\,\ju\omega_0 t}\delta(\omega-\omega_0)
\end{IEEEeqnarray}
where $A$ is a constant.

\end{enumerate}








\color{black}

\subsection{Summary}
In all these examples, there is at least one analysis that has an associated IS which perfectly estimates the true IS. Thus, demonstrating that if components are identified correctly, one can provide an exact time-frequency localization. The monocomponent analysis always produced a perfect estimate of the IS because with the exception of the unit impulse signal, all examples were monocomponent AM--FM signals. More generally, when the template is well matched to the signal synthesis, the estimated IS was generally better than if the template is not well matched. Additionally, when the template is not matched to the signal synthesis, all analyses spread energy across the time-frequency plane with varying degrees of localization. 

As noted in Section \ref{sec:ISA}, an infinite number of instantaneous spectra and parameter sets map to the same signal with \eqref{eq:HSintOmega}. Therefore, one cannot use $z(t)$ to help choose among the parameter sets and thus some criterion is required in order to obtain the true IS. As pointed out by Ville, ``The instantaneous spectrum may only be determined, physically, to some approximation. But another question is presented to us: approximation \textit{to what}?\cite{ville}''. \color{black} This is illustrated with the sinusoidal AM example where the ``correct'' IS is tied to the parameter set used to synthesize the signal. Moreover, with knowledge of ground truth (i.e.~the component set used for signal synthesis), the ISA framework provides a means to construct \color{black} the true IS and quantify the approximation error of the estimated IS through 
\begin{IEEEeqnarray}{rCl} 
    \mathrm{error}(t,\omega) &=&  \mathcal{S}(t,\omega)-\hat{\mathcal{S}}(t,\omega)
\end{IEEEeqnarray}
where $\hat{\mathcal{S}}(t,\omega)$ is an IS estimate as in Figure \ref{fig:ISAsummary}. Note that such a comparison is not considered in the usual treatment of time-frequency analysis, because most analyses do not specify the true (synthesis) IS. Moreover, as we have shown, an IS estimate can come from any of the common analyses or from a different analysis.

As pointed out by Meyer \cite{meyer} while reflecting on the work by Ville \cite{ville}:
\begin{quote}
    The set of all time-frequency atoms is a collection of elementary signals much too large to provide a unique representation of a signal as a linear combination of time-frequency atoms. Each signal admits an infinite number of representations, and this leads us to choose the best among them according to some criterion. This criterion might be the one suggested by Ville: The decomposition of a signal in time-frequency atoms is related to a synthesis, and this synthesis ought logically to be done in accordance with an analysis. [...] However, Ville did not explain how the results of the analysis could lead to an effective synthesis.
\end{quote}
In this work, we have shown for several special cases of the quadratic chirplet, how to relate an analysis to a synthesis. Moreover, because our work begins with the true IS $\mathcal{S}(t,\omega;\mathscr{S})$ synthesized from time-frequency atoms from which the signal $z(t;\mathscr{S})$ is projected, we can assess the estimation error of any IS which arises from analysis.

\section{Conclusions} \label{sec:conclusion}
In this paper, we build upon our prior instantaneous spectral analysis (ISA) theory to unify common signal analyses with time-frequency atoms. The analyses under consideration included time domain analysis, frequency domain analysis, synchrosqueeezed short-time Fourier transform, fractional Fourier analysis, and synchrosqueeezed short-time fractional Fourier transform. In particular, we defined a quadratic chirplet component and atom with two parameters: an effective duration, $\alpha$ and a chirp rate, $r$. Appropriate choices for $\alpha$ and $r$ specify a template associated with one of the analyses considered, thus unifying these analyses. Depending on the analysis, the signal is either projected onto the family of components or the signal is filtered using the family of components as impulse responses for each channel. Each projection or channelized signal is demodulated to form an atom and the atoms are superimposed to form an IS. This process was repeated for each analysis, leading to closed form expressions for the IS corresponding to the various analyses under consideration. In the case of time domain analysis, we defined a limiting form of the quadratic chirplet atom, termed the Shannon atom, to obtain the atom corresponding to a unit impulse.

For several example signals we specified a component triplet, constructed the IS, and synthesized the signal. This IS serves as ground truth and can be used as a reference when comparing to estimated ISs---an advantage unique to ISA when compared to other time-frequency analysis methods. To demonstrate analysis, we apply the expressions developed for the IS corresponding to each common analysis method, to generate estimates of the IS. With knowledge of ground truth, we show that for these examples there is at least one analysis method for which the true IS can be estimated exactly. Generally, there are an infinite number of ISs, including those which can be parameterized by $\alpha$ and $r$, all of which project the signal $z(t)$. Without synthesis information, we cannot say that one IS is a better estimate than another.





\appendices



\ifCLASSOPTIONcaptionsoff
  \newpage
\fi

\bibliographystyle{IEEEtran}
\bibliography{main}

@Preamble{ " \newcommand{\noop}[1]{} " }

@book{carmona1998practical,
  title={Practical Time-Frequency Analysis: Gabor and Wavelet Transforms, with an Implementation in S},
  author={Carmona, Ren{\'e} and Hwang, Wen-Liang and Torresani, Bruno},
  volume={9},
  year={1998},
  publisher={Academic Press}
}

@book{hlawatsch2013time,
  title={{Time-Frequency Analysis}},
  author={Hlawatsch, Franz and Auger, Fran{\c{c}}ois},
  year={2013},
  publisher={John Wiley \& Sons}
}

@article{moca2021time,
  title={{Time-Frequency Super-Resolution with Superlets}},
  author={Moca, Vasile V and B{\^a}rzan, Harald and Nagy-D{\u{a}}b{\^a}can, Adriana and Mureșan, Raul C},
  journal={{Nat.~Commun.}},
  volume={12},
  number={1},
  pages={337},
  year={2021},
  publisher={Nature Publishing Group UK London}
}

@book{Ozaktas2001fractionalfourier,
  title={{The Fractional Fourier Transform: with Applications in Optics and Signal Processing}},
  author={H~M.~Ozaktas and Z.~Zalevsky and M.~A.~Autay },
  year={2001},
  publisher={Wiley}
}

@incollection{ozaktas1999introduction,
  title={{Introduction to the Fractional Fourier Transform and its Applications}},
  author={Ozaktas, Haldun M and Kutay, M Alper and Mendlovic, David},
  booktitle={{Adv.~ Imaging Electron Phys.}},
  volume={106},
  pages={239--291},
  year={1999},
  publisher={Elsevier}
}

@article{shi2020novel,
  title={{Novel Short-Time Fractional Fourier Transform: Theory, Implementation, and Applications}},
  author={Shi, Jun and Zheng, Jiabin and Liu, Xiaoping and Xiang, Wei and Zhang, Qinyu},
  journal={{IEEE Trans.~Signal Process.}},
  volume={68},
  pages={3280--3295},
  year={2020},
  publisher={IEEE}
}

@article{namias1980fractional,
  title={{The Fractional Order Fourier Transform and its Application to Quantum Mechanics}},
  author={Namias, Victor},
  journal={IMA J.~Appl.~Math.},
  volume={25},
  number={3},
  pages={241--265},
  year={1980},
  publisher={Oxford University Press}
}

@article{mcbride1987namias,
  title={{On Namias's Fractional Fourier Transforms}},
  author={McBride, AC and Kerr, FH},
  journal={IMA J.~Appl.~Math.},
  volume={39},
  number={2},
  pages={159--175},
  year={1987},
  publisher={Oxford University Press}
}

@inproceedings{almeida1993introduction,
  title={{An Introduction to the Angular Fourier Transform}},
  author={Almeida, Lu{\'\i}s B},
  booktitle={{Proc.~IEEE Int.~Conf.~Acoust.~Speech Signal Process.~(ICASSP)}},
  volume={3},
  pages={257--260},
  year={1993}
}

@article{condon1937immersion,
  title={{Immersion of the Fourier Transform in a Continuous Group Of Functional Transformations}},
  author={Condon, EU},
  journal={{Proc.~Natl.~Acad.~Sci.}},
  volume={23},
  number={3},
  pages={158--164},
  year={1937},
  publisher={National Acad Sciences}
}

@article{kodera1978analysis,
  title={{Analysis of Time-Varying Signals with Small BT Values}},
  author={K.~Kodera and R.~Gendrin and C.~DeVilledary},
  journal={{IEEE Trans.~Acoust., Speech, Signal Process.}},
  volume={26},
  number={1},
  pages={64--76},
  year={1978},
  publisher={IEEE}
}

@article{auger1995improving,
  title={{Improving the Readability of Time-Frequency and Time-Scale Representations by the Reassignment Method}},
  author={F.~Auger and P.~Flandrin},
  journal={{IEEE Trans.~Signal Process.}},
  volume={43},
  number={5},
  pages={1068--1089},
  year={1995},
  publisher={IEEE}
}

@article{meignen2019synchrosqueezing,
  title={{Synchrosqueezing Transforms: from Low- to High-Frequency Modulations and Perspectives}},
  author={S.~Meignen and T.~Oberlin and D.~Pham},
  journal={Comptes Rendus Physique},
  volume={20},
  number={5},
  pages={449--460},
  year={2019},
  publisher={Elsevier}
}

@book{boashash2015time,
  title={{Time-Frequency Signal Analysis and Processing: a Comprehensive Reference}},
  author={Boashash, Boualem},
  year={2015},
  publisher={Academic press}
}

@book{flandrin2018explorations,
  title={{Explorations in Time-Frequency Analysis}},
  author={Flandrin, Patrick},
  year={2018},
  publisher={Cambridge University Press}
}

@book{grochenig2001foundations,
  title={{Foundations of Time-Frequency Analysis}},
  author={Gr{\"o}chenig, Karlheinz},
  year={2001},
  publisher={Springer Science \& Business Media}
}

@article{ISA2018_Sandoval,
           title = {{The Instantaneous Spectrum: A General Framework for 
                    Time-Frequency Analysis}},
           author = {S.~Sandoval and P.~L.~{De~Leon}},
           journal = {{IEEE Trans.~Sig.~Process.}},
           volume = {66},
           year = {2018},
           month = {Nov},
           pages = {5679-5693} 
}

@book{bremaud2002mathematical,
  title={{Mathematical Principles of Signal Processing: Fourier and Wavelet Analysis}},
  author={Br{\'e}maud, Pierre},
  year={2002},
  publisher={Springer}
}

@book{lim1987advanced,
  title={{Advanced Topics in Signal Processing}},
  author={J.~S.~Lim  and A.~V.~Oppenheim},
  year={1987},
  publisher={Prentice-Hall, Inc.}
}

@article{kodera1976new,
  title={{A New Method for the Numerical Analysis of Non-Stationary Signals}},
  author={K.~Kodera and C.~DeVilledary and R.~Gendrin},
  journal={Phys.~Earth Planet.~Inter.},
  volume={12},
  number={2-3},
  pages={142--150},
  year={1976},
  publisher={Elsevier}
}

@article{auger1994and,
  title={{The Why and How of Time-Frequency Reassignment}},
  author={F.~Auger and P.~Flandrin},
  journal={{Proc.~IEEE Sym.~Time-Frequency Time-Scale Anal.}},
  pages={197--200},
  year={1994},
}

@article{auger2013time,
  title={{Time-Frequency Reassignment and Synchrosqueezing: An Overview}},
  author={F.~Auger and P.~Flandrin and Y.~Lin and S.~McLaughlin and S.~Meignen and T.~Oberlin and H.~Wu},
  journal={{IEEE Signal Process.~Mag.}},
  volume={30},
  number={6},
  pages={32--41},
  year={2013}
}

@article{mann1995chirplet,
  title={{The Chirplet Transform: Physical Considerations}},
  author={S.~Mann and S.~Haykin},
  journal={{IEEE Trans.~Sig.~Process.}},
  volume={43},
  number={11},
  pages={2745--2761},
  year={1995},
  publisher={IEEE}
}

@article{sandoval2022isa,
  title={{ISA.jl: Instantaneous spectral Analysis in Julia}},
  author={Sandoval, Steven and Alshammari, Hasan and Dalal, Mamta},
  journal={SoftwareX},
  volume={20},
  pages={101239},
  year={2022},
  publisher={Elsevier}
}

@article{sandoval2022recasting,
  title={{Recasting the (Synchrosqueezed) Short-Time Fourier Transform as an Instantaneous Spectrum}},
  author={Sandoval, Steven and De Leon, Phillip L},
  journal={Entropy},
  volume={24},
  number={4},
  pages={518},
  year={2022},
  publisher={MDPI}
}

@book{mallat1999wavelet,
  title={{A Wavelet Tour of Signal Processing}},
  author={S.~Mallat},
  year={1999},
  publisher={Academic press}
}

@book{stankovic2014time,
  title={{Time-Frequency Signal Analysis with Applications}},
  author={L.~Stankovi{\'c} and M.~Dakovi{\'c} and T.~Thayaparan},
  year={2014},
  publisher={Artech house}
}

@book{flandrin1998time,
  title={{Time-Frequency/Time-Scale Analysis}},
  author={P.~Flandrin},
  volume={10},
  year={1998},
  publisher={Academic press}
}

@inproceedings{ASRU2015,
  title={{Hilbert Spectral Analysis of Vowels using Intrinsic Mode Functions}},
  author={S.~Sandoval and P.~L.~{De Leon} and J.~M.~Liss},
  booktitle={{IEEE Workshop on Automatic Speech Recognition and Understanding (ASRU)}},
  pages={1-5},
  year={2015}
}

@article{bedrosian1962analytic,
	title={{The Analytic Signal Representation of Modulated Waveforms}},
	author={E.~Bedrosian},
	journal={{Proc.~IRE}},
	volume={50},
	number={10},
	pages={2071-2076},
	year={1962},
	publisher={IEEE}
}

@book{bracewell1980fourier,
  title={{The Fourier Transform and Its Applications}},  author={R.~Bracewell},
  year={1980},
  publisher={McGraw-Hill},
  page = {359}
}

@book{boashash2003time,
  title={{Time Frequency Signal Analysis and Processing}},
  editor={B.~Boashash},
  year={2003},
  publisher={Elsevier}
}

@book{cohen1995time,
  title={{Time-Frequency Analysis}},
  author={L.~Cohen},
  year={1995},
    publisher={Prentice Hall}
}

@Article{ville,
   author  =  "J. Ville",
   title   = {{Theorie et Applications de la Notion de Signal Analytique}},
   journal = "Cables et Transmission",
   year    =  "1948",
   volume  = "2a",
   pages   = "61-74"}

@book{hoskins2009delta,
  title={{Delta Functions: Introduction to Generalised Functions}},
  author={Hoskins, Roland F},
  year={2009},
  publisher={Elsevier},
  edition = {Second}
}

@book{robinson1974non,
  title={{Non-Standard Analysis}},
  author={Robinson, Abraham},
  year={1974},
  publisher={Princeton University Press}
}

@ARTICLE{gaborTOC, 
author={D.~Gabor}, 
journal={{J.~Inst.~Electr.~Eng.~3}}, 
title={{Theory of Communication. Part 1: The Analysis of Information}},
year={1946}, 
month={Nov.}, 
volume={93}, 
number={26}, 
pages={429-441}
}

@inproceedings{xia1999analytic,
  title={{On Analytic Signals with Nonnegative Instantaneous Frequency}},
  author={X.~G.~Xia and L.~Cohen},
booktitle={{Proc.~IEEE Int.~Conf.~Acoust.~Speech Signal Process.}}, 
  year={1999},
  pages={1329-1332}
}

@ARTICLE{Allen1977ShortTime, 
author={J.~B.~Allen and L.~Rabiner}, 
journal={{Proc.~IEEE}}, 
title={{A Unified Approach to Short-Time Fourier Analysis and Synthesis}}, 
year={1977}, 
month={Nov.}, 
volume={65}, 
number={11}, 
pages={1558-1564}
}

@book{papandreou2002applications,
	title={{Applications in Time-Frequency Signal Processing}},
	editor={A.~Papandreou-Suppappola},
	year={2002},
	publisher={CRC press}
}

@book{meyer,
  title={{Wavelets: Algorithms \& Applications}},
  author={Y.~Meyer},
  publisher={Philadelphia: SIAM (Society for Industrial and Applied Mathematics},
  year={1993}
}

\section*{Biography}

\begin{IEEEbiography}[{\includegraphics[width=1in,height=1.25in,clip,keepaspectratio]{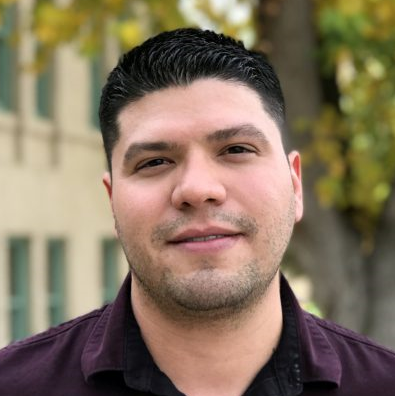}}]{Steven Sandoval} received the B.S.~Electrical Engineering and M.S.~Electrical Engineering from New Mexico State University in 2007 and 2010 respectively, and the Ph.D.~degree in Electrical Engineering from Arizona State University in 2016. Currently, he is at the Klipsch School of Electrical and Computer Engineering at New Mexico State University. His research interests include audio and speech processing, time-frequency analysis, and machine learning.\end{IEEEbiography}

\begin{IEEEbiography}[{\includegraphics[width=1in,height=1.25in,clip,keepaspectratio]{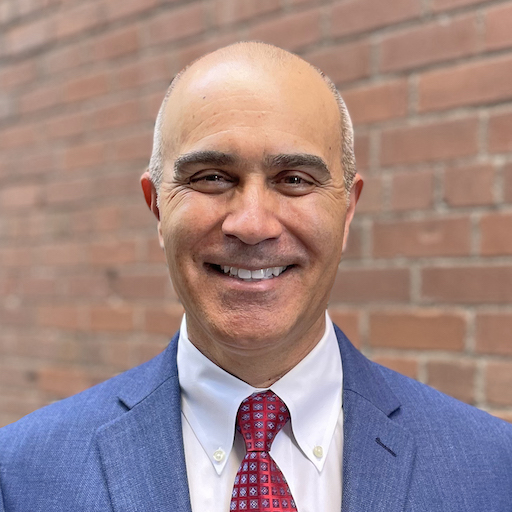}}]{Phillip L.~De Leon} (SM'03) received the B.S.~Electrical Engineering and the B.A.~in Mathematics from the University of Texas at Austin, in 1989 and 1990 respectively and the M.S.~and Ph.D.~degrees in Electrical Engineering from the University of Colorado at Boulder, in 1992 and 1995 respectively. From 1996-2022, De Leon served as Professor in the Klipsch School of Electrical and Computer Engineering at New Mexico State University and in 2015, was awarded the Klipsch Distinguished Professorship. From 2016-2019, De Leon served as Associate Dean of Research in the College of Engineering and from 2019-2022, served as Associate Vice President for Research and Chief Science Officer. Since 2022, De Leon has served as Professor in the Department of Electrical Engineering and as Associate Vice Chancellor for Research and Chief Research Officer at the University of Colorado Denver. His research interests include audio and speech processing, machine learning, and time-frequency analysis.\end{IEEEbiography}

\clearpage

\thispagestyle{empty}


\thispagestyle{empty}

\end{document}